\documentclass[preprintnumbers,amsmath,amssymb,floatfix,nofootinbib]{revtex4}

\usepackage{amsmath,amsfonts,bbm,euscript,hyperref}
\usepackage{graphicx}
\usepackage{dcolumn}
\usepackage{bm}
\usepackage{epsfig}
\epsfclipon

\newcommand{\lsim}{\mbox{\raisebox{-.6ex}{~$\stackrel{<}{\sim}$~}}}
\newcommand{\gsim}{\mbox{\raisebox{-.6ex}{~$\stackrel{>}{\sim}$~}}}

\def\identity{\mathbbm{1}}

\def\sH{\mathcal{H}}

\begin{document}

\preprint{UMN-TH-3107/12, MAD-TH-12-05}

\title{Gravity waves and non-Gaussian features from particle production in a sector gravitationally coupled to the inflaton}

\author{Neil Barnaby$^1$, Jordan Moxon$^2$, Ryo Namba$^1$, Marco Peloso$^1$, Gary Shiu$^{2,3,4}$, Peng Zhou$^2$}

\affiliation{
$^1$  School of Physics and Astronomy,
University of Minnesota, Minneapolis, 55455, USA\\
$^2$ Department of Physics, University of Wisconsin, 
Madison, WI 53706, USA \\
$^3$ Institute for Advanced Study, Hong Kong University of Science and Technology, Hong Kong  \\
$^4$ Institute for Theoretical Physics, University of Amsterdam, Amsterdam, the Netherlands}

\date{\today}

\begin{abstract} 
We study the possibility that particle production during inflation could source observable gravity waves on scales relevant for Cosmic Microwave Background experiments.  A crucial constraint on such scenarios arises because particle production can also source inflaton perturbations, and might ruin the usual predictions for a nearly scale invariant spectrum of nearly Gaussian curvature fluctuations.  To minimize this effect, we consider two models of particle production in a sector that is only gravitationally coupled to the inflaton.  For a single instantaneous burst of massive  particle production, we find that localized features in the scalar spectrum and bispectrum might be observable, but gravitational wave signatures are unlikely to be detectable (due to the suppressed quadrupole moment of non-relativistic quanta) without invoking some additional effects.  We also consider a model with a rolling pseudoscalar that leads to a continuous production of relativistic gauge field fluctuations during inflation.  Here we find that gravitational waves from particle production can actually \emph{exceed} the usual inflationary vacuum fluctuations in a regime where non-Gaussianity is consistent with observational limits.  In this model observable B-mode polarization can be obtained for \emph{any} choice of inflaton potential, and the amplitude of the signal is not necessarily correlated  with the scale of inflation.
\end{abstract}

\maketitle

\begin{widetext}
\tableofcontents\vspace{5mm}
\end{widetext}

\section{Introduction} 
\label{sec:introduction}

Inflation is currently the standard paradigm for solving the flatness, horizon and the unwanted topological relic problems in standard big bang cosmology.  In addition, the quantum fluctuation of the inflaton field provides the seeds of inhomogeneity that seem to fit very well with the current observation on the Cosmic Microwave Background (CMB) and Large Scale Structure (LSS).  A particularly important probe of the physics of inflation is provided by primordial gravitational waves (tensor fluctuations).  A number of observational searches for gravitational wave perturbations are proposed or are currently underway; these may be probed through the B-mode polarization of the CMB \cite{Baumann:2008aq,Bock:2009xw,Farhang:2011ud} or, on smaller scales, through interferometers such as LIGO \cite{LIGO}, VIRGO \cite{VIRGO},  DECIGO \cite{Kawamura:2011zz}, Einstein Telescope \cite{ET}, or LISA \cite{LISA}.

During inflation, gravitational wave fluctuations are inevitably generated by quantum fluctuations of the tensor part of the metric.  These have an amplitude controlled by $\frac{H^{2}}{M_{p}^{2}}$, where $H$ is the Hubble scale and $M_p\approx 2.4\cdot 10^{18}\,\mathrm{GeV}$.  The gravitational wave signal from vacuum fluctuations is detectable only when the inflaton field range is trans-Planckian \cite{Lyth:1996im}, which might be challenging to realize in a controlled effective field theory.  

Additional sources of Gravitational Waves (GW), which are uncorrelated with the usual quantum vacuum fluctuations, may be present in the early universe (see, for instance, \cite{Binetruy:2012ze} for a recent review).  Two broad categories of mechanisms are:
\begin{enumerate}
 \item {\bf Models involving phase transitions:} For example, in first order phase transitions, vacuum bubble collisions \cite{GWfromPT, Chialva:2010jt} and the subsequent turbulence \cite{GWfromPTturb} could source GW.  The generation and decay of cosmic strings can also give rise to large GW \cite{GWfromCStr}. Along similar lines, the self-ordering of a scalar field after a second order phase transition has also been considered \cite{GWfrom2ndPT}.
 \item {\bf Models involving particle production:} Generically the inflaton should be expected to couple to some additional degrees of freedom, as would seem to be  necessary for successful reheating \cite{Kofman:1994rk,Kofman:1997yn,Podolsky:2005bw,Barnaby:2009wr,Braden:2010wd}.  In this case, there is a natural possibility that the time-dependence of the inflaton condensate during inflation leads to the production of some other degrees of freedom which may, in turn, provide an important new source of GW \cite{Cook:2011hg,Senatore:2011sp,Barnaby:2011qe}.  A variety of different models have been proposed; see below for more discussion.\footnote{For gravitational waves from particle production at the \emph{end} of inflation, see \cite{Easther:2006vd,GarciaBellido:2007dg,GarciaBellido:2007af,Dufaux:2007pt,Dufaux:2008dn}; effects on the scalar fluctuations were discussed in \cite{Barnaby:2006cq,Barnaby:2006km,Chambers:2007se, Bond:2009xx}.}
\end{enumerate}
Given the importance of gravitational waves as a probe of inflation, it is important to understand if such mechanisms could be competitive with the usual spectrum of GW from vacuum fluctuations.~\footnote{A GW signal may also be left at the largest scales as an imprint of a pre-inflationary era if inflation had only a minimal duration \cite{Gumrukcuoglu:2008gi}.}

Any mechanism which is being invoked to source gravitational waves might also source scalar metric perturbations.  Therefore, one must take care not to spoil the usual prediction of a nearly scale invariant spectrum of Gaussian scalar curvature perturbations.  Sometimes this concern is evaded by restricting attention to effects that take place on the small scales relevant for interferometers where the scalar fluctuations are not strongly constrained.  In this work we will mostly be interested in models of particle production during inflation, where gravitational waves are sourced on CMB scales.  To test the feasibility of such scenarios, it is crucial to study also the spectrum and bispectrum of the scalar fluctuations, to ensure that these are consistent with observations.

Models of particle production during inflation have received considerable attention in the literature; see for example  \cite{Cook:2011hg,Senatore:2011sp,Barnaby:2011qe,Berera:1995ie,Gupta:2002kn,Moss:2007cv,Chung:1999ve,Romano:2008rr,Barnaby:2009mc,Barnaby:2009dd,Barnaby:2010ke,Barnaby:2010sq,Anber:2009ua,Barnaby:2010vf,Barnaby:2011vw,Sorbo:2011rz,Green:2009ds,LopezNacir:2011kk}.  One class of models involves instantaneous bursts of particle production, leading to localized features in the cosmological perturbations.  Early studies focused on the production of fermion \cite{Chung:1999ve} or scalar \cite{Romano:2008rr} particles, neglecting the feed-back of the produced quanta on the perturbations of the inflaton.  In Ref.~\cite{Barnaby:2009mc} it was however shown that this feed-back effect actually dominates observables for the case of scalar particle production.  Ref.~\cite{Barnaby:2009mc} considered a simple model where scalar $\chi$ particles are produced through a coupling $g^2\varphi^2 \chi^2$ to the inflaton, $\varphi$, using both lattice field theory simulations and also analytical methods.  Models of this same type were subsequently analyzed in the context of trapped inflation \cite{Green:2009ds}.  More recently, models of instantaneous vector particle production have been studied in connection with GW at interferometer scales \cite{Cook:2011hg}. Several scenarios of GW from scalar and string production were instead studied in \cite{Senatore:2011sp}.

Another possibility is that particle production occurs continuously during inflation.  This is quite natural in the context of axion inflation \cite{Barnaby:2010vf}.  Even in the simplest models of inflation driven by a single axion in slow roll on a smooth flat potential, pseudoscalar couplings $\varphi F\tilde{F}$ to gauge fields are ubiquitous.  This interaction leads to a continuous tachyonic production of gauge field fluctuations during inflation.  There are a host of interesting phenomenological signatures: observable equilateral non-Gaussianity \cite{Barnaby:2010vf,Barnaby:2011vw,Barnaby:2011pe}, GW at interferometer scales \cite{Cook:2011hg,Barnaby:2011qe}, and excess power at small scales \cite{Meerburg:2012id,Chluba:2012we}.  Additionally, backreaction effects in such models can assist inflation by dissipating the kinetic energy of the inflaton \cite{Anber:2009ua,Anber:2012du}, analogously to warm inflation \cite{Berera:1995ie} and trapped inflation \cite{Green:2009ds}.~\footnote{The pseudo-scalar interaction is also used to \cite{Adshead:2012kp} to dissipate the inflaton kinetic energy from its classical interaction to a non-abelian vector field.}  In~\cite{Seery:2008ms,Caldwell:2011ra,Barnaby:2012tk} a model of gauge field production through the scalar coupling $I^2(\varphi) F^2$ was considered, in connection with primordial non-Gaussianity and, perhaps, magnetogenesis (this last application is problematic \cite{Demozzi:2009fu,Barnaby:2012tk}). 

Above we discussed some simple models of particle production during inflation where quanta are produced by a \emph{direct coupling} between the inflaton and some additional degrees of freedom.  In such a scenario the produced particles interact with the inflaton through couplings that are typically much stronger than gravitational.  Hence, these produced particles will tend to source scalar curvature fluctuations much more efficiently than GW.  It is not surprising, therefore, that the most stringent CMB constraints on such models often come from features and non-Gaussianity in the scalar fluctuations, rather than from GW.  

In this work, we investigate the possibility to source a significant GW signal on CMB scales without ruining the spectrum of curvature fluctuations.  To minimize the impact on the inflaton fluctuations, we assume that particle production takes place in a ``hidden'' sector that is only gravitationally coupled to the inflaton.\footnote{Our scenario differs from the curvaton model \cite{Lyth:2002my}, in which the inflaton provide the energy density to support inflation while a light curvaton field provides the seed of scalar perturbation. The GW spectrum in the curvaton model has been considered at tree-level \cite{Fonseca:2011aa,Nakayama:2009ce} and at 1-loop level \cite{Bartolo:2007vp}, with possible detectability at future interferometer experiments \cite{Bartolo:2007vp}. See \cite{vector-curvaton} for  examples of vector curvaton.} We compute for the first time the scalar perturbations induced by the gravitational coupling, and we compare their phenomenological impact with that of the produced GW.  Since gravitational interactions are unavoidable, the case we investigate can be thought of as a kind of ``best case'' scenario for the production of GW while minimizing the effect on the scalar fluctuations.  Concretely, we will focus on two models involving the production of spin-$1$ particles:
\begin{enumerate}
 \item {\bf Model I:} Particle production takes place in a hidden sector where a local $U(1)$ invariance is spontaneously broken by the expectation value of some complex scalar $\psi$.  Here gauge fields are produced at an isolated moment - namely, when the mass of the gauge field crosses zero during the evolution of $\psi$ -   and we have localized features in the scalar spectrum and bispectrum, in addition to a localized feature in the tensor spectrum.  For a single isolated burst of production, we find that observational constraints on the scalar perturbations exclude any interesting effect in GW.  However, in a concrete model there may be many bursts of particle production and their resonance could enhance the effect.  We note that this model could lead to interesting phenomenological signatures in the scalar fluctuations; localized features in the spectrum and bispectrum are discussed. 
 \item {\bf Model II:} Particle production takes place in a hidden sector where a rolling pseudoscalar sources gauge field fluctuations continuously.  Here we find that GW from particle production can be competitive with the vacuum fluctuations, or even larger than the vacuum fluctuations, \emph{without} violating observational bounds on non-Gaussianity.  In this model the primordial tensor spectrum can be detectable for \emph{any} choice of inflaton potential (that is, also for small field inflation).  In this model, GW from particle production are chiral.  We show that parity violation in the tensor sector can be almost maximal, which provides a distinctive observable signature of this scenario; see also \cite{Sorbo:2011rz,Anber:2012du}. 
 \end{enumerate}

In summary: we find that the possibility to source interesting GW from particle production is rather model dependent.  Even if particle production occurs in a hidden sector, coupled only gravitationally to the inflationary one, this is not a sufficient condition to ensure that observable GW can be sourced without ruining the scalar spectrum.  On the other hand, we find that in some models GW from particle production can actually \emph{exceed} the usual vacuum fluctuations.  The reason for the smaller GW production in Model I with respect to Model II is that in Model I the gauge quanta are highly non-relativistic after their production, which suppresses their quadrupole moment. We verified that the same suppression takes place if, instead of gauge fields, the particles produced in this mechanism have spins $0$ or $\frac{1}{2}$.~\footnote{We find that the results agree in all three cases, with only order unity differences coming from counting the number of degrees of freedom and also from spin statistics.  These order unity effects can be encoded in a very simple formula, which we present.} As we mentioned, a greater GW effect may be obtained for multiple instances of particle production, or for a more complicated evolution of the gauge field mass, or if the massive particles decay into massless ones short after they are produced \cite{Senatore:2011sp}.

Therefore, due to the possibility of GW from particle production, a measurement of primordial B-modes does not necessarily constitute a measurement of the scale of inflation. Nor does a detectable B-mode signal necessarily require super-Planckian excursions in field space. Fortunately, as we discuss below, the produced GW may be distinguished from the vacuum signal, either by their localization at some given wavelength (assuming that Model I can be modified so to enhance the GW signal), or by their violation of parity.

In terms of observational prospects of gravitational waves on CMB scales, Model II is certainly more promising, as the gauge quanta are automatically relativistic and their quadrupole moment is not suppressed. While the focus of the present work is to examine the signatures and constraints of gravitational waves produced during inflation as a result of hidden sector particle production, the UV sensitivity of inflation motivates us to ask whether such models can be realized in string theory.
Indeed, the 4D low energy spectrum of string theory contains a myriad of axion-like fields, e.g., those  arise from the reduction of antisymmetric form fields on cycles of the internal space.
These closed string axions couple to $U(1)$ gauge fields on the worldvolume of D-branes via 
$\varphi F \tilde{F}$ couplings.
The axion decay constant $f$ one typically finds in string theory constructions is of the order of the GUT scale $M_{GUT} \sim 10^{16}$ GeV (see, e.g., \cite{Banks:2003sx,Svrcek:2006yi}) which sits comfortably within the allowed window for consistency of our model (see eqs.~(\ref{total_bnd}) and (\ref{bnd2-f})). It is not difficult to arrange the inflaton to have no direct coupling to the axion and the gauge field.
We outline some ideas to  realize Model II in string theory in the concluding section.
We leave, however, a detailed study of  these and other string theory embeddings  for future work.

This paper is organized as follows.  In Section \ref{sec:formal} we develop a very general formalism which can be used to compute the effects of particle production on the scalar and tensor cosmological perturbation in a variety of models.  In Section \ref{sec:model1} we apply this formalism to Model I, discussed above.  In Section \ref{sec:model2} we instead apply our formalism to Model II, discussed above.  In Section \ref{sec:conclusions}, we conclude.  In Appendix \ref{app-longA}, we discuss the production of the longitudinal mode in Model I.  This is the first time the longitudinal mode has been accounted for in a model of production of massive vector fields during inflation.  In Appendix \ref{app-mod1-sourcez}, we discuss some details of the computation of the scalar perturbations for Model I.  In Appendix \ref{app:modelII}, we perform an analogous computation for Model II. In Appendix \ref{app-fermions}, we discuss the production of massive fermion fields during inflation, and the amount of GW that they generate.
\section{Formal expressions for observable correlators}
\label{sec:formal}

We assume that quanta of a vector field are produced (either at some discrete moment $t_*$, or throughout inflation) in a sector only gravitationally coupled to the inflaton. We are interested in computing scalar and tensor cosmological perturbations sourced by the vector quanta. The vector field 
can be decomposed as
\begin{equation}
 A_0 = 0 \;\;\;,\;\;\; A_i =  \int \frac{d^3 k}{\left( 2 \pi \right)^{3/2}}  {\rm e}^{i \vec k \cdot \vec x} \, {\tilde A}_i \left( \tau ,\, \vec{k} \right) \;\;\;,\;\;\;  {\tilde A}_i  \left( \tau ,\, \vec{k} \right) =  
 \sum_{\lambda = \pm} \epsilon^{(\lambda)}_i  \left( {\hat k} \right)  \left[  \, a_\lambda \left( \vec{k} \right) \, A_\lambda \left( k \right) +  a_\lambda^\dagger \left( - \vec{k} \right) \, A_\lambda^* \left( k \right)   \right] 
\label{A-deco}
 \end{equation}
where we have included only the transverse vector polarization (in  Section \ref{sec:model1}
we actually study also the  longitudinal polarization, as it  is present, and may be relevant, in that model). Here $\vec{\epsilon}_\lambda$ are circular polarization vectors satisfying  $\vec{k}\cdot \vec{\epsilon}^{\;(\pm)}  \left( \vec{k} \right) = 0$, $\vec{k} \times \vec{\epsilon}^{\;(\pm)} \left( \vec{k} \right) = \mp i k \vec{\epsilon}^{\;(\pm)}  \left( \vec{k} \right)$,
$\vec{\epsilon}^{\;(\pm)}  \left( -\vec{k} \right) = \vec{\epsilon}^{\;(\pm)}  \left( \vec{k} \right)^*$, and normalized according to \\$\vec{\epsilon}^{\,(\lambda)} \left( \vec{k} \right)^* 
\cdot \vec{\epsilon}^{\,(\lambda')} \left( \vec{k} \right) = \delta_{\lambda \lambda'}$.  

In this section, we introduce the formalism for the correlators of the scalar and tensor perturbations sourced by this vector vector, under the assumption that the latter is only gravitationally coupled to the inflationary  sector, and that it enters quadratically in the energy-momentum tensor. The discussion is divided in three parts; in Subsection \ref{subsec:formal-deco} we present the structure of the equations for the scalar and tensor perturbations in presence of the source; in Subsection \ref{subsec:formal-sol} we present the formal solutions, and the structures of the correlators; in Subsection \ref{subsec:formal-pheno}  we show how these expressions enter in observable quantities.

\subsection{Decompositions, and sourced equations for $\zeta ,\,  h_\lambda$}
\label{subsec:formal-deco}

We compute scalar cosmological perturbations in the spatially flat gauge $\delta g_{ij,{\rm scalar}} = 0$. 
In this gauge, the curvature perturbation $\zeta$ is (up to negligible corrections, as we will see in the next two sections) $\zeta = - \frac{H}{\dot{\varphi}^{(0)}} \, \delta \varphi$, where $\varphi^{(0)}$ and $\delta \varphi$ are the unperturbed and perturbed part of the inflaton, respectively.~\footnote{Throughout this work we denote physical time by $t$ and conformal time by $\tau$. 
Derivative wrt physical (conformal) time are denoted by a dot (prime). We denote the scale factor and the Hubble rate by $a$ and $H$, respectively. We denote by $M_p$ the reduced Planck mass. We use the mostly positive $-,+,+,+$ signature everywhere in the paper apart from the model of fermionic production (Subsection \ref{subsec:GW-comparison} and Appendix \ref{app-fermions}) where we switch to the opposite signature.}  We decompose
\begin{equation}
\delta \varphi =  \int \frac{d^3 k}{\left( 2 \pi \right)^{3/2}}  {\rm e}^{i \vec k \cdot \vec x} \,  \frac{Q_\varphi}{a}
\label{phi-deco}
\end{equation}

As we show in the next sections, the effect of the vector fields on the inflaton modes can be encoded in the approximate equation
\begin{equation}
\left[ \partial_\tau^2 + \left( k^2 - \frac{a''}{a} \right) \right] Q_\varphi \left( \tau ,\, \vec{k} \right) \simeq J_\varphi \left( \tau ,\, \vec{k} \right) 
\label{Qf-eq-formal}
\end{equation}
where the source is formally of the type
\begin{equation}
J_\varphi  \left( \tau ,\, \vec{k} \right) \equiv  \int \frac{ d^3 p }{ \left( 2 \pi \right)^{3/2} } \; {\hat O}_{\varphi,ij} \left( \tau ,\, \vec{k} ,\, \vec{p} \right) \, {\tilde A_i} \left( \tau ,\, \vec{p} \right)  {\tilde A_j} \left( \tau ,\, \vec{k} - \vec{p} \right) 
\label{Jf-formal}
\end{equation}
where ${\hat O}$ is a model-dependent operator.

In the tensor sector, the gravity waves are encoded in $g_{ij} = a^2 \left( \delta_{ij} + h_{ij} \right)$, where the modes $h_{ij}$ are transverse and traceless. We introduce the canonical modes
\begin{equation}
 \frac{M_p a}{2} h_{ij} =  \int \frac{d^3 k}{\left( 2 \pi \right)^{3/2}} \, {\rm e}^{i \vec{k} \cdot \vec{x}}  \, \sum_{ \lambda =  \pm } \, \Pi_{ij,\lambda} \left( {\hat k} \right) \,  Q_\lambda \left( \vec{k} \right) \;\;\;,\;\;\;
\Pi_{ij,\lambda} \left( {\hat k} \right) = \epsilon^{(\lambda)}_i \left( {\hat k} \right)  \epsilon^{(\lambda)}_j \left( {\hat k} \right) 
\label{formal-hc-def}
\end{equation}
which  obey the  equation
\begin{equation}
\left[ \partial_\tau^2 + \left( k^2 - \frac{a''}{a} \right) \right]  Q_\lambda \left( \tau ,\, \vec{k} \right) = J_\lambda \left( \tau ,\, \vec{k} \right)
\label{hc-eq-formal}
\end{equation}

The source is obtained by starting from the transverse and traceless spatial part of the energy momentum tensor and projecting along the $\lambda$ polarization with $\Pi_{ij,\lambda}$. One finds
\begin{equation}
J_\lambda \left( \tau ,\, \vec{k} \right) = \Pi_{ij,\lambda}^{*} \left( {\hat k} \right) \int \frac{d^3 x}{\left( 2 \pi \right)^{3/2}} \, {\rm e}^{-i \vec{k} \cdot \vec{x}} \, \frac{a}{M_p} \, T_{ij} \left( \tau ,\, \vec{x} \right)
\label{Jlambda-formal}
\end{equation}
We note that, the multiplication by $ \Pi_{ij,\lambda}^{*} $ automatically projects on the transverse and traceless part of $T_{ij}$.  If (as in the cases that we will consider) the energy-momentum tensor is quadratic in the gauge fields, we recover an expression formally identical to (\ref{Jf-formal}), with a different (and model-dependent) operator $ {\hat O}_{\lambda,ij} \left( \tau ,\, \vec{k} ,\, \vec{p} \right) $. 

In summary, we have formally identical equations for the scalar and tensor canonical modes

\begin{eqnarray}
&&
\left[ \partial_\tau^2 + \left( k^2 - \frac{a''}{a} \right) \right] Q_X \left( \tau ,\, \vec{k} \right) \simeq J_X \left( \tau ,\, \vec{k} \right) \nonumber\\
&&
J_X  \left( \tau ,\, \vec{k} \right) \equiv  \int \frac{ d^3 p }{ \left( 2 \pi \right)^{3/2} } \; {\hat O}_{X,ij} \left( \tau ,\, \vec{k} ,\, \vec{p} \right) \, {\tilde A_i} \left( \tau ,\, \vec{p} \right)  {\tilde A_j} \left( \tau ,\, \vec{k} - \vec{p} \right)  \;\;\;,\;\;\; X = \left\{ \varphi ,\, \lambda=+ ,\, \lambda=- \right\}
\label{pbm-formal}
\end{eqnarray}
The operators ${\hat O}_X$ depend on what $X$ is, and also on the model. In all cases, they are invariant  under the simultaneous $i \leftrightarrow j$ and $\vec{p} \rightarrow \vec{k} - \vec{p}$ operations.

\subsection{Formal solutions and correlators}
\label{subsec:formal-sol}

The equation (\ref{pbm-formal}) is formally solved by
\begin{equation}
Q_X \left( \tau ,\, \vec{k} \right) =  Q_{X,{\rm v}}  \left( \tau ,\, \vec{k} \right) 
+  Q_{X,{\rm s}}  \left( \tau ,\, \vec{k} \right) 
\label{formal-sol1}
\end{equation}
where $ Q_{X,{\rm v}}  $ is the standard  vacuum solution, obtained in absence of the vector source; the sourced term is instead
\begin{eqnarray}
 Q_{X,{\rm s}}  \left( \tau ,\, \vec{k} \right) & = &  \int^\tau d \tau' G_k \left( \tau ,\, \tau' \right) \, J_X \left( \tau' ,\, \vec{k} \right) \nonumber\\
G_k \left( \tau ,\, \tau' \right) & \simeq & \frac{1}{k^3 \tau \tau'} \left[ k \tau' \, \cos \left( k \tau' \right) - \sin \left( k \tau' \right) \right] \;\;\;,\;\;\; - k \, \tau \ll 1
\label{formal-sol2}
\end{eqnarray}
where the Green function is only given  in the super-horizon regime $ - k \, \tau \ll 1$. The two terms in (\ref{formal-sol1}) are uncorrelated.

We are interested in the two and three point correlators of the sourced solutions. For the two point correlator, we find
\begin{equation}
\left\langle Q_{X,{\rm s}} \left( \tau ,\, \vec{k}_1 \right)  Q_{Y,{\rm s}} \left( \tau ,\, \vec{k}_2 \right) \right\rangle =  \int^\tau d \tau_1 G_{k_1} \left( \tau ,\, \tau_1 \right)   \int^\tau d \tau_2 G_{k_2} \left( \tau ,\, \tau_2 \right) \, \left\langle J_X \left( \tau_1 ,\, \vec{k}_1 \right)  J_Y \left( \tau_2 ,\, \vec{k}_2 \right) \right\rangle 
\label{formal-QQ}
\end{equation}
where
\begin{eqnarray} 
\left\langle J_X \left( \tau_1 ,\, \vec{k}_1 \right)  J_Y \left( \tau_2 ,\, \vec{k}_2 \right) \right\rangle 
& = & 2 \int \frac{d^3 p_1 d^3 p_2}{\left( 2 \pi \right)^3} \, 
{\hat O}_{X,ij} \left( \tau_1 ,\, \vec{k}_1 ,\, \vec{p}_1 \right) 
{\hat O}_{Y,lm}  \left( \tau_2 ,\, \vec{k}_2 ,\, \vec{p}_2 \right) \nonumber\\
&&\quad\quad\quad \times
\left\langle {\tilde A}_i \left( \tau_1 ,\, \vec{p}_1 \right)  {\tilde A}_m \left( \tau_2 ,\, \vec{k}_2 - \vec{p}_2 \right) \right\rangle \,
\left\langle {\tilde A}_j \left( \tau_1 ,\, \vec{k}_1 - \vec{p}_1 \right)  {\tilde A}_l \left( \tau_2 ,\, \vec{p}_2 \right) \right\rangle
\label{formal-JJ1}
\end{eqnarray}
In writing this expression, we Wick decomposed the $\langle A^4 \rangle$ correlator coming from $\langle J^2 \rangle$, and then used the symmetry of the ${\hat O}_{Y,lm} $ operator.

From (\ref{A-deco}) we then see that the two point correlator of the gauge field is formally of the type
\begin{equation}
\left\langle {\tilde A}_i \left( \tau_1 ,\, \vec{q}_1 \right)  {\tilde A}_j \left( \tau_2 ,\, \vec{q}_2 \right) \right\rangle 
= \sum_\sigma {\cal P}_{ij}^{(\sigma)} \left( {\hat q}_1 \right)  
{\cal D}_{(0,0)}^{(\sigma)} \left[ \tau_1 ,\, \tau_2 ;\, q_1 \right] \, \delta^{(3)} \left( \vec{q}_1 + \vec{q}_2 \right)
\label{formal-AA}
\end{equation}
where
\begin{equation}
{\cal P}_{ij}^{(\pm)} \left( {\hat p} \right) \equiv 
\epsilon_i^{(\pm)} \left( \vec{p} \right) \, \epsilon_j^{(\pm)*} \left( \vec{p} \right)
= \frac{1}{2} \left( \delta_{ij} - {\hat p}_i \, {\hat p}_j \right) \mp \frac{i}{2} \epsilon_{ijk} \, {\hat p}_k
\label{identity-pol}
\end{equation}
The index ${(0,0)}$ on ${\cal D}$ indicates that no time derivative is acting on the modes ${\tilde A}$ on the l.h.s. of (\ref{formal-AA}), and it is introduced for later convenience. Using the hermiticity of $A_\mu \left( x \right)$, one can show that  ${\cal D}_{(0,0)}^{(\sigma)} \left[ \tau_1 ,\, \tau_2 ;\, q \right] = {\cal D}_{(0,0)}^{(\sigma) *} \left[ \tau_2 ,\, \tau_1 ;\, q \right] $. We show in the following sections that, in both models we study, ${\cal D}_{(0,0)}^{(\sigma)}$ becomes real and symmetric under $\tau_1 \leftrightarrow \tau_2$ for all times after the particle production.

Inserting (\ref{formal-AA}) into (\ref{formal-JJ1}) we obtain
\begin{eqnarray}
\left\langle J_X \left( \tau_1 ,\, \vec{k}_1 \right)  J_Y \left( \tau_2 ,\, \vec{k}_2 \right) \right\rangle 
& = & 2 \, \delta^{(3)} \left( \vec{k}_1 + \vec{k}_2 \right)  \int \frac{d^3 p}{\left( 2 \pi \right)^3} \, \sum_{\sigma,\sigma'} {\cal P}_{im}^{(\sigma)} \left( {\hat p} \right) {\cal P}_{lj}^{(\sigma')} \left(  \widehat{p - k_1} \right)     \nonumber\\
&&\quad\quad
{\hat O}_{X,ij} \left( \tau_1 ,\, \vec{k}_1 ,\, \vec{p} \right) \, 
{\hat O}_{Y,lm}  \left( \tau_2 ,\, - \vec{k}_1 ,\, \vec{p} - \vec{k}_1 \right) \, 
{\cal D}_{(0,0)}^{(\sigma)} \left[ \tau_1 ,\, \tau_2 ;\, p \right] \, 
{\cal D}_{(0,0)}^{(\sigma')} \left[ \tau_1 ,\, \tau_2 ;\, \vert \vec{p} - \vec{k}_1  \vert \right] \nonumber\\
\label{formal-JJ2}
\end{eqnarray}

Finally, we only need to compute the connected three point correlator of $Q_\varphi$. Proceeding as for the two point function we obtain
\begin{equation}
\left\langle 
\prod_{i=1}^3 
Q_{\varphi,{\rm s}}  \left( \tau ,\, \vec{k}_i \right)  \right\rangle = 
  \int^\tau d \tau_1 G_{k_1} \left( \tau ,\, \tau_1 \right)   \int^\tau d \tau_2 G_{k_2} \left( \tau ,\, \tau_2 \right) 
\int^\tau d \tau_3 G_{k_3} \left( \tau ,\, \tau_3 \right) \, \left\langle 
\prod_{i=1}^3 J_\varphi \left( \tau_i ,\, \vec{k}_i \right)  
\right\rangle 
\label{formal-QQQ}
\end{equation}
with
\begin{eqnarray}
 \left\langle \prod_{i=1}^3 J_\varphi \left( \tau_i ,\, \vec{k}_i \right)  \right\rangle 
& = & 8 \, \delta^{(3)} \left( \vec{k}_1 + \vec{k}_2 + \vec{k}_3 \right)  \int \frac{d^3 p}{\left( 2 \pi \right)^{9/2}} 
\; \sum_{\sigma,\sigma',\sigma''} 
{\cal P}_{im}^{(\sigma)} \left( {\hat p} \right) 
{\cal P}_{nj}^{(\sigma')} \left( \widehat{p - k_1} \right)  
{\cal P}_{lo}^{(\sigma'')} \left( \widehat{p + k_2} \right)   \nonumber\\
&&\quad\quad \quad\quad
{\hat O}_{\varphi,ij} \left( \tau_1 ,\, \vec{k}_1 ,\, \vec{p} \right) \, 
{\hat O}_{\varphi,lm}  \left( \tau_2 ,\,  \vec{k}_2 ,\, \vec{p} + \vec{k}_2 \right) \, 
{\hat O}_{\varphi,no}  \left( \tau_3 ,\,  \vec{k}_3 ,\, \vec{p} - \vec{k}_1 \right) \nonumber\\
&&\quad\quad \quad\quad
{\cal D}_{(0,0)}^{(\sigma)} \left[ \tau_1 ,\, \tau_2 ;\, p \right] \, 
{\cal D}_{(0,0)}^{(\sigma')} \left[ \tau_1 ,\, \tau_3 ;\, \vert \vec{p} - \vec{k}_1  \vert \right] 
{\cal D}_{(0,0)}^{(\sigma'')} \left[ \tau_2 ,\, \tau_3 ;\, \vert \vec{p} + \vec{k}_2 \vert \right] 
\label{formal-JJJ}
\end{eqnarray}

\subsection{Phenomenology}
\label{subsec:formal-pheno}

From the two point scalar correlator we obtain the power spectrum
\begin{equation}
\frac{k_1^3}{2 \pi^2} 
\left\langle \zeta \left( \vec{k}_1 \right) \,  \zeta \left( \vec{k}_2 \right) \right\rangle
=  P_{\zeta}  \; \delta^{(3)} \left( \vec{k}_1 + \vec{k}_2 \right) \;\;\;,\;\;\;
\end{equation}
The two solutions (\ref{formal-sol1}) are incoherent and therefore their power spectra add up. Recalling the standard slow roll result for the vacuum mode, and using the two expressions (\ref{formal-QQ}) and 
(\ref{formal-JJ2}), we obtain
\begin{eqnarray}
P_\zeta & = &  P_{\zeta,v} +  P_{\zeta,s} \nonumber\\
P_{\zeta,v} & = & \frac{H^4}{4 \pi^2 \dot{\varphi}^{(0) 2}} \nonumber\\
P_{\zeta,s}  & = & \frac{k_1^3}{\pi^2} \,  \frac{H^2}{a^2 \dot{\varphi}^{(0)2}} \, 
 \int^\tau d \tau_1 G_{k_1} \left( \tau ,\, \tau_1 \right) 
 \int^\tau d \tau_2 G_{k_1} \left( \tau ,\, \tau_2 \right) 
 \int \frac{d^3 p}{\left( 2 \pi \right)^3} \, \sum_{\sigma,\sigma'} {\cal P}_{im}^{(\sigma)} \left( {\hat p} \right) {\cal P}_{lj}^{(\sigma')} \left(  \widehat{p - k_1} \right)     \nonumber\\
&&\quad\quad\quad
{\hat O}_{\varphi,ij} \left( \tau_1 ,\, \vec{k}_1 ,\, \vec{p} \right) \, 
{\hat O}_{\varphi,lm}  \left( \tau_2 ,\, - \vec{k}_1 ,\, \vec{p} - \vec{k}_1 \right) \, 
{\cal D}_{(0,0)}^{(\sigma)} \left[ \tau_1 ,\, \tau_2 ;\, p \right] \, 
{\cal D}_{(0,0)}^{(\sigma')} \left[ \tau_1 ,\, \tau_2 ;\, \vert \vec{p} - \vec{k}_1  \vert \right] 
\label{formal-Pz}
\end{eqnarray}

For the tensor mode, we show below that modes of different polarizations are uncorrelated. Starting from the decomposition (\ref{formal-hc-def}), the power in each polarization is
\begin{equation}
\frac{k_1^3}{2 \pi^2} \frac{4}{a^2 M_p^2}
\left\langle Q_\lambda \left( \vec{k}_1 \right) \,  Q_\lambda \left( \vec{k}_2 \right) \right\rangle
=  P_\lambda  \; \delta^{(3)} \left( \vec{k}_1 + \vec{k}_2 \right) \;\;\;,\;\;\;
\label{formal-Plambda-def}
\end{equation}
Also in this case, the power of the vacuum and the sourced modes add up
\begin{eqnarray}
P_\lambda & = &  P_{\lambda,v} +  P_{\lambda,s} \nonumber\\
P_{\lambda,v} & = & \frac{H^2}{ \pi^2 \, M_p^2} \nonumber\\
P_{\lambda,s}  & = & \frac{k_1^3}{\pi^2} \,  \frac{4}{a^2 \, M_p^2} \, 
 \int^\tau d \tau_1 G_{k_1} \left( \tau ,\, \tau_1 \right) 
 \int^\tau d \tau_2 G_{k_1} \left( \tau ,\, \tau_2 \right) 
 \int \frac{d^3 p}{\left( 2 \pi \right)^3} \, \sum_{\sigma,\sigma'} {\cal P}_{im}^{(\sigma)} \left( {\hat p} \right) {\cal P}_{lj}^{(\sigma')} \left(  \widehat{p - k_1} \right)     \nonumber\\
&&\quad\quad\quad
{\hat O}_{\lambda,ij} \left( \tau_1 ,\, \vec{k}_1 ,\, \vec{p} \right) \, 
{\hat O}_{\lambda,lm}  \left( \tau_2 ,\, - \vec{k}_1 ,\, \vec{p} - \vec{k}_1 \right) \, 
{\cal D}_{(0,0)}^{(\sigma)} \left[ \tau_1 ,\, \tau_2 ;\, p \right] \, 
{\cal D}_{(0,0)}^{(\sigma')} \left[ \tau_1 ,\, \tau_2 ;\, \vert \vec{p} - \vec{k}_1  \vert \right] 
\label{formal-Ph}
\end{eqnarray}

The tensor-to-scalar ratio $r$ is defined as
\begin{equation}
r = \frac{\sum_\lambda P_\lambda}{P_\zeta}
\label{def-r}
\end{equation}
If the sourced term is absent, one recovers the standard vacuum slow roll result $r_{\rm v} = \frac{8 \, \dot{\varphi}^{(0) 2}}{H^2 \, M_p^2} = 16 \epsilon$. However, we see that the sourced contribution can modify this result.

Finally, we are interested in the bi-spectum of the scalar modes, defined as
\begin{equation}
\left\langle \zeta \left( \vec{k}_1 \right) \zeta \left( \vec{k}_2 \right) \zeta \left( \vec{k}_3 \right) \right\rangle =
B_\zeta \left( \vec{k}_i \right) \, \delta^{(3)} \left( \vec{k}_1 + \vec{k}_2 + \vec{k}_3 \right)
\end{equation}
We disregard the contribution from the vacuum modes, as it is unobservable for standard slow roll inflation. Using (\ref{formal-QQQ}) and (\ref{formal-JJJ}) we obtain
\begin{eqnarray}
B_\zeta \left( \vec{k}_i \right) & = &  8 \int^\tau d \tau_1 G_{k_1} \left( \tau ,\, \tau_1 \right)   \int^\tau d \tau_2 G_{k_2} \left( \tau ,\, \tau_2 \right) \int^\tau d \tau_3 G_{k_3} \left( \tau ,\, \tau_3 \right) \;
 \int \frac{d^3 p}{\left( 2 \pi \right)^{9/2}} 
\; \sum_{\sigma,\sigma',\sigma''} 
{\cal P}_{im}^{(\sigma)} \left( {\hat p} \right) 
{\cal P}_{nj}^{(\sigma')} \left( \widehat{p - k_1} \right)  
{\cal P}_{lo}^{(\sigma'')} \left( \widehat{p + k_2} \right)   \nonumber\\
&& \!\!\!\!\!\!\!\!  \!\!\!\!\!\!\!\!  \!\!\!\!\!\!\!\!  \!\!\!\!\!\!\!\!
{\hat O}_{\varphi,ij} \left( \tau_1 ,\, \vec{k}_1 ,\, \vec{p} \right) \, 
{\hat O}_{\varphi,lm}  \left( \tau_2 ,\,  \vec{k}_2 ,\, \vec{p} + \vec{k}_2 \right) \, 
{\hat O}_{\varphi,no}  \left( \tau_3 ,\,  \vec{k}_3 ,\, \vec{p} - \vec{k}_1 \right) 
{\cal D}_{(0,0)}^{(\sigma)} \left[ \tau_1 ,\, \tau_2 ;\, p \right] \, 
{\cal D}_{(0,0)}^{(\sigma')} \left[ \tau_1 ,\, \tau_3 ;\, \vert \vec{p} - \vec{k}_1  \vert \right] 
{\cal D}_{(0,0)}^{(\sigma'')} \left[ \tau_2 ,\, \tau_3 ;\, \vert \vec{p} + \vec{k}_2 \vert \right] \nonumber\\
\label{formal-Bz}
\end{eqnarray}

In the following sections, we compute the bispectrum for equilateral configurations (which is where it is peaked in the models that we consider), and use it to define an equilateral nonlinear parameter \cite{Barnaby:2011vw}
\begin{equation}
f_{NL,{\rm equil. \; eff.}} \left( k \right) = \frac{10}{9  \left( 2 \pi \right)^{5/2} } \,  \frac{k^6}{P_\zeta^2 \left( k \right)} \, B_\zeta \vert_{k_1 = k_2 = k_3 = k}
\label{formal-fnl}
\end{equation}

\section{Model I: Vector produced by non-adiabatic change of its mass}
\label{sec:model1}

We consider a model where the sector in which particle production takes place has a local U(1) invariance spontaneously broken by the expectation value of a complex scalar $\Psi$:
\begin{equation}
 S = \int d^4x \sqrt{-g} \left[\frac{^{}}{_{}}\right. \frac{M_p^2}{2} R
-\underbrace{\frac{1}{2}(\partial\varphi)^2 - V(\varphi) }_{\mathrm{inflaton}\hspace{2mm}\mathrm{sector}} 
  - \underbrace{   \vert \left( \partial_\mu - i e A_\mu \right) \Psi \vert^2 - U \left( \vert \Psi \vert \right) 
- \frac{1}{4} F^2   }_{\mathrm{hidden}\hspace{2mm}\mathrm{sector}} 
 \left.\frac{^{}}{_{}}\right] \, .
\label{model1}
\end{equation}
The field $\varphi$ is the inflaton field, which is assumed to be only gravitationally coupled to the $\Psi-A_\mu$ sector. We work in the unitary gauge $\Psi = \frac{\psi}{\sqrt{2}}$, where $\psi$ is real. We assume that the background value $\psi^{(0)} \left( t \right)$ crosses zero at the time $t_*$ during inflation. Close to this time, the gauge field mass can be approximated by
\begin{equation}
m = e \, \psi^{(0)} \simeq e \dot{\psi}_*^{(0)} \left( t - t_* \right) \equiv \dot{m}_*  \left( t - t_* \right) 
\label{vector-mass}
\end{equation}
For definiteness, we take $ \dot{m}_* > 0$. The (comoving) frequency of a gauge mode, $\omega = \sqrt{k^2 + a^2 \, m^2}$ varies nonadiabatically ($\omega' > \omega^2$) when the mass vanishes. As we will show later, the gauge field modes that dominate the observational signatures have $k \sim \sqrt{\dot{m}_*}$. For such modes, the frequency changes nonadiabatically in the time interval $~\sim t_* \pm \frac{1}{ \sqrt{\dot{m}_*}}$, provided that the expansion of the universe can be disregarded in this time interval.
This is the case for $\frac{1}{ \sqrt{\dot{m}_*}} \ll \frac{1}{H}$. In the remainder of this Subsection we 
concentrate on the production during this interval, and disregard the expansion of the universe (we use physical and conformal time interchangeably). We can eliminate any reference to the scale factor by normalizing $a \left( t_* \right) = 1$. In the following Subsections the expansion is taken into account.

The non-adiabatic change of the frequency causes non-perturbative production of the gauge modes. To compute this, we decompose the vector field as in (\ref{A-deco}). We further decompose the mode functions in positive and negative frequency modes
\begin{eqnarray}
&& A_\lambda \left( k \right) = \alpha_k \left( \tau  \right) f_k \left( \tau  \right) +  \beta_k \left( \tau  \right) f_k^* \left( \tau  \right) \;\;\;,\;\;\; f_k \equiv \frac{{\rm e}^{- i \int^\tau d \tau' \, \omega \left( \tau' \right)} }{ \sqrt{2 \omega \left( \tau \right)} } \nonumber\\
&&  A_\lambda' \left( k \right) = - i \omega \left[ \alpha_k \left( \tau  \right) f_k \left( \tau  \right) -  \beta_k \left( \tau  \right) f_k^* \left( \tau  \right) \right]
\label{A-deco2}
\end{eqnarray}
(the second line is the decomposition of the modes of the conjugate momentum to $A_i$). The decomposition (\ref{A-deco})  disregards the longitudinal vector mode; we compute this mode in Appendix \ref{app-longA}, and we discuss its effects below in this section. We have suppressed the index $\lambda$ in this decomposition, since the Bogolyubov coefficients are the same for both helicities. For $t \ll t_*$, the mode is in the adiabatic vacuum $\alpha = 1$ (up to an arbitrary phase); the quantity $\vert \beta \vert^2$ is the occupation number of the gauge modes.

The gauge field modes satisfy $A'' + \omega^2 A = 0$. With the approximated expression (\ref{vector-mass}), this equation can be analytically solved in terms of two parabolic cylinder functions. The linear combination satisfying the proper initial condition $\alpha = 1$ gives (up to an arbitrary phase) 
\cite{Kofman:1997yn,Cook:2011hg}
\begin{equation}
\alpha_k \left( t \gg t_* \right) \simeq  \sqrt{1+{\rm e}^{-\frac{\pi k^2}{\dot{m}_*}}} \;\;\;,\;\;\;
\beta_k  \left( t \gg t_* \right) \simeq -  {\rm e}^{-\frac{\pi k^2}{2 \, \dot{m}_*}} 
\label{mod1-ablate}
\end{equation}
Namely, one finds a ${\rm O } \left( 1 \right)$ occupation number for modes up to $k \simeq  \sqrt{\dot{m}_*}$, while the production is exponentially suppressed at higher momenta. An analogous result is obtained for scalar \cite{Kofman:1997yn} and fermion \cite{Chung:1999ve} fields produced by a mass varying as in (\ref{vector-mass}). This result is valid provided that
\begin{equation}
H^2 \ll \dot{m}_* \ll \sqrt{6} e M_p H
\label{mod1-bck-cond}
\end{equation}
where (as we mentioned previously) the first condition ensures that the expansion of the universe can be disregarded during the interval of particle production, while the second condition imposes that the kinetic energy of $\psi$ is negligible with respect to the inflaton energy density.

Having the mode functions, we can now compute the correlators (\ref{formal-AA}). The correlators need to be regularized. The regularization can be performed   by normal ordering with respect to 
the time dependent annihilation creation operators 
\begin{equation}
{\bar a}_\lambda \left( \tau ,\, \vec{k} \right) \equiv \alpha_k \left(  \tau \right) a_\lambda \left( \vec{k} \right) + \beta_k^* \left( \tau  \right) a_\lambda^\dagger \left( - \vec{k} \right) \;\;\;,\;\;\;
{\bar a}_\lambda^\dagger \left( \tau ,\, - \vec{k} \right) \equiv \beta_k \left( \tau  \right) a_\lambda \left( \vec{k} \right) + \alpha_k^* \left( \tau  \right) a_\lambda^\dagger \left( - \vec{k} \right) 
\label{abar-def}
\end{equation}
These are the ``physical'' operators of the system, as one can show that they diagonalize the Hamiltonian at all times. Using (\ref{A-deco}) and  (\ref{A-deco2}) it is immediate to show that, in terms of these operators,
\begin{equation}
{\tilde A}_i \left( \tau ,\, \vec{k} \right) = \sum_\lambda \epsilon_i^{(\lambda)} \left( {\hat k} \right) \left[
f_k \left( \tau  \right) {\bar a}_\lambda \left( \tau ,\, \vec{k} \right) +  
f_k^* \left( \tau  \right) {\bar a}_\lambda^\dagger \left( \tau ,\, - \vec{k} \right) \right] 
\end{equation}
We then compute $\left\langle : {\tilde A}_i \left( \tau ,\, \vec{k} \right)  {\tilde A}_i \left( \tau' ,\, \vec{k}' \right) : \right\rangle$ by normal ordering with respect to the ${\bar a}$, and ${\bar a}^\dagger$, but by recalling that the vacuum state is annihilated by the original time-independent $a$ (we are working in the Heisenberg picture, in which the states are constant).~\footnote{This procedure is conventionally adopted in Bogolyubov computations, although it is not always spelled out;  indeed, the initial Hamiltonian, without normal ordering, can be cast in the form ${\hat H} = \int d^3 k \,  \omega_k \, {\hat N}$, where the counting operator is  ${\hat N} =  \frac{ {\bar a} {\bar a}^\dagger +   {\bar a}^\dagger {\bar a} }{2}$. One then defines normal ordering wrt the barred operators,  $:{\hat N}: = {\bar a}^\dagger \, {\bar a}$, and evaluates $\langle :{\hat N}: \rangle $ using (\ref{abar-def}) and recalling that the vacuum is annihilated by the  original time-independent $a$; only in this way one obtains  $\langle :{\hat N}: \rangle = \vert \beta \vert^2$. We employ the exact same procedure for evaluating the correlators ${\cal D}$.} Proceeding in this way, and casting the result as in (\ref{formal-AA}), we obtain
\begin{eqnarray}
 {\cal D}^{(\sigma)}_{(0,0)} \left[ \tau ,\,  \tau' ;\, k \right] &=& 
  \alpha_k \left( \tau  \right) \beta_k^* \left( \tau'  \right) f_k \left( \tau \right) f_k \left( \tau' \right)+  \beta_k \left( \tau  \right) \alpha_k^* \left( \tau'  \right) f_k^* \left( \tau \right) f_k^* \left( \tau' \right) \nonumber\\
&& 
 +   \beta_k \left( \tau  \right) \beta_k^* \left( \tau'  \right) f_k^*  \left( \tau \right) f_k \left( \tau' \right)+  \beta_k^*   \left( \tau  \right) \beta_k \left( \tau'  \right) f_k \left( \tau \right) f_k^* \left( \tau' \right) 
\end{eqnarray}

Clearly, the result is $\sigma-$independent, since both helicities are produced in the same amount. 
After the particle production, this expression simplifies into
\begin{equation}
{\cal D}^{(\sigma)}_{(0,0)} \left[ \tau ,\,  \tau' ;\, k \right] = \alpha_k \, \beta_k^* \,  f_k \left( \tau \right) f_k \left( \tau' \right) + \alpha_k^* \, \beta_k \,  f_k^* \left( \tau \right) f_k^* \left( \tau' \right) + \vert \beta_k \vert^2 \left[  f_k^*  \left( \tau \right) f_k \left( \tau' \right)+   f_k \left( \tau \right) f_k^* \left( \tau' \right) \right]
\label{mod1-D00}
\end{equation}
where $\alpha$ and $\beta$ assume the asymptotic values (\ref{mod1-ablate}). As we anticipated, this expression is real and symmetric under $\tau \leftrightarrow \tau'$.

The same result (\ref{mod1-D00}) is obtained if one computes  $\left\langle  {\tilde A}_i \left( \tau ,\, \vec{k} \right)  {\tilde A}_i \left( \tau' ,\, \vec{k}' \right)  \right\rangle$ without  normal ordering, and then subtracts  the term that one would have in absence of particle production ($\alpha = 1 ,\, \beta=0$), as done in
\cite{Barnaby:2009mc}. Therefore these two regularizations are equivalent after the particle production has taken place.

For further convenience, we extend the definition (\ref{formal-AA}) to
\begin{equation}
\left\langle :
\left( \frac{\partial }{ \partial \tau_1 } \right)^a {\tilde A}_i \left( \tau_1 ,\, \vec{q}_1 \right)  
\left( \frac{\partial }{ \partial \tau_2 } \right)^b {\tilde A}_j \left( \tau_2 ,\, \vec{q}_2 \right) : \right\rangle 
\equiv \sum_\sigma {\cal P}_{ij}^{(\sigma)} \left( {\hat q}_1 \right)  
{\cal D}_{(a,b)}^{(\sigma)} \left[ \tau_1 ,\, \tau_2 ;\, q_1 \right] \, \delta^{(3)} \left( \vec{q}_1 + \vec{q}_2 \right)
\end{equation}

We compute these quantities by using (\ref{A-deco2}). After  $\alpha$ and $\beta$  have assumed the asymptotic values (\ref{mod1-ablate}), the correlators of our interest become
\begin{eqnarray}
{\cal D}^{(\sigma)}_{(1,0)} \left[ \tau ,\,  \tau' ;\, k \right] & = & - i \omega_k \left( \tau \right) \left\{ \alpha_k \, \beta_k^* \,  f_k \left( \tau \right) f_k \left( \tau' \right) - \alpha_k^* \, \beta_k \,  f_k^* \left( \tau \right) f_k^* \left( \tau' \right) + \vert \beta_k \vert^2 \left[ - f_k^*  \left( \tau \right) f_k \left( \tau' \right)+   f_k \left( \tau \right) f_k^* \left( \tau' \right) \right] \right\}
\nonumber\\
{\cal D}^{(\sigma)}_{(0,1)} \left[ \tau ,\,  \tau' ;\, k \right] & = &  - i \omega_k \left( \tau' \right) \left\{ \alpha_k \, \beta_k^* \,  f_k \left( \tau \right) f_k \left( \tau' \right) - \alpha_k^* \, \beta_k \,  f_k^* \left( \tau \right) f_k^* \left( \tau' \right) + \vert \beta_k \vert^2 \left[  f_k^*  \left( \tau \right) f_k \left( \tau' \right) -   f_k \left( \tau \right) f_k^* \left( \tau' \right) \right] \right\}
\nonumber\\
{\cal D}^{(\sigma)}_{(1,1)} \left[ \tau ,\,  \tau' ;\, k \right] & = & - \omega_k \left( \tau \right) \,  \omega_k \left( \tau' \right) \left\{ \alpha_k \, \beta_k^* \,  f_k \left( \tau \right) f_k \left( \tau' \right) + \alpha_k^* \, \beta_k \,  f_k^* \left( \tau \right) f_k^* \left( \tau' \right) - \vert \beta_k \vert^2 \left[  f_k^*  \left( \tau \right) f_k \left( \tau' \right)+   f_k \left( \tau \right) f_k^* \left( \tau' \right) \right] \right\} \nonumber\\
\label{mod1-Dab}
\end{eqnarray}
We note that in the regime in which (\ref{mod1-D00}) and (\ref{mod1-Dab}) are valid, 
$\partial_{\tau} {\cal D}^{(\sigma)}_{(0,0)} \left[ \tau ,\,  \tau' ;\, k \right] = {\cal D}^{(\sigma)}_{(1,0)} \left[ \tau ,\,  \tau' ;\, k \right] $, and analogously for the other terms. Namely, in this regime, time derivatives can be equivalently taken before or after evaluating the correlator. One can verify that (for the regularized correlators) this is not the case during the particle production, when $\alpha$ and $\beta$ are still functions of time.

\subsection{Scalar perturbations sourced by the vector modes}
\label{subsec:zeta1}

We are interested in the phenomenological consequences of the vector field production in the model 
(\ref{model1}). As mentioned in the Introduction, we work under the assumption that the $\Psi-A_\mu$ sector has a negligible energy density with respect to the inflationary sector, and that the two sectors are coupled only gravitationally. Under these assumptions, the main signature of the particle production is encoded in how the gauge quanta enter in the gravitational equations.  In this subsection we study the effect on the scalar metric and inflaton perturbations; the effect  on the tensor modes is computed in the next subsection. 

We divide the computation in $3$ parts; in the first one, we present the master equation for the density perturbation $\zeta$; in the second and the third part  we compute, respectively, the power spectrum and the bispectrum of the part of $\zeta$ sourced by the vector quanta.

\subsubsection{Master equation for $\zeta$}

 We perform computations in the spatially flat gauge $\delta g_{ij,{\rm scalar}} = 0$. We then decompose the fields into background $+$ first order $+$ second order perturbations; for example, for the inflaton field we write
\begin{equation}
\varphi = \varphi^{(0)} +  \varphi_1 +  \varphi_2  + \dots
\label{phi-deco2}
\end{equation}
(dots denotes perturbations of higher order, that we ignore). Under our working assumption that 
$\Psi$ gives a negligible contribution both to the background energy density, and to the cosmological perturbations (which is certainly the case, if $\rho_\Psi$ is sufficiently small), the gauge curvature perturbations $\zeta$ in this gauge is
\begin{equation}
\zeta = - \frac{H}{\dot{\phi}} \, \left(  \varphi_1 +  \varphi_2 \right)
\end{equation}

We are interested in a master equation for  $ \zeta$. We follow the same steps outlined in Section 5 of 
 \cite{Barnaby:2011vw}. We start from the inflaton equation of motion
\begin{equation}
\partial_\mu \left( \sqrt{-g} \,  g^{\mu \nu} \, \partial_\nu \, \varphi \right) - \sqrt{-g} \, \frac{d V}{d \varphi}    = 0
\label{formal-phieq}
\end{equation}
Formally, only perturbations of the inflaton and of the metric enter in this equation. However,  one can 
use the gravitational equations to express the modes $\delta g_{00}$ and $\delta g_{0i}$  in terms of the other perturbations, 
\begin{equation}
\delta g_{00} = \delta g_{00} \left[ \varphi_i ,\, A_\mu ,\, h_{ij} \right] \;\;\;,\;\;\;
\delta g_{0i} = \delta g_{0i} \left[ \varphi_i ,\, A_\mu ,\, h_{ij} \right] \;\;,
\label{formal-adm}
\end{equation}
and then insert these expressions back into (\ref{formal-phieq}). The explicit expressions for (\ref{formal-adm}) can be most directly obtained from the Einstein equations.
\footnote{We note that only the tensor mode $h_{ij}$ enters in the spatial perturbations $\delta g_{ij}$, since  $\delta g_{ij,{\rm scalar}} = 0$ in our gauge, and since we can disregard vector metric perturbations as in the standard case.}  The vector field modes enter in these equations through their contribution to the  energy momentum tensor
\begin{equation}
T_{\mu \nu, {\rm gauge}} = F_{\mu \alpha} \, F_\nu^{\;\;\alpha} + m^2 \, A_\mu A_\nu +
g_{\mu \nu} \left( - \frac{1}{4} F^2 - \frac{m^2}{2} A^2 \right)
\label{Tmunu-mod1}
\end{equation}
where $m^2 = e^2 \psi^{(0) \,2} \left( t \right)$.~\footnote{We remind that we are working in the unitary gauge, ($\Psi$ real) and disregarding perturbations of $\Psi$. In principle, the gauge quanta also source perturbations $\delta \Psi$, which are then gravitationally coupled to the inflationary sector. We do not expect that this effect is more important than the direct gravitational coupling of the gauge modes to the inflationary sector.} 

Equivalently, we can eliminate the  $\delta g_{00}$ and $\delta g_{0i}$ modes from the action  using the  so called energy and momentum constraints in the ADM formalism. Clearly, this is equivalent to using
(\ref{formal-adm}) at the level of the equations. Both methods are presented in  \cite{Barnaby:2011vw}. The net result is that (\ref{formal-phieq}) can be put into a master  equation in terms of $\varphi_i$, $A_\mu$, and $h_{ij}$. The master equation, in absence of gauge fields, was first derived in  \cite{Malik:2006ir}. 

As one can readily employ the same exact steps outlined in  \cite{Barnaby:2011vw} to the current model, we do not repeat those computations here. There are only two changes that need to be taken into account when one repeats  the computations of   \cite{Barnaby:2011vw} for the present model.  Firstly, in the model of  \cite{Barnaby:2011vw} the vector field had a direct coupling to the inflaton $\Delta {\cal L} = - \frac{\alpha}{4 f} \varphi F {\tilde F}$; here the coupling is absent, so we simply formally take $\frac{\alpha}{f} = 0$ in all the steps of
 \cite{Barnaby:2011vw}. Secondly, the vector field was taken to be massless in  \cite{Barnaby:2011vw}, while it is massive here.  This amounts in the additional contributions  proportional to the vector mass given in (\ref{Tmunu-mod1}), or equivalently, in the extra term in the cubic action for the perturbations in the ADM formalism
 \begin{equation}
 \Delta S = - \frac{1}{2} \int d^4 x N a^2 m^2 A_i^2
 \end{equation}
where $N$ is the lapse factor.

As the energy-momentum tensor (\ref{Tmunu-mod1}) is quadratic in the gauge field modes, the first order metric and inflaton perturbations are not affected by the gauge field. Therefore the master equation at first order reproduces the standard  first order expression
\begin{equation}
\left( \partial_\tau^2 - \nabla^2 \right) \varphi_1  + 2 {\cal H}  \varphi_1'  \simeq 0
\label{mod1-eqphi1}
\end{equation}
where $\nabla^2 \equiv \partial_i \partial_i$ and ${\cal H} \equiv \frac{a'}{a}$. The approximation symbol arises because on the left hand side of  this expression  we have disregarded a ``mass term'' for $\varphi_1$ which is proportional to the slow roll parameters
(the same is true for eq. (\ref{mod1-eqphi2}))
\begin{equation}
\epsilon \equiv \frac{M_p^2}{2} \left( \frac{V_{,\varphi}}{V} \right)^2 \ll 1 \;\;,\;\;
\eta \equiv M_p^2 \frac{V_{,\varphi \varphi}}{V}  \ll 1 
\label{slow}
\end{equation}
The reason for this is that we are not interested in slow roll corrections to our results.

At next order in perturbations theory we find instead
\begin{eqnarray}
\left( \partial_\tau^2 - \nabla^2 \right)  \varphi_2  + 2 {\cal H}  \varphi_2'  \simeq
  - \frac{ \varphi^{(0)'} a^2}{2 M_p^2 {\cal H}}
\left\{ \frac{E^2+B^2}{2} + \frac{1}{a^4} \nabla^{-2} \partial_\tau \left[ a^4 \vec{\nabla} \cdot \left( \vec{E} \times \vec{B} \right) \right]  \right\}   - \frac{ \varphi^{(0)'} }{2 M_P^2 a^2 \, {\cal H}} \, \frac{a^2 \, m^2}{2} \, A_i A_i + \dots 
\label{mod1-eqphi2}
\end{eqnarray}
Dots denote terms which are proportional to squares of the first order inflaton and tensor metric perturbations, namely they are of ${\rm O } \left( \varphi_1^2 \right)$ and of  ${\rm O } \left( h_{1,ij}^2 \right)$ (we recall that  $\delta \Psi$ can be disregarded); these terms are explicitly given in  \cite{Malik:2006ir}. The ``electric'' and ``magnetic'' fileds are defined in analogy to the electromagnetic expressions, namely
\begin{equation}
E_i = - \frac{1}{a^2} A_i' \;\;,\;\; B_i = \frac{1}{a^2} \epsilon_{ijk} \partial_j A_k
\end{equation}

The expression  (\ref{mod1-eqphi2}) should be compared with eq. (5.15) of  \cite{Barnaby:2011vw}. Also here we have disregarded the inflaton ``mass term'', which is given instead in   \cite{Barnaby:2011vw} at leading order in slow roll approximation. The expression in  \cite{Barnaby:2011vw} had a term resulting from the direct inflaton-vector field coupling, that  is absent in the present model.~\footnote{This provides the main difference between the scalar field cosmological perturbations obtained in \cite{Barnaby:2011vw} and in the present work; the computations of  \cite{Barnaby:2011vw} were done under the working assumption that the direct interaction is stronger than the gravitational coupling, so that all the other terms were disregarded. In this work we compute instead the more complicated contribution of the gravitational interactions.} On the other hand, the final term in (\ref{mod1-eqphi2}) - proportional to the vector mass - is absent in  \cite{Barnaby:2011vw}. Finally, we note the presence of a typo (a missing ${\cal H}$) in  eq. (5.15) of  \cite{Barnaby:2011vw}. 

\subsubsection{Scalar source and power spectrum}

Identifying the decompositions (\ref{phi-deco}) and (\ref{phi-deco2}), the two equations (\ref{mod1-eqphi1}) and (\ref{mod1-eqphi2}) can then be combined into a unique linear non-homogeneous differential equation for $Q_\varphi$, which is formally of the type (\ref{Qf-eq-formal}), in terms of the two sources
\begin{equation}
J \left[ A_\mu^2 \right] + {\tilde J} \left[ \varphi_1^2 ,\, h_{1,ij}^2 \right] 
\label{mod1-sourceJJt}
\end{equation}
Namely, the source $J$ is obtained from the terms explicitly written on  right hand side of 
(\ref{mod1-eqphi2}), while ${\tilde J}$ is obtained from the terms denoted with dots.  The two sources are uncorrelated, and therefore the two particular solutions sourced by them can be obtained independently. The solution sourced by ${\tilde J}$ is the one emerging from  cosmological second order perturbation theory in absence of gauge fields. For standard scalar field slow roll inflation, this term provides a negligible contribution to the power spectrum, and unobservable non-Gaussianity. Therefore, we disregard this term in this work. In Appendix \ref{app-mod1-sourcez} we show that 
\begin{equation}
J \simeq  \frac{\dot{\phi}}{2 M_p^2 H a} \int{\frac{d^3 p}{\left( 2 \pi \right)^{3/2}}} \left[  \hat{k}_i \hat{k}_j \left( M^2 - \partial_\tau^{(1)} \,   \partial_\tau^{(2)} \right)  - M^2 \, \delta_{ij} \right] {\tilde A}_i \left( \tau ,\, \vec{p} \right) {\tilde A}_j \left( \tau ,\, \vec{k} -  \vec{p} \right) 
\;\;\;,\;\;\; M \equiv a \, m = a \, e \, \psi^{(0)}
\label{mod1-source-z}
\end{equation}
where we are using the notation
\begin{equation}
\partial_\tau^{(1)} \,   \partial_\tau^{(2)} \;  f \; g  \equiv \partial_\tau f \; \partial_\tau g 
\label{12-not} 
\end{equation}
It is worth noting that the source does not diverge as $k \rightarrow 0$; this is not immediate from (\ref{mod1-eqphi2}), since one term contains an inverse laplacian. See  Appendix \ref{app-mod1-sourcez} for details.

From (\ref{mod1-source-z}) we see that, for this model, the source is formally of the type (\ref{Jf-formal}), with
\begin{equation}
{\hat O}_{\varphi,ij} \left( \tau ,\, \vec{k} ,\, \vec{p} \right) =  \frac{\dot{\phi}}{2 M_p^2 H a}  \left[  \hat{k}_i \hat{k}_j \left( M^2 - \partial_\tau^{(1)} \,   \partial_\tau^{(2)} \right)  - M^2 \, \delta_{ij} \right] 
\label{mod1-Ophi}
\end{equation}

We insert this operator in (\ref{formal-Pz}); we also use the fact that ${\cal D}_{(0,0)}^{(\sigma)} $ is actually helicity independent in this model, so that the sum over the helicity $\sigma$ is limited to
$\sum_\sigma {\cal P}_{im}^{(\sigma)} \left( {\hat p} \right) = \delta_{im} - {\hat p}_i \, {\hat p}_m$ - see  
(\ref{identity-pol}) - and analogously for the sum over $\sigma'$. Moreover, in the integrand we approximate $\vec{p} - \vec{k} \simeq \vec{p}$.~\footnote{The reason for this is that, as we shall see,  the signal from particle production is maximal at $k \sim H$ (namely, for modes of the size of the horizon when particle production occurs; recall that the scale factor is normalized to one at the moment of particle production). We shall also see  that the source integrand is peaked at $p \sim \sqrt{\dot{m}_*}$. We therefore have $k \ll p$ due to (\ref{mod1-bck-cond}).} We obtain 
\begin{eqnarray}
P_{\zeta,{\rm s}} \left( k \right) & \simeq & \frac{k^3}{4 \pi^2 a^2 M_p^4} \, \int_{\tau_*}^\tau d \tau_1 \frac{G_k \left( \tau ,\, \tau_1 \right) }{a \left( \tau_1 \right)}  \int_{\tau_*}^\tau d \tau_2 \frac{G_k \left( \tau ,\, \tau_2 \right) }{a \left( \tau_2 \right)} \;  \int{\frac{d^3 p}{\left( 2 \pi \right)^3}}  \;
\Bigg\{ \left[ 1 + \left( {\hat k} \cdot {\hat p} \right)^4 \right] M^2 \left( \tau_1 \right)  M^2 \left( \tau_2 \right) 
\delta_0^a \, \delta_0^b \nonumber\\
&&  \quad\quad\quad\quad 
+  \left(   {\hat k} \cdot {\hat p} \right)^2  \left[ 1 - \left(   {\hat k} \cdot {\hat p} \right)^2 \right]
\left[ M^2 \left( \tau_1 \right)  \delta_0^a \, \delta_1^b  + 
 M^2 \left( \tau_2 \right)  \delta_1^a \, \delta_0^b  \right] 
+ \left[ 1 - \left(   {\hat k} \cdot {\hat p} \right)^2 \right]^2 
\delta_1^a \, \delta_1^b   \Bigg\} \,
{\cal D}_{(a,b)} \left[ \tau_1 ,\, \tau_2 ;\, p \right]^2 \nonumber\\
\label{mod1-pz-firstexp}
 \end{eqnarray}
where we have omitted the suffix $(\sigma)$ from the correlators (since they are $\sigma-$independent in this model). By inserting the expressions (\ref{mod1-D00}) and (\ref{mod1-Dab}) for the correlators, and by performing the angular integrals, we obtain
\begin{eqnarray}
P_{\zeta,{\rm s}} \left( k \right) & \simeq & \frac{k^3}{4 \pi^2 a^2 M_p^4} \, \int_{\tau_*}^\tau d \tau_1 \frac{G_k \left( \tau ,\, \tau_1 \right) }{a \left( \tau_1 \right)}  \int_{\tau_*}^\tau d \tau_2 \frac{G_k \left( \tau ,\, \tau_2 \right) }{a \left( \tau_2 \right)} \;  
\frac{M \left( \tau_1 \right) M \left( \tau_2 \right)}{2 \pi^2} \int d p \, p^2 \nonumber\\
&& \!\!\!\!\!\!\!\!
\Bigg\{
\vert \beta_p \vert^2 \left[ \vert \alpha_p \vert^2 + \vert \beta_p \vert^2 \right]
+ \frac{2 \vert \beta_p \vert^2}{3} \, {\rm Re } \left[ \alpha_p^* \beta_p \left( {\rm e}^{i \gamma \left( \tau_1 \right)} +  {\rm e}^{i \gamma \left( \tau_2 \right)}  \right) \right]
+ \frac{11}{15} \, {\rm Re } \left[ \alpha_p^{*2} \beta_p^2 {\rm e}^{i \left[ \gamma \left( \tau_1 \right) +  \gamma \left( \tau_2 \right) \right]} + \vert \beta_p \vert^4  {\rm e}^{i \left[ \gamma \left( \tau_1 \right) -  \gamma \left( \tau_2 \right) \right]} \right] \bigg\} \nonumber\\
\label{mod1-pz-intermediate}
\end{eqnarray}
where $\gamma \left( \tau \right) \equiv 2 \int_{\tau_*}^\tau d \tau' \, M \left( \tau' \right)$, and where the Bogolyubov coefficients are evaluated after the particle production, and given in  (\ref{mod1-ablate}).
To obtain this expression we have disregarded $p$ with respect to $M$ inside the mode functions $f_p$ present inside the correlators. This is appropriate since, as we shall see, the integrand is peaked at $p \sim \sqrt{\dot{m}_*} \ll M$.

The curly parenthesis in (\ref{mod1-pz-intermediate}) contains several terms proportional to the fast oscillating phase ${\rm e}^{i \gamma}$. All these terms give a negligible contribution to the final result. To see this, we can  separate the momentum and time integrals in this expression, and obtain
\begin{eqnarray}
P_{\zeta,{\rm s}} \left( k \right) & \simeq & \frac{k^3}{8 \pi^4 a^2 M_p^4} \int d p p^2 \left\{ 
\vert \beta_p \vert^2 \left[ \vert \alpha_p \vert^2 + \vert \beta_p \vert^2 \right] {\cal T}_k^2
+ \frac{4}{3} \vert \beta_p \vert^2 {\cal T}_k \, {\rm Re } \left( \alpha_p^* \beta_p {\cal E}_k \right)
+ \frac{11}{15} \vert \beta_p \vert^4 \vert {\cal E}_k \vert^2 + \frac{11}{15} \, {\rm Re } \left( \alpha_p^{* 2} \beta_p^2 {\cal E}_k^2 \right) \right\} \nonumber\\
\label{mod1-sep-tp}
\end{eqnarray}
where
\begin{equation}
{\cal T}_k \equiv \int_{\tau_*}^\tau d \tau' G_k \left( \tau ,\, \tau' \right) \, m \left( \tau' \right) \;\;\;,\;\;\;
{\cal E}_k \equiv \int_{\tau_*}^\tau d \tau' G_k \left( \tau ,\, \tau' \right) \, m \left( \tau' \right) \, {\rm e}^{2 i \int_{\tau_*}^{\tau'} d \tau'' M \left( \tau'' \right)}
\label{def-TE}
\end{equation}

Using the expression (\ref{formal-sol2}) for the Green function, the expression
\begin{equation}
m \left( \tau \right) = \dot{m}_* \, \left( t - t_* \right) = \frac{\dot{m}_*}{H} \, {\rm ln } \left( \frac{-\tau_*}{-\tau} \right) = - \frac{\dot{m}_*}{H} \, {\rm ln } \left( -H \tau \right) 
\label{mod1-mtau}
\end{equation}
(in the last equality we have used the fact that $a \left( \tau_* \right) = 1$) we find, in the super-horizon $- k \tau \ll 1$ limit (in practice, we introduce the integration variable $y' = - k \tau'$, and we integrate it from $0$ to $-k \tau_*$, rather than from $-k \tau$ to $-k \tau_*$),
\begin{equation}
{\cal T}_k \simeq \frac{ a\left( \tau \right) \dot{m}_* }{ 27 \, H^3 } \, _2F_3 \left( \frac{3}{2} ,\, \frac{3}{2}  ;\, \frac{5}{2} ,\, \frac{5}{2} ,\, \frac{5}{2} ;\; \frac{- k^2}{4 H^2} \right) 
\end{equation}
where the generalized hypergeometric function evaluates to $\sim 1$ for $k \ll H$ and to $\sim \frac{27 \pi}{2} \frac{H^3}{k^3} \left( {\rm ln } \frac{k}{H} - 0.423 \right)$ for $k \gg H$. 

The time integral in the exponent of ${\cal E}$ can be done analytically, and, leads to
\begin{equation}
{\cal E}_k   =  \frac{a \left( \tau \right) \, \dot{m}_*}{k^3} \, \int_{-H \tau}^1 \frac{d y}{y} \left[ \frac{k}{H} y \, \cos \left( \frac{k}{H} y \right) - \sin \left( \frac{k}{H} y \right) \right] \, \ln \left( y \right) \left\{  \cos \left[ \frac{\dot{m}_*}{H^2} \, \ln^2 y \right] + i \, \sin \left[ \frac{\dot{m}_*}{H^2} \, \ln^2 y \right] \right\}
\end{equation}

Since $\dot{m}_* \gg H$, the term in curly parenthesis is rapidly oscillating unless $y \equiv \frac{\tau'}{\tau_*} \simeq 1$. This however means that the integrand is peaked at the moment in which the particle production is taking place. In our computation we used the asymptotic values  (\ref{mod1-ablate}) for 
$\alpha ,\, \beta$ (valid once particle production has completed). Therefore we  only provide  an upper bound on ${\cal E}_k$. It is easy to verify that instead the integrand of  ${\cal T}_k$ is dominated by times at which  (\ref{mod1-ablate}) hold.

Provided that $k \ll \sqrt{\dot{m}_*}$ (which is true in the range of our interest, since the occupation number of the gauge field is exponentially suppressed in the opposite regime), we can keep in the integral only the fast oscillating term, and the log term, and obtain the estimate
\begin{eqnarray}
\vert {\cal E}_k \vert & < & \left\vert \frac{a \left( \tau \right) \, \dot{m}_*}{k^3} \, 
\left[ \frac{k}{H}  \, \cos \left( \frac{k}{H}  \right) - \sin \left( \frac{k}{H}  \right) \right] 
 \, \int_{-H \tau}^1 d y \, \ln \left( y \right) \, \left\{  \cos \left[ \frac{\dot{m}_*}{H^2} \, \ln^2 y \right] + i \, \sin \left[ \frac{\dot{m}_*}{H^2} \, \ln^2 y \right] \right\} \right\vert \nonumber\\
& \simeq & \frac{a \left( \tau \right) H^2}{2 k^3 } \, \left\vert 
 \frac{k}{H} \,  \cos \frac{k}{H} - \sin \frac{k}{H}  \right\vert 
\label{estimate-E}
\end{eqnarray}
for the upper bound. We indeed see that $\frac{\vert {\cal E}_k \vert}{{\cal T}_k} < \frac{H^2}{\dot{m}_*} \ll 1$.

This confirms that the oscillatory terms in (\ref{mod1-pz-intermediate}) - or (\ref{mod1-sep-tp}) - provide a negligible contribution to the final result. Performing the momentum integral we finally obtain
\begin{equation}
P_{\zeta,{\rm s}} \left( k \right)  \simeq  
\frac{2+\sqrt{2}}{46,656 \pi^5} \, \frac{k^3 \, \dot{m}_*^{7/2}}{H^6 M_p^4} \, \left[ _2F_3 \left( \frac{3}{2} ,\, \frac{3}{2}  ;\, \frac{5}{2} ,\, \frac{5}{2} ,\, \frac{5}{2} ;\; \frac{- k^2}{4 H^2} \right)  \right]^2  \;\;\;,\;\;\; k \ll \sqrt{ \dot{m}_* }
\label{mod1-Pz}
\end{equation}
while the result is exponentially suppressed and uninteresting at larger momenta. The exponential suppression of the spectrum is due to the fact that the occupation numbers of the vector particles are exponentially suppressed at such large momenta, see eq. (\ref{mod1-ablate}). We discuss this result in Subsection \ref{subsec:Pheno1}.

\subsubsection{Scalar bispectrum}

We now evaluate the formal expression (\ref{formal-Bz}) for the bispectrum in this model. This expression contains various intermediate quantities that have been given above. Explicitly, we write the green functions  in eq. (\ref{formal-sol2}), the projection operators in eq. (\ref{identity-pol}), the operator ${\cal O}_\varphi$ in eq. (\ref{mod1-Ophi}), and the correlators in eqs. (\ref{mod1-D00}) and (\ref{mod1-Dab}). The computation follows the same steps presented in the previous Subsection for $P_\zeta$.
Also in this case we use the approximation $k \ll p \ll M$, and we find that the terms with a fast oscillating phase in the final time integral can be disregarded. We obtain
\begin{eqnarray}
B \left( k_1 ,\, k_2 ,\, k_3 \right) & \simeq & \frac{1}{2 M_p^6 a \left( \tau \right)^3} \, {\cal T}_{k_1} \, 
 {\cal T}_{k_2} \,  {\cal T}_{k_3} \, \int \frac{d^3p}{\left( 2 \pi \right)^{9/2}} \vert \beta \vert^4 \left( 3 \vert \alpha \vert^2 + \vert \beta \vert^2 \right) \nonumber\\
 & \simeq & \frac{27+8 \sqrt{6}}{22,674,816 \, \pi^{9/2}} \, \frac{\dot{m}_*^{9/2}}{H^9 M_p^6}
  \, \prod_{i=1}^3  \,  _2F_3 \left( \frac{3}{2} ,\, \frac{3}{2}  ;\, \frac{5}{2} ,\, \frac{5}{2} ,\, \frac{5}{2} ;\; \frac{- k_i^2}{4 H^2} \right)  \;\;\;,\;\;\;  k_i \ll \sqrt{ \dot{m}_* }
 \end{eqnarray}

This corresponds to the effective equilateral nonlinear parameter (\ref{formal-fnl})
\begin{equation}
f_{NL,{\rm equil. \; eff.}} \left( k \right) \simeq 2.1 \cdot 10^7 \frac{k^6 \, \dot{m}_*^{9/2}}{H^9 M_p^6} \, 
 \left[ _2F_3 \left( \frac{3}{2} ,\, \frac{3}{2}  ;\, \frac{5}{2} ,\, \frac{5}{2} ,\, \frac{5}{2} ;\; \frac{- k^2}{4 H^2} \right)  \right]^3  \;\;\;,\;\;\; k \ll \sqrt{ \dot{m}_* }
\end{equation}
where we have used the numerical value $P_\zeta \simeq 2.5 \cdot 10^{-9}$ for the power spectrum.
Analogous to the sourced part of the power spectrum (\ref{mod1-Pz}), the bispectrum and the nonlinear parameter are exponentially suppressed and uninteresting at larger momenta.

\subsection{Gravity waves sourced by the vector modes}
\label{subsec:GW1}

We now compute the amount of gravity waves sourced by the produced gauge fields. Inserting  the energy momentum tensor (\ref{Tmunu-mod1}) for this model in the general expression (\ref{Jlambda-formal}), we see that the source is formally of the type (\ref{pbm-formal}) with
\begin{equation}
{\cal O}_{ij,\lambda}= \frac{\Pi_{mn,\lambda}^*  \left( {\hat k} \right) }{a M_p} \left[ \delta_{mi} \delta_{nj} \left( - \partial_\tau^{(1)} 
 \partial_\tau^{(2)} + M^2 \right) + \epsilon_{mai} \epsilon_{nbj} p_a \left( k - p \right)_b \right]
\label{mod1-Olambda}
 \end{equation}
where we recall our notation (\ref{12-not}).

It is instructive to compare this expression with the analogous operator in the source of the scalar perturbations, given in  (\ref{mod1-Ophi}). In writing  (\ref{mod1-Ophi}) we disregarded terms proportional to spatial momenta $\vec{p}$ and $\vec{p}-\vec{k} \simeq \vec{p}$ with respect to time derivatives and the mass $M$. The reason for this is that $k \ll p \ll M$, as we explained   after  (\ref{mod1-pz-intermediate}). As a consequence, time derivatives acting on a mode also give $\partial_\tau A \sim M A \gg p A$, and should be retained in the operator, when compared to spatial momenta. Disregarding the momentum terms  is correct for the scalar source. However, we see that in (\ref{mod1-Olambda}) the time derivatives enter with an opposite sign to $M^2$. As a consequence, the  dominant contributions from the first two terms in (\ref{mod1-Olambda}) cancel against each other, and therefore we need to keep the complete structure. We now show explicitly how the cancellation arises.

We insert the operator (\ref{mod1-Olambda}) in the formal expression (\ref{formal-Ph}) for the power in the tensor modes. The other intermediate quantities entering in  (\ref{formal-Ph}) are the Green functions, given  in  (\ref{formal-sol2}), the projection operators, given in  (\ref{identity-pol}),  and the correlators, given in  (\ref{mod1-D00}) and (\ref{mod1-Dab}). We obtain~\footnote{We remind that the tensor power is obtained from $\langle h_\lambda  h_\lambda \rangle$. We have verified that   $\langle h_\lambda  h_\lambda' \rangle \propto \delta_{\lambda \lambda'}$.}
\begin{eqnarray}
P_{\lambda,{\rm s}}  & = & \frac{2 k^3}{3 \pi^4 a^2 M_p^4}  \, \int_{\tau_*}^\tau d \tau_1 \frac{G_k \left( \tau ,\, \tau_1 \right) }{a \left( \tau_1 \right)}  \int_{\tau_*}^\tau d \tau_2 \frac{G_k \left( \tau ,\, \tau_2 \right) }{a \left( \tau_2 \right)} \;  \int d p \, p^2 \;
\Bigg\{\frac{7}{5} \left[ M \left( \tau_1 \right)^2 \delta_{a0} - \delta_{a1} \right]
 \left[ M \left( \tau_2 \right)^2 \delta_{b0} - \delta_{b1} \right] \nonumber\\
 && \quad\quad  \quad\quad  \quad\quad   \quad\quad  \quad\quad   + 
 p^2 \; \delta_{b0}  \left[ M \left( \tau_1 \right)^2 \delta_{a0} - \delta_{a1} \right] + 
 p^2 \; \delta_{a0}  \left[ M \left( \tau_2 \right)^2 \delta_{b0} - \delta_{b1} \right] + 
\frac{7}{5} p^4 \delta_{a0} \delta_{b0} \Bigg\}  \, {\cal D}_{(a,b)} \left[ \tau_1 ,\, \tau_2 ;\, p \right]^2 \nonumber\\
\label{mod1-ph-firstexp}
 \end{eqnarray}
where we have disregarded $k$ as compared to $p$.

This expression is analogous to eq. (\ref{mod1-pz-firstexp}) for $P_\zeta$, with the difference that here we have already performed the trivial angular integrals of $d^3 p$. Terms without $p$ in the curly parenthesis are obtained from the square of the first term $\propto \left( - \partial_\tau^{(1)}  \partial_\tau^{(2)} + M^2 \right) $ in (\ref{mod1-Olambda}) ( recall that two ${\cal O}_\lambda$ enter in $P_\lambda$ ) 
. We note that the structure  $ - \partial_\tau^{(1)}  \partial_\tau^{(2)} + M^2  $ is preserved in (\ref{mod1-ph-firstexp}) since the suffix $0$ ($1$) on ${\cal D}$ indicates that ${\tilde A}$ 
(${\tilde A'}$) is present in the correlator. The term  $\propto p^4$  in the curly parenthesis is obtained from square of the other term in  (\ref{mod1-Olambda}). The terms  $\propto p^2$ are the mixed terms. 
The different coefficients ($\frac{7}{5}$ vs $1$) follow from the angular integrals.

The result (\ref{mod1-ph-firstexp}) is a sum of squares of correlators. In each square, most terms present fast oscillating phases; as we shall see, these terms give a negligible contribution to the final result once the time integrals are performed. This is analogous to $\vert {\cal T} \vert \gg \vert {\cal E} \vert $ in eq. (\ref{def-TE}). Squaring the expressions  (\ref{mod1-D00}) and (\ref{mod1-Dab}) we obtain
\begin{eqnarray}
{\cal D}_{(0,0)} \left[ \tau_1 ,\, \tau_2 ;\, p \right]^2 &=& \frac{\vert \beta_p \vert^2 \left( \vert \alpha_p \vert^2 + \vert \beta_p \vert^2 \right)}{2 \omega_p \left( \tau_1 \right)  \omega_p \left( \tau_2 \right) } + {\rm oscillatory \; phases} \nonumber\\
{\cal D}_{(1,0)} \left[ \tau_1 ,\, \tau_2 ;\, p \right]^2 &=& \omega_p^2 \left( \tau_1 \right)  {\cal D}_{(0,0)} \left[ \tau_1 ,\, \tau_2 ;\, p \right]^2 +  {\rm oscillatory \; phases} \nonumber\\
{\cal D}_{(0,1)} \left[ \tau_1 ,\, \tau_2 ;\, p \right]^2 &=& \omega_p^2 \left( \tau_2 \right)  {\cal D}_{(0,0)} \left[ \tau_1 ,\, \tau_2 ;\, p \right]^2 +  {\rm oscillatory \; phases} \nonumber\\
{\cal D}_{(1,1)} \left[ \tau_1 ,\, \tau_2 ;\, p \right]^2 &=& \omega_p^2 \left( \tau_1 \right) \omega_p^2 \left( \tau_2 \right)  {\cal D}_{(0,0)} \left[ \tau_1 ,\, \tau_2 ;\, p \right]^2 +  {\rm oscillatory \; phases} 
\label{mod1-nonosc-gw}
\end{eqnarray}
We recall that $\omega_p \left( \tau_i \right) = \sqrt{M \left( \tau_i \right)^2 + p^2}$, and that momentum integrand has its support in the region $p \ll M \left( \tau_i \right) $. We see, however, that the 
${\rm O } \left( M^4 \right)$ and ${\rm O } \left( M^2 \right)$ parts of the curly parenthesis in (\ref{mod1-ph-firstexp}) cancel for the non-oscillatory contributions (\ref{mod1-nonosc-gw}). 

Using the full expressions  (\ref{mod1-D00}) and (\ref{mod1-Dab}) for the correlators, we obtain
\begin{eqnarray}
P_{\lambda,{\rm s}}  & = & \frac{2 k^3}{3 \pi^4 a^2 M_p^4}  \, \int_{\tau_*}^\tau d \tau_1 \frac{G_k \left( \tau ,\, \tau_1 \right) }{a \left( \tau_1 \right)}  \int_{\tau_*}^\tau d \tau_2 \frac{G_k \left( \tau ,\, \tau_2 \right) }{a \left( \tau_2 \right)} \;  \int d p \, p^2 \; \nonumber\\
 &&   
\Bigg\{ 
\frac{2 p^4}{5 \, \omega_p \left( \tau_1 \right) \omega_p \left( \tau_2 \right)} \vert \beta_p \vert^2 \left( \vert \alpha_p \vert^2 + \vert \beta_p \vert^2 \right)
  - \frac{4}{5} p^2 \vert \beta_p \vert^2 \left[ \frac{M^2 \left( \tau_1 \right)}{\omega_p \left( \tau_1 \right) \omega_p \left( \tau_2 \right)} \, {\rm Re } \left( \alpha_p^* \beta_p \, {\rm e}^{i \gamma \left( \tau_1 \right)} \right) + \tau_1 \leftrightarrow \tau_2 \right] \nonumber\\
&&  
   + \frac{2}{5} \, \frac{7 M^2 \left( \tau_1 \right) M^2 \left( \tau_2 \right) + 6 p^2 \left[ M^2 \left( \tau_1 \right) M^2 \left( \tau_2 \right) \right] + 6 p^4}{\omega_p \left( \tau_1 \right)  \omega_p \left( \tau_2 \right)  }
\left[ {\rm Re } \left( \alpha_p^{* 2} \beta_p^2 {\rm e}^{i \left[ \gamma \left( \tau_1 \right) +  \gamma \left( \tau_2 \right) \right]} \right) + \vert \beta_p \vert^4 {\rm Re } \left( {\rm e}^{i \left[ \gamma \left( \tau_1 \right) -  \gamma \left( \tau_2 \right) \right]} \right) \right] \Bigg\}  \nonumber\\
\label{mod1-ph-secondstexp}
 \end{eqnarray}
where $\gamma \left( \tau_i \right)$ are the oscillatory phases defined immediately after (\ref{mod1-pz-intermediate}). The first term in the curly parenthesis is obtained from the non-oscillatory parts (\ref{mod1-nonosc-gw}), and we see that it is indeed of ${\rm O } \left( p^4 \right)$. As we shall see, this is the term that dominates the final result.

Eq. (\ref{mod1-ph-secondstexp}) is therefore characterized by a part without oscillatory phase plus a part with one oscillatory phase plus a part with two oscillatory phases. For each part we only take the leading prefactor in the $p \ll M \left( \tau_i \right)$ regime. This expression then rewrites 
\begin{eqnarray}
P_{\lambda,{\rm s}} & \simeq & \frac{2 k^3}{3 \pi^4 a^2 M_p^4}   \int d p \, p^2 \left\{
\frac{2}{5} p^4 \vert \beta_p \vert^2 \left( \vert \alpha_p \vert^2 + \vert \beta_p \vert^2 \right) \tilde{\cal T}_k^2 - \frac{8}{5} p^2 \vert \beta_p \vert^2 \tilde{\cal T}_k {\rm Re } \left( \alpha_p^* \beta_p \tilde{\cal E}_k \right) 
+ \frac{14}{5} {\rm Re} \left( \alpha_p^{*2} \beta_p^2 \tilde{\cal E}_k^2 \right) + \frac{14}{5} \vert \beta_p \vert^4 \vert \tilde{\cal E}_k \vert^2 \right\} \nonumber\\
\label{mod1-ph-thirdexp}
 \end{eqnarray}
where
\begin{equation}
\tilde{\cal T}_k \equiv \int_{\tau_{\rm min}}^\tau d \tau' \frac{G_k \left( \tau ,\, \tau' \right)}{a \left( \tau' \right) \omega_p \left( \tau' \right)} \;\;\;,\;\;\;
\tilde{\cal E}_k \equiv  \int_{\tau_{\rm min}}^\tau d \tau' \frac{G_k \left( \tau ,\, \tau' \right)}{a \left( \tau' \right) \omega_p \left( \tau' \right)} 
\, M^2 \left( \tau' \right) \, {\rm e}^{2 i \int_{\tau_*}^{\tau'} d \tau'' M \left( \tau'' \right)}
\label{def-TE-tilde}
\end{equation}

We evaluate the time integrals after the particle production ($\tau > \tau_{\rm min}$), so that $\alpha_p$ and $\beta_p$ can indeed be taken as constant. Particle production is completed when $\omega_p^2 > \dot{\omega_p}$, which, in the support region $p \lsim \sqrt{\dot{m}_*}$ of the momentum integral, gives
$t_{\rm min} \simeq t_* + \frac{1}{\sqrt{\dot{m}_*}}$. This corresponds to the conformal time
$\tau_{\rm min} \simeq - \frac{1}{H} \, {\rm exp } \left( - \frac{H}{\sqrt{\dot{m}_*}} \right)$. For $\tau > \tau_{\rm min}$, we can approximate $\omega_p \simeq M \simeq a \, m$, with $m$ given in (\ref{mod1-mtau}). We then obtain the expression 
\begin{eqnarray}
\tilde{\cal T}_k &\simeq& \frac{H^2 a \left( \tau \right)}{\dot{m}_* \, k^3} \, 
\int_{-H \tau}^{{\rm exp} \left( - \frac{H}{\sqrt{\dot{m}_*}} \right)} dy' \left[ \frac{k}{H} y' \, \cos \left( \frac{k}{H} y' \right) - \sin \left( \frac{k}{H} y' \right) \right] \, \frac{y'}{\ln \left( y' \right)} \nonumber\\
&\simeq&  \frac{H^2 a \left( \tau \right)}{\dot{m}_* \, k^3} \, \left[ \frac{k}{H}  \, \cos \left( \frac{k}{H}  \right) - \sin \left( \frac{k}{H}  \right) \right] \, \int_{0}^{{\rm exp}  \left( - \frac{H}{\sqrt{\dot{m}_*}} \right)} 
 \frac{d y'}{\ln \left( y' \right)} \simeq   \frac{H^2 a \left( \tau \right)}{\dot{m}_* \, k^3} \, \left[   \sin \left( \frac{k}{H}  \right)  -   \frac{k}{H}  \, \cos \left( \frac{k}{H}  \right)      \right] \, {\rm ln } \left( \frac{\sqrt{\dot{m}_*}}{H} \right) 
\nonumber\\
\label{mod1-ttilde}
 \end{eqnarray}
where the integration variable in the first expression is $y' = - H \tau'$.  In going from the first to the second line we have used the fact that the integrand is peaked at the asymptotically early times, while in the final expression we have used $H \ll \sqrt{\dot{m}_*}$. We note that the result is only logarithmically sensitive to the difference between $\tau_{\rm min}$ and $\tau_*$.

Under the approximation $\omega \simeq M$, the integral $\tilde{\cal E}_k$ coincides with the integral 
 ${\cal E}_k$ defined in (\ref{def-TE}) (we actually notice that $\vert \tilde{\cal E}_k \vert \lsim \vert {\cal E}_k \vert $). We obtained an upper bound for this quantity in  (\ref{estimate-E}). The relative contribution of the two time integrals to (\ref{mod1-ph-thirdexp}) can therefore be estimated as
 \begin{equation}
\left\vert \frac{p^2 \, \tilde{\cal T}_k }{  \tilde{\cal E}_k } \right\vert_{p \simeq \sqrt{\dot{m}_*}} >  {\rm ln } \left( \frac{\sqrt{\dot{m}_*}}{H} \right) \gg 1
\end{equation} 
This shows that the oscillatory integral $\tilde{\cal E}_k$ can be disregarded in our estimate of $P_{\lambda,{\rm s}}$,
\begin{equation}
P_{\lambda,{\rm s}} \big\vert_{{\rm from \; } A_T} \simeq  \frac{4 k^3}{15 \pi^4 a^2 M_p^4} 
\tilde{\cal T}_k^2  \int d p \,  p^6 \vert \beta_p \vert^2 \left( \vert \alpha_p \vert^2 + \vert \beta_p \vert^2 \right) 
\label{mod1-Plambda-beforpint}
\end{equation}
Using the results (\ref{mod1-ablate}), we finally obtain
\begin{equation}
P_{\lambda,{\rm s}} \big\vert_{{\rm from \; } A_T} \simeq  \frac{\left( 8 + \sqrt{2} \right) \, H^4 \, \dot{m}_*^{3/2}}{32 \, \pi^7 \, M_p^4 \, k^3} \,
\left[ \sin \left( \frac{k}{H} \right) - \frac{k}{H} \, \cos \left( \frac{k}{H} \right) \right]^2  \, {\rm ln }^2 \left( \frac{\sqrt{\dot{m}_*}}{H} \right) \;\;\;,\;\;\; k \ll \sqrt{\dot{m}_*}
\label{mod1-Plambda}
\end{equation}
while the result is exponentially suppressed and uninteresting at larger momenta. We discuss this result in Subsection \ref{subsec:Pheno1}.

In these last two expressions we have emphasized that in this computation we have considered only the gravity waves sourced by the transverse modes of the gauge field. In Appendix \ref{app-longA} we present the full computation, including also the longitudinal vector polarization. We find that the expression 
(\ref{mod1-Plambda-beforpint}) is replaced by 
\begin{equation}
P_{\lambda,{\rm s}} 
\big\vert_{{\rm from \; } A_T \; {\rm and \; } A_L}
 \simeq  \frac{2 k^3}{15 \pi^4 a^2 M_p^4} 
\tilde{\cal T}_k^2  \int d p \,  p^6 \left[ 2 \vert \beta_p \vert^2 \left( \vert \alpha_p \vert^2 + \vert \beta_p \vert^2 \right) +   \vert \beta_p^L \vert^2 \left( \vert \alpha_p^L \vert^2 + \vert \beta_p^L \vert^2 \right) \right]
\label{Plamda-TT-LL}
\end{equation}
where $\alpha^L$ and $\beta^L$ are the Bogolyubov coefficients of the longitudinal mode. Namely, we see that, if they are produced in the same amount, the longitudinal quanta contribute to the gravity waves as one transverse polarization. This results in a factor $\frac{3}{2}$ multiplying the final result (\ref{mod1-Plambda}).

\subsection{Phenomenology}
\label{subsec:Pheno1}

\begin{figure}[htbp]
\begin{center}
\includegraphics[width=0.4\textwidth]{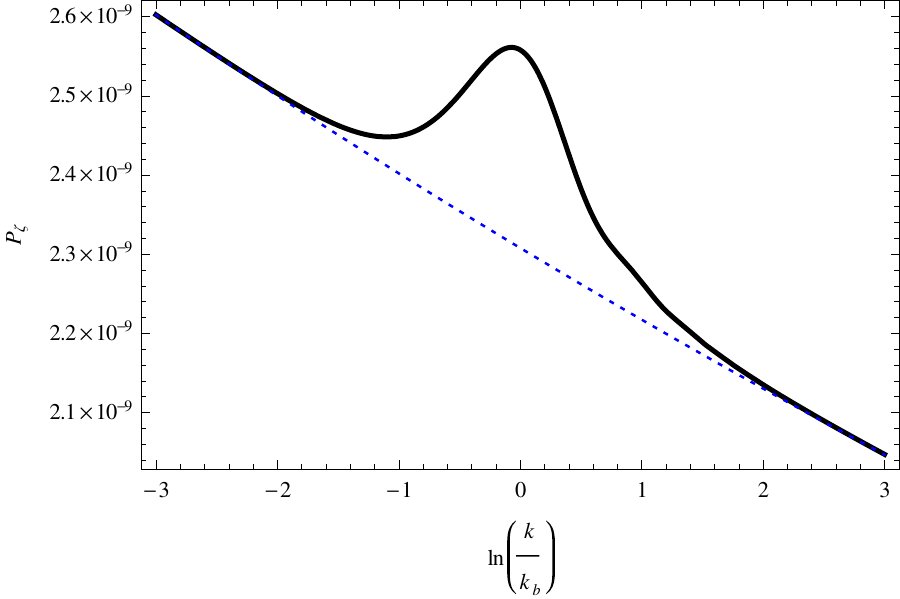}
\includegraphics[width=0.4\textwidth]{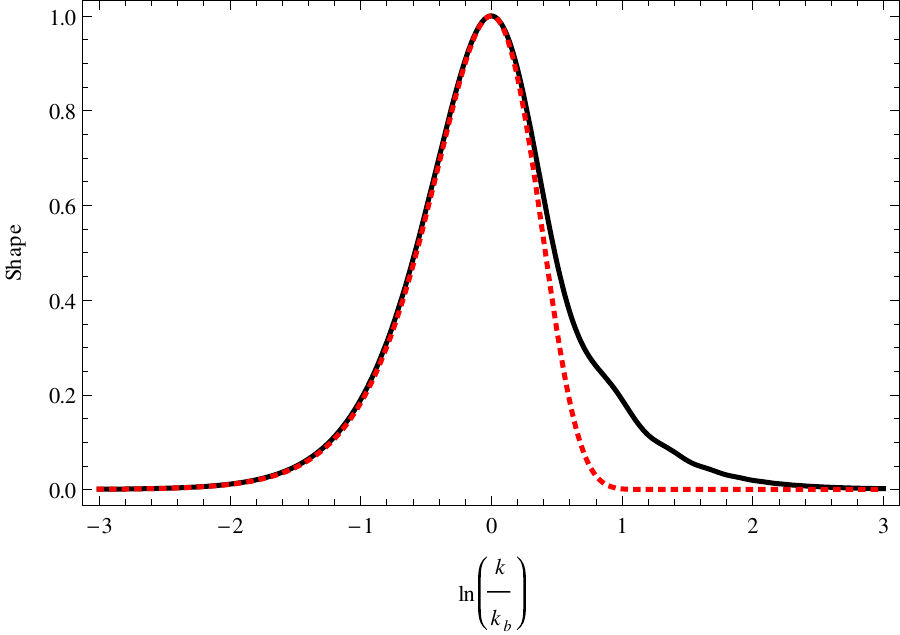}
\end{center}
\caption{Left panel: the total observable power spectrum in the model (\ref{model1}) for a representative choice of parameters, illustrating the appearance of a localized bump feature.  The solid black curve is the total spectrum while the dashed blue line gives the spectrum of the vacuum fluctuations for comparison.  Right panel: a comparison of the shape function $S_b$ (solid black curve) and the fitting function $S_{\mathrm{fit}}$ (dashed red curve).}
\label{fig:feature}
\end{figure}

The observable spectrum of curvature fluctuations in the model (\ref{model1}) is given by the sum of the sourced contribution and the usual nearly scale-invariant contribution from the vacuum fluctuations; see section \ref{sec:formal}.  The sourced part of the spectrum in the model (\ref{model1}) leads to a localized ``bump'' feature in the primordial power spectrum for scales leaving the horizon at the moment $t=t_*$, when the gauge fields are produced.  We have illustrated this feature in the left panel of Fig.~\ref{fig:feature} for an arbitrary choice of parameters.  The most stringent observational constraints on the model (\ref{model1}) come from non-observation of such localized bump features.  These observational constraints are the subject of this subsection.~\footnote{In this discussion, we disregard the contribution from the longitudinal vector mode to the source of tensor and scalar perturbations. As discussed at the end of the previous Subsection, the longitudinal mode changes the result for the gravity wave spectrum by at most a factor  $\frac{3}{2}$ (if it is produced in equal amount to each tensor mode). We expect an analogous enhancement for  $\zeta$. The current discussion can be easily modified to account for these additional  ${\rm O} \left( 1 \right)$ factors; all of our conclusions are unchanged.}

We can write the total observable power spectrum in the following form:
\begin{equation}
\label{model1:Ptot}
 P_{\zeta}(k) = 
  \underbrace{   \mathcal{P} \left(\frac{k}{k_p}\right)^{n_s-1}}_{ = P_{\zeta,\mathrm{v}}(k)} 
  + \underbrace{A_{b} \, S_{b}\left[\frac{k}{k_{b}}\right]}_{ = P_{\zeta,\mathrm{s}}(k) } \, .
\end{equation}
The spectrum of the vacuum fluctuations are characterized as usual: $k_p$ is the pivot scale (taken to be $0.002 \, \mathrm{Mpc}^{-1}$, consistent with \cite{Hinshaw:2008kr}) while the amplitude and tilt are given by
\begin{equation}
\label{vac_characterization}
  \mathcal{P} \equiv \frac{H^2}{8\pi^2 \epsilon M_p^2} \, , \hspace{5mm} n_s = 1 + 2\eta - 6\epsilon \, .
\end{equation}
Here all quantities are understood to be evaluated at the moment when the pivot scale left the horizon.  The sourced contribution in (\ref{model1:Ptot}) describes a localized bump-like feature that we characterize by an amplitude $A_b$, a location $k_b$, and a ``shape function'' $S_b$.  The shape and amplitude are given by:
\begin{equation}
\label{bump_characterization}
 A_b \equiv 3.2 \cdot 10^{-2} \, \mathcal{P}^2 \, \epsilon^2 \, \frac{\dot{m}_*^{7/2}}{H^7} \, ,
 \hspace{5mm}
 S_b [x] = 4.7 \, x^3 \, \left[_{2}F_{3} \left( \frac{3}{2},\frac{3}{2};\frac{5}{2},\frac{5}{2},\frac{5}{2}; -5.5\, x^2\right)  \right]^2 \, .
\end{equation}
The shape function has been constructed so that the global maximum is $S_b(x=1) = 1$, meaning that the feature in (\ref{model1:Ptot}) reaches its maximum value, $A_b$, when $k=k_b$.  For this work we assume that particle production could have taken place at any moment during inflation, so the location of the feature is treated as arbitrary.\footnote{Concretely $k_b \approx 4.67 H$ if the scale factor is normalized to one at  $t=t_*$.}

The bump-like feature (\ref{bump_characterization}) is very similar to the one that would be generated due to instantaneous production during inflation of \emph{scalar} particles directly coupled to the inflaton.  The signatures of scalar particle production during inflation, including important rescattering effects, were fully derived in \cite{Barnaby:2009mc}.  The phenomenology of the bump-like feature in the curvature spectrum  was subsequently studied in \cite{Barnaby:2009dd}.  (See \cite{Barnaby:2010ke} for a discussion of non-Gaussian signatures and \cite{Barnaby:2010sq} for a review.)  To derive observational constraints we follow closely the analysis in \cite{Barnaby:2009dd} and replace the somewhat complicated shape function (\ref{bump_characterization}) by the following simple fitting function
\begin{equation}
\label{Sfit}
 S_{\mathrm{fit}}(x) =  4.5 \, x^3 \, e^{-1.5 x^2}  \, ,
\end{equation}
which, again, is normalized to have maximal value unity at $x=1$.  In the right panel of Fig.~\ref{fig:feature} we show that the simple formula (\ref{Sfit}) provides an adequate description of the feature.

In~\cite{Barnaby:2009dd}, a variety of data sets were used to perform a detailed analysis of the observational constraints on localized features with shape (\ref{Sfit});  we refer the reader to that paper for a discussion of the methodology.  See also \cite{Chantavat:2010vt} for forecast constraints from Large Scale Structure data, and see \cite{Chluba:2012we} for a discussion of both current and forecasted constraints from measurements of the CMB energy spectrum.  In the present work we are mostly interested in a feature localized on CMB scales, in which case the likelihood contours presented in \cite{Barnaby:2009dd} are approximately flat and the observational bound can be roughly summarized as:
\begin{equation}
\label{bnd1}
  \frac{A_b}{\mathcal{P}} \lsim 0.1 \hspace{5mm}\Rightarrow\hspace{5mm} \frac{\dot{m}_*^{1/2}}{H} \lsim 1.2\, \mathcal{P}^{-1/7} \, \epsilon^{-2/7}
\end{equation}
Since the sourced part needs to be subdominant in the power spectrum,  $\mathcal{P} = 2.5\cdot 10^{-9}$; then, the bound (\ref{bnd1}) can be expressed as
\begin{equation}
\label{bnd2}
 \frac{\dot{m}_*^{1/2}}{H} \lsim 20 \, \epsilon^{-2/7} \, .
\end{equation}
(Recall that $\dot{m}_*^{1/2} \gg H$ and $\epsilon \ll 1$ are both required for theoretical consistency.)  We will see shortly that the observational bound (\ref{bnd2}) excludes the possibility of having any interesting effect from particle production in the tensor spectrum.

So far we have discussed the bump-like feature in the spectrum of curvature perturbations.  However, there will also be a corresponding localized feature in the \emph{bispectrum}.  This kind of localized non-Gaussianity is very far from scale invariant and hence quite different from the  bispectrum templates that are most often used in data analysis.  To get a very rough sense of the amplitude of non-Gaussianity in our model, we have computed the effective ($k$-dependent) $f_{NL}^{\mathrm{equil}}(k)$ parameter, which exhibits a bump-like structure.  This parameter assumes the maximal value:
\begin{equation}
\label{fNLmax}
 \left. f_{NL}^{\mathrm{equil.eff}}(k) \right|_{\mathrm{max}} \approx 1.9\cdot 10^{-10} \, \epsilon^3 \, \frac{\dot{m}_*^{9/2}}{H^9} 
 \lsim 92 \, \epsilon^{3/7} \;\;\;,\;\;\; {\rm at } \; k \simeq 1.2 \, k_b
\end{equation}
where $\mathcal{P} = 2.5\cdot 10^{-9}$ has been used in the first equality and the upper bound comes from imposing (\ref{bnd2}).  It should be emphasized that existing observational constraints (or forecasts) on $f_{NL}^{\mathrm{equil}}$ \emph{cannot} be applied directly to (\ref{fNLmax}).  A dedicated search for this type of localized non-Gaussianity would be interesting, however, it is beyond the scope of this paper.  (See also \cite{Barnaby:2010ke,Battefeld:2011ut} for a discussion about localized non-Gaussianities from scalar particle production during inflation.)  In the event of a detection of a bump in the power spectrum, localized non-Gaussian features could play an important role to falsify (or support) models of particle production during inflation.

Using our result (\ref{mod1-Plambda}), the total primordial tensor spectrum in the model (\ref{model1}) is given by
\begin{equation}
\label{PGWtot}
 P_{\mathrm{GW}}(k) = \sum_\lambda \left[ \frac{^{}}{_{}} P_{\lambda,v} + P_{\lambda,s} \right] 
 = \frac{2 H^2}{\pi^2 M_p^2} \left[  1 + 4.17 \cdot 10^{-4} \, \frac{H^2}{M_p^2} \, \frac{\dot{m}_*^{3/2}}{H^3} \ln^2\left( \frac{\dot{m}_*^{1/2}}{H} \right) S_{\mathrm{GW}}\left[ x \right]  \right] \, .
\end{equation}
where in the shape function 
\begin{equation}
 S_{\mathrm{GW}}\left[ x \right] \equiv \frac{0.0226}{x^3} \left[ \sin \left( 4.67 \, x \right) - 4.67 \, x \, \cos \left( 4.67 \, x \right) \right]^2 \, ,
 \end{equation}
the scale $x$ is normalized as in the scalar shape function (\ref{bump_characterization}). The shape function is maximized at $x = \frac{k}{k_b} \simeq 0.53$, where it evaluates to $\simeq 1$.

 To estimate the amplitude of the gravitational wave signal that can be obtained from particle production, we evaluate the tensor-to-scalar ratio on scales where the second term in (\ref{PGWtot}) is maximized.  Assuming the observational constraint (\ref{bnd2}) is satisfied we can approximate $P_\zeta\approx P_{\zeta,v} = 2.5\cdot 10^{-9}$ and we find:
\begin{equation}
\label{rtotmax}
 \left. \frac{}{} r(k) \right|_{\mathrm{max}} \approx 16 \epsilon \left[ 1 + 8.2 \cdot 10^{-11} \, \epsilon \, \frac{\dot{m}_*^{3/2}}{H^3} \ln^2\left( \frac{\dot{m}_*^{1/2}}{H} \right)  \right] \;\;\;,\;\;\; {\rm at } \; k \simeq 0.53 \, k_b
\end{equation}
Using (\ref{bnd2}) and $\epsilon \lsim 0.006$ (corresponding to $r_{\rm vac} \lsim 0.1$) it is straightforward to show that the sourced contribution to (\ref{rtotmax}) is always $\lsim 10^{-6}$, which is undetectably small.  We conclude that an observationally interesting signature in gravitational waves cannot be obtained in the model (\ref{model1}).

Before concluding, we note that the estimates presented here should be interpreted as lower bounds on the efficiency of particle production effects.  We have considered only a single instance where $\psi=0$, leading to a single burst of vector particle production.  However, in a concrete model one could expect $\psi$ to undergo damped oscillations about the minimum of its potential, passing through zero several times before its kinetic energy is dissipated due to Hubble friction or backreaction effects.  In such a scenario resonance effects would be expected to enhance the occupation number of the produced gauge fields by some factor $E$ that could, in principle, be $\gg 1$.  The enhancement $n_k \rightarrow E n_k$ leads to a factor of $E^2$ in the sourced contribution to the tensor spectrum, so that the relevant term in (\ref{rtotmax}) becomes
\begin{equation}
 \left. \frac{}{} r(k) \right|_{\mathrm{max,sourced}} \approx 1.3\cdot 10^{-9} \, E^2 \epsilon^2 \, \frac{\dot{m}_*^{3/2}}{H^3} \ln^2\left( \frac{\dot{m}_*^{1/2}}{H} \right) \, ,
\end{equation}
where we still assume that the scalar spectrum is dominated by the vacuum fluctuations.  The sourced part of the scalar spectrum also gets enhanced by a factor of $E^2$, so the bound (\ref{bnd2}) becomes stronger:
\begin{equation}
  \frac{\dot{m}_*^{1/2}}{H} \lsim 20 (\epsilon E)^{-2/7} \, .
\end{equation}
Combining these results we find:
\begin{equation}
  \left. \frac{}{} r(k) \right|_{\mathrm{max,sourced}} \lsim 10^{-5} \, \left( E \epsilon \right)^{8/7} \, \ln^2\left[20 (E \epsilon )^{-2/7}\right] \, .
\end{equation}
The value $E \gsim 200 \, \epsilon^{-1}$ gives $r_{\rm s} \sim 0.01$ at the bump. It would be interesting to study under which conditions this value can be reached in a concrete model.

\subsection{Comparison with GW sourced by modes of different spins}
\label{subsec:GW-comparison}

In Subsection \ref{subsec:GW1}, we have computed the power in gravity waves sourced by vector fields produced in the model (\ref{model1}). Schematically, the sourced part of the  power spectrum is   ${\cal P}_\lambda \propto \int d^3 p \, \langle T T \rangle$, where $T$ is the traceless-transverse spatial part of the energy-momentum tensor of source (the gauge fields, in this case) with appropriate contraction.
The spatial part of the energy momentum tensor contain dominant terms that scale as $T_{ij} \sim M^2 
A_i A_j$ in the $M \gg p$ limit (we recall that $p$ and $M$ are, respectively, the  momentum and mass of the quanta sourcing the gravity waves). One could therefore conclude that ${\cal P}_\lambda \propto \int d p \, p^2 \, M^4$. However, we showed that the dominant terms cancel against each other. Also the next to leading term in a $\frac{p^2}{M^2}$ Taylor expansion of the integral cancel, and one is left with  ${\cal P}_\lambda \propto \int d p \, p^6$, see eq. (\ref{mod1-Plambda-beforpint}).

In Appendix \ref{app-fermions} we performed the analogous computation using fermion fields rather than vector fields as sources. In this case the spatial part of the energy-momentum tensor has terms of the type $T_{ij} \sim {\bar \chi} \gamma_i p_j \chi$, and one may conclude that  ${\cal P}_\lambda \propto \int d p \, p^4 \, M^2$ (the factor $M^2$ coming from the different normalization of the fermion wave function with respect to the vector one, compare the function $f$ in (\ref{A-deco2}) and in 
(\ref{fermi-deco})). Also in this case there is however a cancellation, resulting in $\langle \chi^2 \rangle \propto p^2$, see eq. (\ref{fermi-C11-C12-res}), and in ${\cal P}_\lambda \propto \int d p \, p^6$, see eq. (\ref{fermi-Plambda-beforpint}).

These two scalings agree with that obtained if the source is a scalar particle. In this case, $T_{ij} \propto p_i p_j \phi^2$, and one immediately has  ${\cal P}_\lambda \propto \int d p \, p^6$ without any cancellation. In fact, from our results (\ref{mod1-Plambda-beforpint}) and (\ref{fermi-Plambda-beforpint}), and from the result for the analogous computation with a scalar source given in  \cite{Cook:2011hg}, we obtain a very general expression for the power spectrum:
\begin{equation}
P_{\lambda,{\rm s}}  \simeq  \frac{2 \, g_s \, k^3}{15 \pi^4 a^2 M_p^4} 
\tilde{\cal T}_k^2  \int d p \,  p^6 \vert \beta_p \vert^2 \left( \vert \alpha_p \vert^2 + \left( - 1 \right)^{2  s} \,
\vert \beta_p \vert^2 \right) 
\label{general-Plambda-beforpint}
\end{equation}
where $s$ is the spin of the sourcing field, and $g_s$ is the number of degrees of freedom of that field: $g_s = 1$ for a scalar, $g_s = 2$ for a vector if the longitudinal mode  is produced in a negligible amount ($g_s = 3$ if it is produced in the same amount as each transverse mode), and $g_s = 4$ for a Dirac fermion.

We see that, apart from the difference in the number of degrees of freedom, and a small difference due to the spin statistics, the different fields in the nonrelativistic regime $M \gg p$ have a comparable quadrupole moment (transverse and traceless projection of $T_{ij}$) and generate a comparable amount of gravity waves.

\section{Model II: Vector produced by a pseudo-scalar interaction}
\label{sec:model2}

In this section we consider the following model
\begin{equation}
\label{S2}
 S = \int d^4x \sqrt{-g} \left[\frac{^{}}{_{}}\right. \frac{M_p^2}{2} R
-\underbrace{\frac{1}{2}(\partial\varphi)^2 - V(\varphi) }_{\mathrm{inflaton}\hspace{2mm}\mathrm{sector}} 
  - \underbrace{ \frac{1}{2}(\partial\psi)^2 - U(\psi) - \frac{1}{4} F^2 - \frac{\psi}{4f} F\tilde{F} }_{\mathrm{hidden}\hspace{2mm}\mathrm{sector}} 
 \left.\frac{^{}}{_{}}\right] \, .
\end{equation}
In addition to a standard inflationary sector, we have introduced a ``hidden'' sector consisting of a light pseudoscalar, $\psi$, and a $U(1)$ gauge field, $A_\mu$, whose energy density is small as compared to that of the inflaton (so that the Friedmann equation takes the usual form $3 H^2 M_p^2 \approx V(\phi)$).  As in section \ref{sec:model1}, the hidden sector in (\ref{S2}) has been introduced so that the production of gauge field fluctuations can provide a new source of inflationary gravitational waves, complementary to the usual quantum vacuum fluctuations of the tensor part of the metric.  Unlike the model of section \ref{sec:model1}, however, we will see that particle production in the theory (\ref{S2}) occurs continuously during inflation, leading to broad-band signatures rather than localized features in the scalar and tensor $n$-point correlation functions.

The coupling $\psi^{(0)}(t) F\tilde{F}$ of the gauge field to the time-dependent pseudoscalar condensate leads to an exponential production of fluctuations $ A_\mu$.  This effect has already been discussed at length in the literature -- see Refs.~\cite{Anber:2009ua,Barnaby:2010vf,Barnaby:2011vw}, for example -- and here we only review the key features that will be necessary for our analysis.  Employing the decomposition (\ref{A-deco}) we find the following linearized equation of motion of the gauge field mode functions
\begin{equation}
\label{pseudo_eom}
 \left[ \partial_\tau^2 + k^2 \pm \frac{2k\xi}{\tau}  \right] A_{\pm}(\tau,k) = 0 \, , \hspace{5mm} \xi \equiv \frac{\dot{\psi}^{(0)}}{2 H f} \, .
\end{equation}
If the pseudoscalar is in an overdamped regime then the parameter $\xi$ can be treated as a constant.  Moreover, we assume that $\dot{\psi}^{(0)} > 0$ so that the ``$+$'' helicity state of the gauge field gets copiously produced while the ``$-$'' state remains in the vacuum and its effect is renormalized away.\footnote{None of our result for the scalar or tensor correlation functions will depend on the choice $\dot{\psi}^{(0)}>0$.}  The properly normalized solutions of (\ref{pseudo_eom}) can be written as \cite{Anber:2009ua}
\begin{equation}
\label{A+sol}
 A_+(\tau,k) \approx \left(\frac{-\tau}{8\xi k}\right)^{1/4} e^{\pi \xi - \sqrt{-2\xi k\tau}} \, , \hspace{5mm} 
 A_+'(\tau,k) \approx \left( \frac{2\xi k}{-\tau} \right)^{1/2} \, A_+(\tau,k)\, .
\end{equation}
This solution is valid only in the phase space interval $\frac{1}{8\xi} \ll -k\tau \ll 2\xi$, where the production of gauge fluctuations is most important.  By restricting ourselves to this regime we effectively cut-off an ultra-violet divergence associated with the usual quantum vacuum fluctuations of the gauge field on sub-horizon scales; see \cite{Barnaby:2011vw} for more discussion.  We have also assumed that $\xi \gsim \mathcal{O}(1)$, so that the phase space of produced fluctuations is non-trivial and  each mode experiences a significant exponential enhancement, $e^{\pi \xi} \gg 1$, near horizon crossing.  (For $\xi < 1$ there is no interesting particle production in the model.) 

We are interested in a scenario where inflation is driven by the potential energy of the $\varphi$ field, so that $3 H^2 M_p^2 \approx V(\varphi)$; the ``hidden'' sector in (\ref{S2}) instead should give a small contribution to the total energy density of the universe.  This requirement imposes several constraints on the model parameters, which we now discuss.  We must first require that the energy density in the produced gauge field fluctuations is smaller than the kinetic energy of $\psi$,
\begin{equation}
\label{small:rho_gauge}
 \frac{1}{2} \langle \vec{E}^2 + \vec{B}^2 \rangle \ll \frac{\dot{\psi}^{(0) 2}}{2} \, ,
\end{equation}
where the ``electric'' and ``magnetic'' fields are $E_i \equiv -\frac{1}{a^2}A_i'$, $B_i \equiv \frac{1}{a^2}\epsilon_{ijl}\partial_jA_l$.  Using the solution (\ref{A+sol}) to evaluate the expectation value \cite{Barnaby:2011vw}, the condition (\ref{small:rho_gauge}) can be written as: 
\begin{equation}
\label{small:rho_gauge2}
 \frac{H^2}{ \dot{\psi}^{(0)}   } \ll 60 \, \xi^{3/2} e^{-\pi\xi}  \, .
\end{equation}
We also require that the energy density of the rolling pseudoscalar can be neglected with respect to that of the inflaton:
\begin{equation}
\label{small:rho_psi}
 \frac{1}{2}\left(\dot{\psi}^{(0)} \right)^2 + U\left(\psi^{(0)}\right) \ll 3 H^2 M_p^2 \, .
\end{equation}
Finally, we require that $\xi$ is adiabatically evolving, $\frac{\dot{\xi}}{H \, \xi} = \frac{\ddot{\psi}^{(0)}}{H \, \dot{\psi}^{(0)}} - \frac{\dot{H}}{H^2} \ll 1$, so that it is appropriate to treat it as nearly constant during the time interval in which each mode of $A$ is relevant (namely, close to horizon crossing, when the mode is produced, and affect cosmological perturbations). As $ \vert \frac{\dot{H}}{H^2} \vert \ll 1$ during inflation, we need to require
\begin{equation}
 \frac{\ddot{\psi}^{(0)}}{H \, \dot{\psi}^{(0)}}  \ll 1 \;\;\Leftrightarrow\;\; m_\psi \ll \frac{3 H}{2}
\label{small:psidot}
\end{equation}
where in the last condition we have approximated $U \left( \psi \right)$ as a quadratic potential, and we have required the evolution of $\psi^{(0)}$ to be in the overdamped regime. Throughout our analysis we will require that the conditions (\ref{small:rho_gauge2}), (\ref{small:rho_psi}), and (\ref{small:psidot})  are simultaneously satisfied (these conditions are discussed at the end of Subsection \ref{subsec:pheno2}).

The background equation for the pseudoscalar reads:
\begin{equation}
 \ddot{\psi}^{(0)} + 3 H \dot{\psi}^{(0)} + U'\left(\psi^{(0)}\right) = \frac{1}{f}\langle \vec{E}\cdot\vec{B} \rangle \, .
\end{equation}
It is interesting to note that the condition (\ref{small:rho_gauge2}) guarantees right hand side of this equation can be disregarded. Indeed  \cite{Barnaby:2011vw}:
\begin{equation}
\label{small:dissipation}
|U'\left(\psi^{(0)}\right)| \gg \frac{1}{f}|\langle \vec{E}\cdot\vec{B} \rangle| 
\;\;\; \Leftrightarrow \;\;\;
 \frac{H^2}{\dot{\psi}^{(0)}} \ll 82 \xi^{3/2} e^{-\pi\xi} \, .
\end{equation}
which is implied by (\ref{small:rho_gauge2}).

The quantity $\mathcal{D}^{(\lambda)}$ introduced in (\ref{formal-AA}) characterizes the two-point function of the produced gauge field fluctuations.  This function, and its derivative, can be written explicitly in terms of the c-number mode functions as
\begin{equation}
 \mathcal{D}_{(0,0)}^{(\lambda)}\left[\tau_1,\tau_2;q\right] \equiv A_\lambda(\tau_1,q)A^*_\lambda(\tau_2,q) \, , \hspace{5mm}
 \mathcal{D}_{(1,1)}^{(\lambda)}\left[\tau_1,\tau_2;q\right] \equiv A_\lambda'(\tau_1,q)A^{*'}_\lambda(\tau_2,q) \, .
\end{equation}
Considering only the ``$+$'' helicity state and using the approximate solution (\ref{A+sol}) we have
\begin{equation}
\label{mod2-D}
 \mathcal{D}_{(0,0)}^{(+)}\left[\tau_1,\tau_2;q\right] \approx \frac{(\tau_1\tau_2)^{1/4}}{\sqrt{8\xi q}} e^{2\pi\xi-\sqrt{2\xi q}\left[\sqrt{-\tau_1}+\sqrt{-\tau_2}\right]} \, ,
\hspace{5mm}  
\mathcal{D}_{(1,1)}^{(+)}\left[\tau_1,\tau_2;q\right] \approx \frac{2\xi q}{\sqrt{\tau_1\tau_2}} \mathcal{D}_{(0,0)}^{(+)}\left[\tau_1,\tau_2;q\right] \, .
\end{equation}

\subsection{Scalar perturbations sourced by the vector modes}
\label{subsec:zeta2}

\subsubsection{The master equation}

The hidden sector in (\ref{S2}) decouples from the inflaton in the limit $M_p\rightarrow \infty$.  However, at finite $M_p$, gravitational couplings will transmit the effects of particle production in the hidden sector to the inflaton perturbations, modifying the usual predictions for the observable curvature fluctuations.  To see this effect we follow closely the analysis of subsection \ref{subsec:zeta1} and derive a master equation for the inflaton perturbations in the model (\ref{S2}).  At linear order in perturbation theory we have recovered the standard result
\begin{equation}
 \left( \partial_\tau^2 - \nabla^2 \right)  \varphi_1  + 2 {\cal H}  \varphi_1'  \simeq 0 \, ,
\label{mod2-eqphi1}
\end{equation}
where we work to leading order in slow roll parameters.  At second order, instead, we have the following equation of motion  
\begin{eqnarray}
\left( \partial_\tau^2 - \nabla^2 \right)  \varphi_2  + 2 {\cal H}  \varphi_2'  \simeq
  - \frac{ \varphi^{(0)'} a^2}{2 M_p^2 {\cal H}}
\left\{\frac{^{}}{_{}}\right. \underbrace{\frac{E^2+B^2}{2}}_{\mathrm{gives}\hspace{1mm}J_1} 
\,\,\,+\,\,\, \frac{1}{a^4} \underbrace{\nabla^{-2} \partial_\tau \left[ a^4 \vec{\nabla} \cdot \left( \vec{E} \times \vec{B} \right) \right]}_{\mathrm{gives}\hspace{1mm}J_2} 
\left.\frac{^{}}{_{}} \right\} + \dots
\label{mod2-eqphi2}
\end{eqnarray}
which is formally equivalent to (\ref{mod1-eqphi2}) with $m^2 \rightarrow 0$.  In (\ref{mod2-eqphi2}) the  $\cdots$ schematically denotes terms involving $\varphi_1^2$, $\psi_1^2$ and $h_{1,ij}^2$ which do not involve the exponential factors $e^{\pi\xi}$ that characterize the gauge field modes (\ref{A+sol}) and may therefore be neglected.

As we did in Subsection \ref{subsec:zeta1}, both the first and second order equations can be combined into the single master equation (\ref{Qf-eq-formal}). The first order mode $\varphi_1$  is  the homogeneous solution of this master equation, while the second order $\varphi_2$ is  the particular solution of this master equation.  The source in the master equation can be written as a sum of two terms
\begin{equation}
\label{mod2-Jvarphi}
 J_\varphi\left(\tau,\vec{k}\right) = J_1\left(\tau,\vec{k}\right) + J_2\left(\tau,\vec{k}\right) \, .
\end{equation}

The source $J_2$ is associated with the second term on the right hand side of (\ref{mod2-eqphi2}), which is non-local in position space.  We show in  Appendix \ref{app:modelII} that this term is actually  infra-red finite, and that the full source can be cast in the form
\begin{equation}
 J_\varphi\left(\tau,\vec{k}\right) = \frac{\varphi'^{(0)} a^3}{4M_p^2 \sH} \int\frac{d^3p}{(2\pi)^{3/2}} \left[ -1 + \frac{(p-|\vec{k}-\vec{p}|)^2}{k^2} \right]
 \left[ \,\, \tilde{E}_i\left(\tau,\vec{p}\right) \tilde{E}_i\left(\tau,\vec{k}-\vec{p}\right) + \tilde{B}_i\left(\tau,\vec{p}\right)\tilde{B}_i\left(\tau,\vec{k}-\vec{p}\right) \,\, \right] \, .
\label{mod2-explicitsource}
\end{equation}
where we have defined the ``electric'' and ``magnetic'' field operators as
$\tilde{E}_i(\tau,\vec{k}) \equiv -\frac{1}{a^2}\tilde{A}_i'\left(\tau,\vec{k}\right)$ and $\tilde{B}_i\left(\tau,\vec{k}\right) \equiv \frac{i}{a^2}\epsilon_{ijl} k_j \tilde{A}_l\left(\tau,\vec{k}\right)$. This expression appears simpler than the corresponding source (\ref{mod1-source-z}) in the previous model, due to the fact that only the modes $A_+$  are relevant here.

In this model the energy density in the ``electric'' field dominates over that in the ``magnetic'' field \cite{Barnaby:2011vw}.  Dropping terms involving $\tilde{B}_i$ we have a source term of the form (\ref{Jf-formal}) where the model-dependent operator can be approximated by
\begin{equation}
\label{mod2-O}
 \hat{\mathcal{O}}_{\varphi,ij}\left(\tau,\vec{k},\vec{p}\right) \approx \frac{\varphi'^{(0)} }{4M_p^2 \sH a} \left[ -1 + \frac{(p-|\vec{k}-\vec{p}|)^2}{k^2} \right]\,\, \delta_{ij}\,\,
 \partial_\tau^{(1)} \partial_\tau^{(2)} \, ,
\end{equation}
where we recall our notation (\ref{12-not}).

\subsubsection{Two-point and three-point correlation functions}

We now proceed to compute the two-point and three-point correlation functions of the gauge invariant curvature perturbation, $\zeta$.  We are only interested in contributions that are sourced by particle production effects, since the vacuum fluctuations in the model (\ref{S2}) are standard.  The sourced contribution to the power spectrum of $\zeta$ is given by (\ref{formal-Pz}).  As discussed above, we only consider the $\sigma=+$ contributions in the sum over helicity states.  Using the explicit expression (\ref{mod2-O}) for the operator $\hat{O}$, along with the identity (\ref{identity-pol}), we find that (\ref{formal-Pz}) can be written as
\begin{eqnarray}
 P_{\zeta,s}(k) &\approx& \frac{k^3}{64\pi^2 M_p^4 a^2} \int \frac{d\tau_1}{a(\tau_1)} \,G_k(\tau,\tau_1) \,\, \int \frac{d\tau_2}{a(\tau_2)}\, G_k(\tau,\tau_2) \nonumber \\
 &&\times \,\, \int \frac{d^3p}{(2\pi)^3} \, \left[ 1 + \frac{p^2-\vec{k}\cdot\vec{p}}{p|\vec{k}-\vec{p}|}  \right]^2 \, 
            \left[ 1 - \frac{(p-|\vec{k}-\vec{p}|)^2}{k^2}         \right]^2 \,\, 
\mathcal{D}_{(1,1)}^{(+)}\left[\tau_1,\tau_2;p\right] \, \mathcal{D}_{(1,1)}^{(+)}\left[\tau_1,\tau_2;|\vec{k}-\vec{p}|\right] \, .
\end{eqnarray}
Next we insert our previous result (\ref{mod2-D}) for $\mathcal{D}_{(1,1)}^{(+)}$.  Since we are interested in computing the spectrum at late times, the explicit expression (\ref{formal-sol2}) for the Green function can be employed.  The sourced power spectrum takes the form
\begin{equation}
 P_{\zeta,s}(k) = \frac{\xi e^{4\pi\xi} H^4}{128\pi^2 M_p^4} \int \frac{d^3q}{(2\pi)^3} q^{1/2} |\hat{k}-\vec{q}|^{1/2} 
 \left[ 1-(q-|\hat{k}-\vec{q}|)^2   \right]^2
 \left[ 1 - \frac{\vec{q}\cdot(\hat{k}-\vec{q})}{q |\hat{k}-\vec{q}|}  \right]^2 \, \mathcal{I}^2\left[q,|\hat{k}-\vec{q}|\right] \, , \label{P-almost} 
\end{equation}
where we have introduced dimensionless variables $\vec{q} \equiv \vec{p} / k$ and $\hat{k} \equiv \vec{k} / k$.  The dimensionless time integral is defined as
\begin{equation}
\label{calI}
 \mathcal{I}\left[a,b\right] \equiv \int_{-k\tau}^{\infty} dz \,\, \frac{\sin z - z\cos z}{z^{1/2}} \,\, e^{-2\sqrt{2\xi z}\left[\sqrt{a}+\sqrt{b}\right]} \, ,
\end{equation}
(notice that $z\equiv -k\tau'$).  In the super-horizon regime, $-k\tau \ll 1$, we can set the lower bound of integration to zero.  Moreover, particle production effects are most interesting in the regime $\xi \gsim \mathcal{O}(1)$, in which case the integral (\ref{calI}) has most of its support in the region $z\ll 1$ where we can approximate $\sin z - z \cos z \approx \frac{z^3}{3}$.  Hence, we have the following analytical approximation
\begin{equation}
 \mathcal{I}\left[a,b\right] \approx \int_{0}^{\infty} dz \,\,\frac{z^{5/2}}{3} \,\, e^{-2\sqrt{2\xi z}\left[\sqrt{a}+\sqrt{b}\right]} =
 \frac{15}{32\sqrt{2} \, \left(\sqrt{a} + \sqrt{b}\right)^7 \, \xi^{7/2} } \, .
\end{equation}
Finally, the momentum integral in (\ref{P-almost}) must be performed numerically, giving the final result
\begin{equation}
\label{mod2:finalPz}
 P_{\zeta,s}(k) \approx 4\cdot 10^{-10} \frac{H^4}{M_p^4} \frac{e^{4\pi\xi}}{\xi^6} \, .
\end{equation}

The computation of the three-point correlation function is completely analogous to the derivation that we have outlined for the source power spectrum.  Here we simply state the final result for the bispectrum
in the equilateral configuration:
\begin{equation}
\label{mod2:finalBz}
 B_\zeta \left( k_1 = k_2 = k_3 \equiv k \right) \approx 2.6\cdot 10^{-13} \frac{H^6}{M_p^6} \frac{e^{6\pi\xi}}{\xi^9} \frac{1}{k^6} \, .
\end{equation}
The reason for considering the equilateral configuration is that, in this model, the source at any moment is dominated by  modes with wavelength comparable to the horizon at that moment. This generates mostly correlations between scalar perturbations of comparable size \cite{Barnaby:2010vf,Barnaby:2011vw}.

\subsection{Gravity waves sourced by the vector modes}
\label{subsec:GW2}

The produced gauge field fluctuations that are described by the mode solution (\ref{A+sol}) carry anisotropic stress/energy and provide a source of gravitational wave fluctuations that is complimentary to the standard quantum vacuum fluctuations from inflation.\footnote{At second order in cosmological perturbation theory tensor fluctuations can also be sourced by bi-linear combinations of the first order scalar fluctuations, $\delta\varphi$ and $\delta\psi$.  Neither of these exhibit the exponential enhancement that characterizes the linear gauge field perturbations -- see equation (\ref{A+sol}) -- therefore we can safely neglect this effect in what follows.}  The computation of the gravitational wave spectrum in the model (\ref{S2}) follows closely what we had for the the curvature perturbation.  Moreover, this computation has in fact already been performed in 
\cite{Barnaby:2010vf,Sorbo:2011rz,Barnaby:2011vw}. 
 Therefore we simply state the final result:
\begin{eqnarray}
\label{mod2:finalPt}
P_+ & = & P_{+,{\rm v}} + P_{+,{\rm s}}  \simeq   \frac{ H^2}{\pi^2 M_p^2} \left[  1 + 8.6\cdot 10^{-7} \, \frac{H^2}{M_p^2} \frac{e^{4\pi\xi}}{\xi^6}   \right] \nonumber\\
P_- & = & P_{-,{\rm v}} + P_{-,{\rm s}}  \simeq   \frac{ H^2}{\pi^2 M_p^2} \left[  1 + 1.8\cdot 10^{-9} \, \frac{H^2}{M_p^2} \frac{e^{4\pi\xi}}{\xi^6}   \right] 
\end{eqnarray}
The first term in the square braces corresponds to the usual contribution from the quantum vacuum fluctuations of the graviton, and the two helicities have equal power.  The second term, on the other hand, corresponds to gravitational wave perturbations that have been sourced by particle production effects in the hidden sector. The production is much more significant for the $h_+$ mode, due to the fact that the source consists of only $A_+$ modes.

\subsection{Phenomenology}
\label{subsec:pheno2}

In this subsection we explore the phenomenology of the scalar and tensor cosmological fluctuations in the model (\ref{S2}).  We first consider the spectrum of curvature fluctuations.  The total observable power spectrum is the sum of (\ref{mod2:finalPz}) and the standard result for the vacuum fluctuations (see the general result in eq.~(\ref{formal-Pz})).  Explicitly we have
\begin{equation}
\label{mod2:totalPz}
 P_\zeta \approx \mathcal{P}\left[ 1 + 2.5\cdot 10^{-6} \, \epsilon^2\mathcal{P} \frac{e^{4\pi\xi}}{\xi^6} \right] \, , 
  \hspace{5mm} \mathcal{P} \equiv \frac{H^2}{8\pi^2\epsilon M_p^2} \, ,
\end{equation}
where the first term in the square braces is the usual spectrum from quantum vacuum fluctuations, while the second term is the sourced contribution coming from gauge field production in the model (\ref{S2}).  

Next, we consider the spectrum gravitational wave fluctuations.  From (\ref{mod2:finalPt}) and (\ref{mod2:totalPz}) we can write the tensor-to-scalar ratio as
\begin{equation}
\label{mod2:r} 
r \equiv \frac{\sum_\lambda P_{\lambda}}{P_\zeta} \approx 
16 \epsilon \frac{  1 + 3.4\cdot 10^{-5}\epsilon\mathcal{P} \frac{e^{4\pi\xi}}{\xi^6} }{  1 + 2.5\cdot 10^{-6} \epsilon^2 \mathcal{P} \frac{e^{4\pi\xi}}{\xi^6} }  \, .
\end{equation}
When both the tensor and scalar spectra are dominated by the vacuum fluctuations we recover the standard result, $r\approx 16\epsilon$.  On the other hand, if the sourced contributions to (\ref{mod2:finalPt}) and (\ref{mod2:totalPz}) dominate then we have $r\approx 218$, independently of model parameters.  The current observational limit is $r \lsim 0.17$ \cite{Keisler:2011aw}, while $r \sim 0.01$ might be detectable with future missions \cite{Baumann:2008aq}.  Since the expression (\ref{mod2:r}) interpolates between $16\epsilon$ and $218$, it follows that an observable signal can be obtained for \emph{any} value of $\epsilon$.   

Let us now discuss the phenomenology of the scalar perturbations. From (\ref{mod2:totalPz}) and (\ref{mod2:r}) we get
\begin{equation}
\frac{ P_{\zeta,s} }{ P_{\zeta,v} } \simeq \frac{r-16 \, \epsilon}{218}
\end{equation}
We therefore see that the scalar power spectrum is dominated by the vacuum part ($P_\zeta \simeq {\cal P}$). Concerning the scalar bispectrum,  the effective nonlinearity parameter is immediately obtained from (\ref{formal-fnl}) and (\ref{mod2:finalBz}):
\begin{equation}
\label{mod2:fNL}
  f_{NL}^{\mathrm{equil.eff}} \approx 1.5\cdot 10^{-9} \epsilon^3\frac{\mathcal{P}^3}{P_\zeta^2}\frac{e^{6\pi\xi}}{\xi^9} \, .
\end{equation}
The current CMB  bound is $-214 < f_{NL}^{\mathrm{equil}} < 266$ \cite{Komatsu:2010fb} while $f_{NL} \sim \mathcal{O}(10)$ might be accessible with Planck \cite{Planck}.  

\begin{figure}[htbp]
\begin{center}
\includegraphics[width=0.45\textwidth]{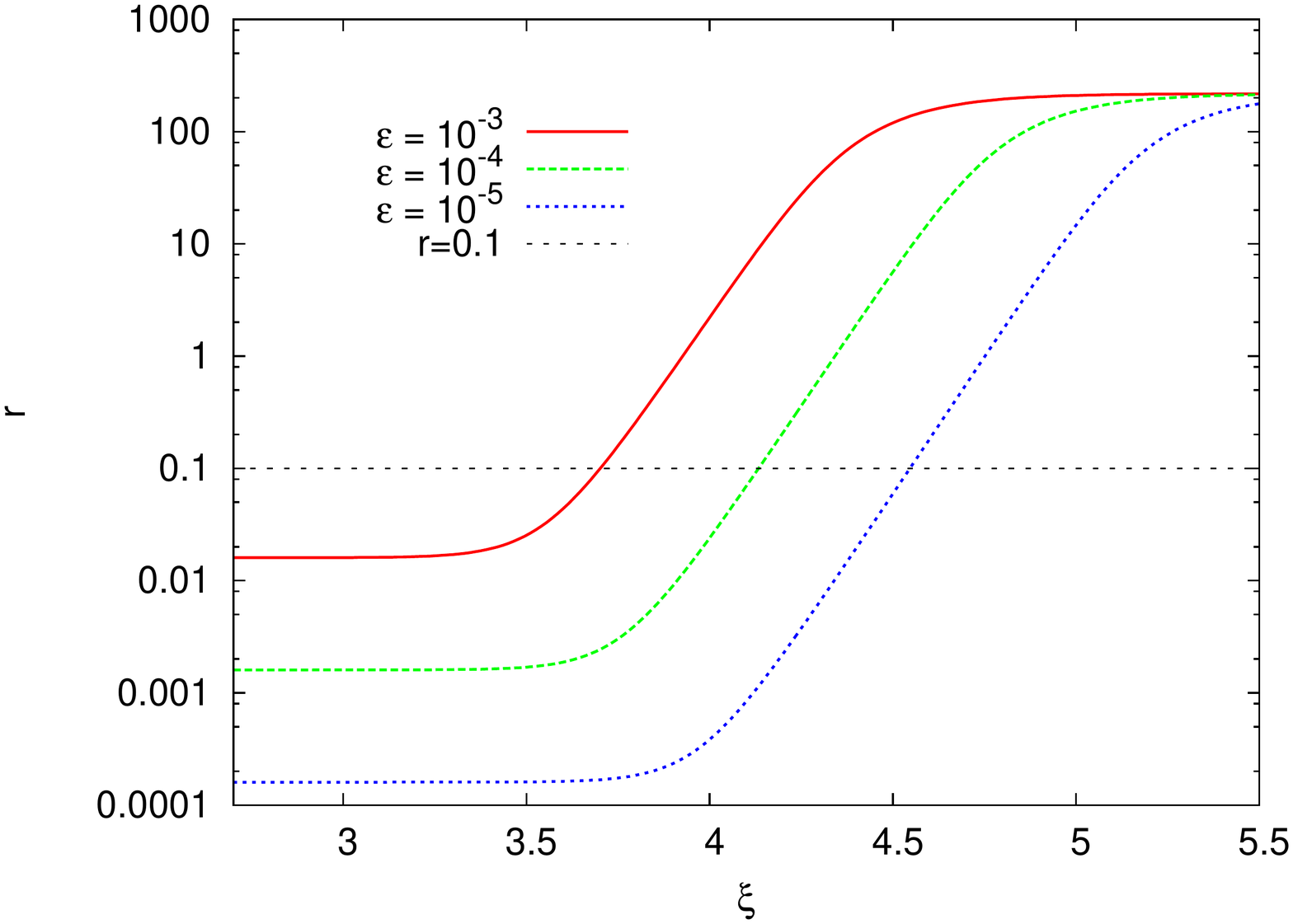}
\includegraphics[width=0.45\textwidth]{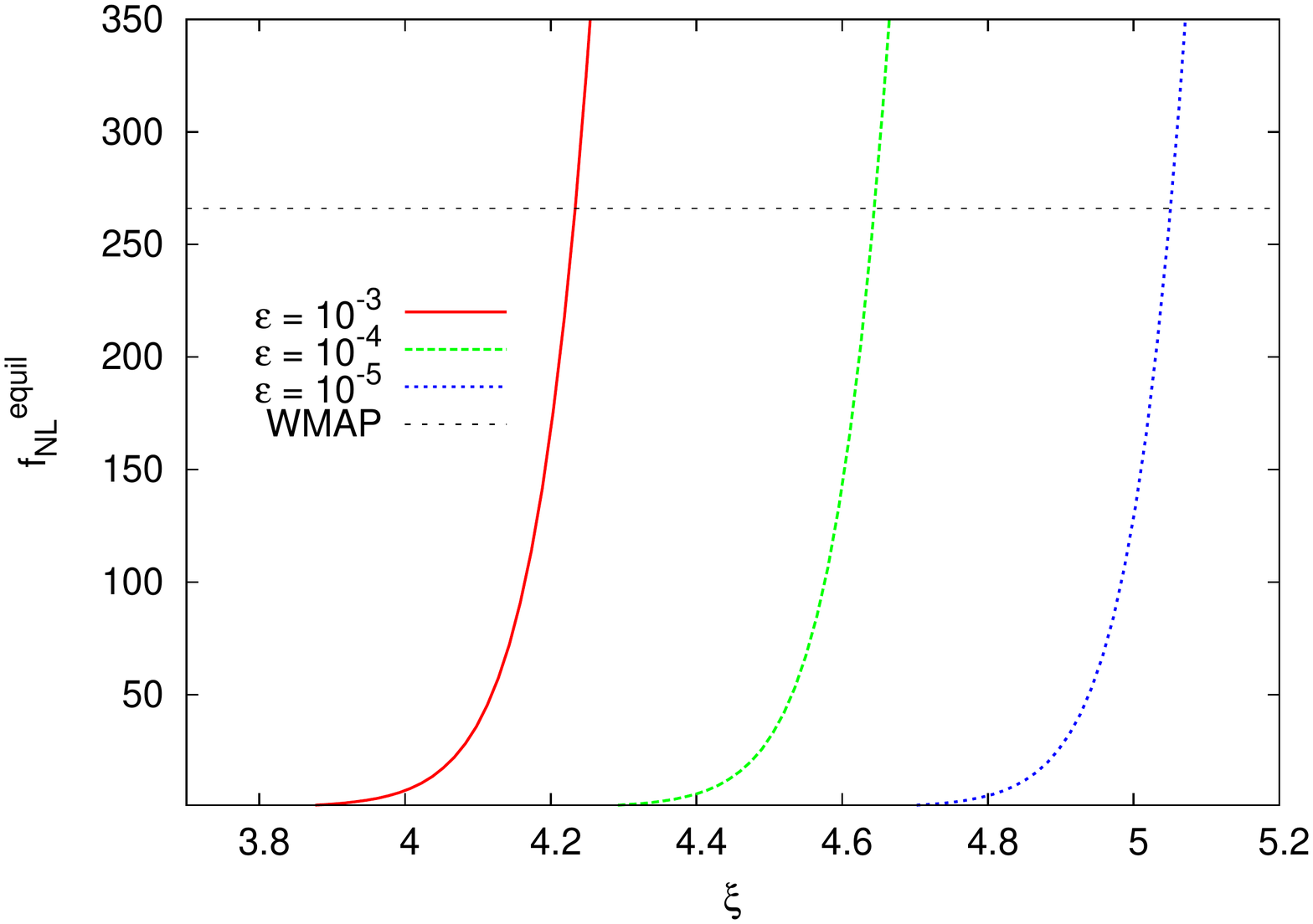}
\end{center}
\caption{Left panel: The tensor-to-scalar ratio as a function of $\xi$, for several illustrative choices of $\epsilon$. The horizontal line corresponds to $r=0.1$, the approximate current observational limit.  Notice that an observable tensor-to-scalar ratio can be achieved for any inflationary potential, by suitably tuning the dynamics in the hidden sector.  Right panel: The effective nonlinearity parameter as a function of $\xi$, for several illustrative choices of $\epsilon$.  The horizontal line corresponds to $f_{NL}=266$, the approximate current observational limit on non-Gaussianity.}
\label{fig:r&fNL}
\end{figure}

To obtain the correct amplitude of density fluctuations, we must impose the normalization condition $P_\zeta = 2.5\cdot 10^{-9}$ on CMB scales.  We can use this condition to eliminate the parameter $\mathcal{P}$ in favour of $\xi$ and $\epsilon$.  Having done so, the key observables $r$ and $f_{NL}^{\mathrm{equil.eff}}$ then depend only on the model-dependent quantities $\xi$ and $\epsilon$, which in turn depend on the inflationary potential and the dynamics of the hidden sector fields.  In Fig.~\ref{fig:r&fNL} we plot our results for $r$ and $f_{NL}$ as a function of $\xi$, for various representative choices of $\epsilon$.

\begin{figure}[htbp]
\begin{center}
\includegraphics[width=0.5\textwidth]{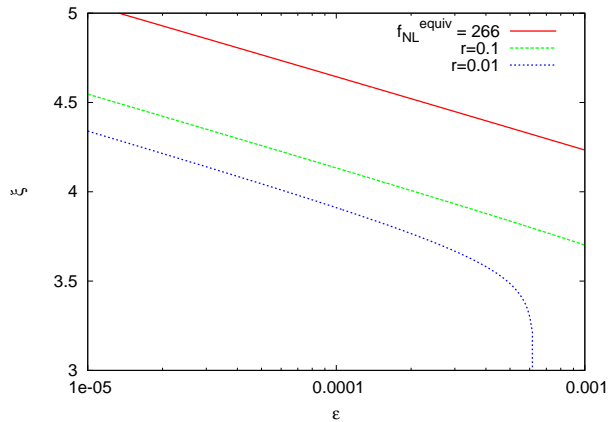}
\end{center}
\caption{Here we plot contours in the $\xi-\epsilon$ plane leading to $f_{NL}=266$ (the current observational bound on non-Gaussianity), $r=0.1$ (the current observational bound on tensor modes), and $r=0.01$ (which may be detectable in the near future).  The region above the solid red line is ruled out by producing too much non-Gaussianity while the region above the dashed green line is ruled out by non-detection of tensor fluctuations.  We see that the non-Gaussianity bound is weaker, meaning that the dominant signature of the model comes from gravitational waves.}
\label{fig:xi-epsilon}
\end{figure}

The observational bound on the tensor-to-scalar ratio forces us into a region of parameter space where non-Gaussianity is undetectably small.  Therefore gravitational wave fluctuations constitute the most interesting phenomenology associated with the model (\ref{S2}).  This is shown in Fig.~\ref{fig:xi-epsilon} where we plot contours in the $\xi-\epsilon$ plane leading to various phenomenologically interesting scenarios. We note that our findings are relevant also for values of $\epsilon$ smaller than those shown in the figure.

As discussed in \cite{Barnaby:2011vw,Sorbo:2011rz,Anber:2012du}, the sourced contribution to the tensor spectrum is chiral; only one helicity state is efficiently sourced by the gauge field fluctuations (\ref{A+sol}).  This effect may be detected through TB and EB correlations in the CMB \cite{Saito:2007kt,Gluscevic:2010vv}. This was first explored by  \cite{Sorbo:2011rz} in the case in which the inflaton is the pseudo-scalar sourcing the vector modes; in this case, the direct inflaton-gauge field coupling is so strong that, typically, the main bound on the gauge field production is given by the sourced scalar perturbations (non-gaussianity \cite{Barnaby:2010vf,Barnaby:2011vw} and, depending on the inflaton potential, increased power at small scales \cite{Barnaby:2011qe,Meerburg:2012id}). To overcome this,   \cite{Sorbo:2011rz} assumed the presence of $\sim 1000$ sourcing gauge fields (this decreases the amount of non-gaussianity), or the curvaton mechanism for the generation of the scalar perturbations. For some values of parameters, the signal can be above the $1 \sigma$ detection line for a cosmic-variance limited experiment   \cite{Sorbo:2011rz}. As we shall now discuss, a more optimistic conclusion is reached if one assumes that the gauge field production occurs in a sector 
 only gravitationally  coupled to the inflaton, as we have studied here.

A measure of the net handedness of the tensor modes is the following quantity:
\begin{equation}
\left\vert  \Delta\chi \right\vert \equiv \left\vert \frac{P_{+} - P_{-}}{P_{+} + P_{-}} \right\vert = \frac{ 3.4 \cdot 10^{-5} \epsilon \mathcal{P} \frac{e^{4\pi\xi}}{\xi^6} }{ 1 + 3.4 \cdot 10^{-5} \epsilon \mathcal{P} \frac{e^{4\pi\xi}}{\xi^6} } \simeq  1 - \frac{16 \, \epsilon}{r} \, , 
\label{r-dxi}
\end{equation}
which interpolates between zero (at small $\xi$, when the vacuum fluctuations dominate the tensor mode spectrum) and unity (at large $\xi$ when the sourced GW dominate the tensor mode spectrum). In the final approximation we have used the fact that, for $r < 0.1$, the scalar power spectrum in this model is dominated by the vacuum modes.

\begin{figure}[htbp]
\begin{center}
\includegraphics[width=0.5\textwidth]{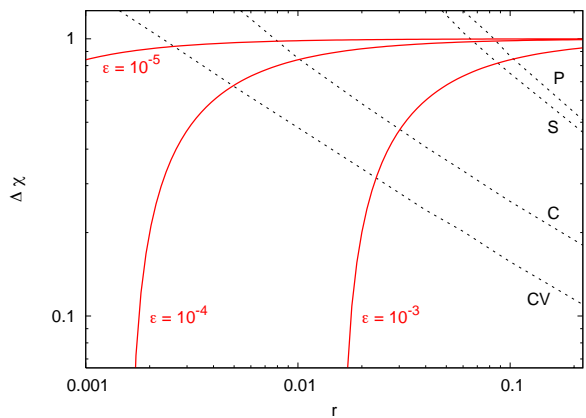}
\end{center}
\caption{Red/solid lines: Predictions for $r$ vs $\Delta \chi$ in the model (\ref{S2}); each line is obtained for a fixed value of $\epsilon$, and for varying $\xi$, with greater $\xi$ corresponding to greater particle production, and therefore larger signal. Black/dotted lines:  $1 \sigma$ detection lines for the Planck (P), SPIDER (S), CMB-Pol (C), and a cosmic-variance limited (CV) experiment.  The signal needs to be above a line to be detectable at $1 \sigma$ by that experiment. These experimental forecasts are 
an approximate copy of the lines shown in  Figure 2 of \cite{Gluscevic:2010vv}.  }
\label{fig:DeltaChi}
\end{figure}

In Figure \ref{fig:DeltaChi} we plot the relation  (\ref{r-dxi}) in the  $r$ vs $\Delta \chi$ plane,  for a few representative values of $\epsilon$; each of the red/solid lines is characterized by a given $\epsilon$, and by varying $\xi$ (growing $\xi$ leads to more gravity wave production, and therefore greater values of $r$ and $\xi$). We stress that arbitrary large values of $r$ in the range shown in the figure can be reached for any value of $\epsilon$. As $\epsilon$ decreases, this requires a greater and greater amount of sourced modes, which in turn leads to a greater and greater $\Delta \chi$. This explain why, for any given obtained $r$, greater $\Delta \chi$ correspond to smaller $\epsilon$. These predictions are superimposed in the figure to $1 \sigma$ detection lines from various experiments; from top to bottom, the lines shown are for the ongoing and forthcoming Planck (P) \cite{Planck} and SPIDER (S) \cite{Crill:2008rd} experiments, for the suggested CMB-Pol experiment (C)  \cite{Baumann:2008aq}, and for a hypothetical cosmic-variance limited experiment (CV).  The signal needs to be above a line to be detectable at $1 \sigma$ by that experiment. These lines are taken by Figure 2 of  \cite{Gluscevic:2010vv}. We observe that, for some values of parameters, the parity-violation could be detected (at least at $1 \sigma$) already by the ongoing / forthcoming Planck and SPIDER experiments.

Before concluding this section, we comment on the constraints (\ref{small:rho_gauge2}) and (\ref{small:rho_psi})  which are necessary for the consistency of our calculation.  We find:
\begin{equation}
\label{total_bnd}
 0.074 \frac{\sqrt{\epsilon \mathcal{P}} e^{\pi \xi}}{\xi^{5/2}} \ll \frac{f}{M_p} \ll \frac{1.2}{\xi}
\,  \sqrt{ 1 - \frac{ U \left( \psi \right) }{ V \left( \varphi \right) } }
\end{equation}
where the first condition, obtained from (\ref{small:rho_gauge2}),  ensures that the energy density of the produced gauge quanta is smaller than the kinetic energy of $\psi$, while the second condition, obtained from  (\ref{small:rho_psi}), ensures that the energy density of $\psi$ is smaller than that of the inflaton. 

The interval in (\ref{total_bnd}) exists for any reasonable choice of model parameters, therefore we can always choose $f/M_p$ such that that various backreaction constraints  are satisfied. The lower bound on the inflationary scale becomes simpler in the limit in which the sourced part of the gravity wave signal dominates over the vacuum one (which is the regime of most interest for our work). Using (\ref{mod2:r}) in this regime, the lower limit in (\ref{total_bnd}) can be expressed as:
\begin{equation}
\frac{f}{M_p} \gg \frac{3.7 \cdot 10^{-3} \, r^{1/4}}{\xi} \sim {\rm O } \left( 10^{-4} \right) 
\;\;\;,\;\;\;  r_{\rm v} <  r_{\rm s} \sim 0.01 - 0.1 
\label{bnd2-f}
\end{equation}
where $r$ has been chosen so that the gravity wave signal is observable in the near future (we note that this requires $\xi \sim 4-5$).

Finally, using (\ref{vac_characterization}), the condition (\ref{small:psidot}) for the adiabatic evolution of $\xi$ can be cast in the form
\begin{equation}
\frac{m_\psi}{M_p} \ll 6.6 \cdot 10^{-4}  \, \sqrt{\epsilon}
\label{mod2-cond-mpsi}
\end{equation}
 For a quadratic inflaton potential, $m_\varphi \simeq 6.4 \cdot 10^{-6} \, M_p$ and $\sqrt{\epsilon} \simeq 0.09$ (for $60$ e-folds of inflation). The condition (\ref{mod2-cond-mpsi}) then rewrites $m_\psi \ll 9 m_\varphi$.

\section{Conclusions}
\label{sec:conclusions}

A large experimental effort is currently taking place to detect gravitational waves from inflation.
The conventional vacuum signal will be detectable only if the scale of inflation is sufficiently high, $V^{1/4} \gsim 10^{16} \, {\rm GeV}$ (corresponding to $r \gsim 0.01$, or $\epsilon \gsim 6 \cdot 10^{-4}$ in single field slow roll inflation). Does this imply that, for a lower scale, this experimental effort will be unsuccessful, or can we hope that an observable gravity wave signal can be obtained from some different mechanism ? And can we distinguish the gravity waves generated by this mechanism from the conventional vacuum ones ?

Recently,  \cite{Cook:2011hg} and \cite{Senatore:2011sp} considered the possible gravity signal from particle production taking place during inflation. A difficulty with this idea is that the same particle production will also source scalar perturbations; the observed scalar perturbations have properties in perfect agreement with standard vacuum fluctuations generated by the simplest inflationary model; in particular, they are gaussian to a very high degree, while there is no reason to expect that this should be the case for a generic model of particle production.~\footnote{In fact, as shown in \cite{Barnaby:2010vf}, particle production is a very simple source of observable non-gaussianity of the scalar perturbations, even for the highly motivated class of natural inflation models.} Therefore, any mechanism of gravity waves from inflation needs to explain why the mechanism has not already manifested itself in the scalar sector. 

To evade this problem,   \cite{Cook:2011hg} studied scenarios in which the particle production takes place (or increases at a sufficient level) only towards the end of inflation, so to generate gravity waves only at scales much smaller than the CMB ones. At these scales, the scalar perturbations are only constrained by primordial black holes (see for instance \cite{Lin:2012gs} for a recent study), which is a significantly weaker bound than those from non-gaussianity at the CMB scales. One of the scenarios of 
 \cite{Cook:2011hg} was further studied in \cite{Barnaby:2011qe} to account for the backreaction of the produced particles on the background inflaton. It was shown in  \cite{Barnaby:2011qe} that in this region the sourced gravity wave signal  can already be observed by the next LIGO stage \cite{LIGO}. The main purpose of the present paper is to study whether an analogous mechanism can produce an observable gravity wave signal at CMB scales, without conflicting with the limits from non-gaussianity in the scalar sector.
 
Let us denote by $X$ the field that is produced during inflation, and that sources the gravity waves. In the models considered in  \cite{Senatore:2011sp,Cook:2011hg} quanta of $X$ are produced by the motion of the inflaton, which we denote by $\varphi$. This implies a direct coupling between the inflaton and the produced quanta. As a consequence, if the inflaton is the source of cosmological perturbations $\zeta$, quanta of $X$ will source $\zeta$ with a stronger than gravitational interaction. On the contrary, the source of gravity waves from $X$ is of gravitational strength. To minimize the relative amount of produced $\zeta$ vs. produced gravity waves, in this work we made the opposite assumption of considering the weakest possible coupling (in standard gravitational theory) between $X$ and the inflaton: namely, we assumed that $\varphi$ and $X$ are coupled only gravitationally. We therefore assumed that particle production occurs in a ``hidden sector'' and in this paper, for the first time, we computed the amount of scalar perturbations $\zeta$ induced by $X$ through a purely gravitational interaction. We then computed the amounts of gravity waves produced by $X$ in these two models (for model II, we quote existing results), and we compared the two effects.

Clearly, there is a large arbitrariness in the choice of the model for particle production, and we do not claim our findings to be exhaustive; in particular, we did not study here the analogous of all the scenarios considered in  \cite{Senatore:2011sp}, where for instance multiple bursts of particle production, and production of strings were also studied. We study two models in which $X$ is a vector field, which is produced by the motion of a field $\psi \neq \varphi$. In model I, the vector field has a mass term $\psi \left( t \right)^2 A^2$, and quanta of $A$ are produced when the classical value of $\psi$ crosses zero. 
In model II the vector is continuously sourced by a pseudo-scalar $\frac{\psi}{f} F {\tilde F}$ interaction. The reasons for considering these two models is that, after the particle production, the vector quanta are highly massive in Model I, while massless in Model II. We showed that in the first case this gives rise to a strong suppression of the gravity wave signal with respect to the amount of scalar perturbations. In the remainder of this concluding section, we summarize our findings in these two models, together with some discussion.

\begin{enumerate}
 \item {\bf Model I:} The main signature of particle production in this model is a bump in the scalar power spectrum, at the scales that exited the horizon when the gauge quanta were produced.~\footnote{This is qualitatively identical to the findings of \cite{Barnaby:2009mc}, where the sourced field is a scalar with  mass depending on the inflaton.} If this bump will be  observed, this mechanism can be supported / disproved by the presence/ absence of an analogous bump in the bispectrum, which we also computed here for the first time. The spectrum of gravity waves produced by the gauge quanta also presents a peak at the same scales, which - if sufficiently high -  could     distinguish them from the vacuum gravity waves.    However, we found that - once the bound from not having observed a large bump in the scalar spectrum is respected - the amount of gravity waves produced in this model is completely unobservable ($r \lsim 10^{-6}$). The relative smallness of the gravity waves vs scalar perturbations produced in the model may come as a surprise, due to the fact that the gauge quanta are coupled gravitationally to both these quantities. The reason for the suppression is due to the fact that the quanta are highly non-relativistic after they are produced. This highly suppresses their quadrupole moment, and the amount of gravity waves that they generate.
\end{enumerate} 

To verify this, we also computed the amount of gravity waves produced if the vector field is replaced by a fermion, with mass $\propto \psi \left( t \right)$. Ref. \cite{Cook:2011hg} computed the amount of gravity waves produced by a scalar with mass  $\propto \psi \left( t \right)$.  From our two results, and from the result of \cite{Cook:2011hg} for the scalar case, we actually obtained the very general formula (\ref{general-Plambda-beforpint}) for the amount of gravity waves produced in all these cases. We see that there is no large enhancement between the different spins, apart from the proportionality of the final result to the number of degrees of freedom in each case. 

We did not compute the amount of scalar perturbations  $\zeta$ sourced in the fermionic case. However, we believe that also in this case the result will be analogous to the one that we have computed, and that therefore our conclusions on the relative importance of gravity waves vs scalar perturbations production apply for produced particles of any spin.

 In this model, the vector field is massive after the production, and it therefore also possesses  a longitudinal component. The results for this component are more model dependent than those of the transverse components: they depend on the specific choice of the potential $U$ for $\psi$ (the results for the transverse component are independent of $U$ provided that it is smaller than the kinetic energy of $\psi$ during particle production). The longitudinal mode may drive the theory out of perturbative regime when $m \propto \psi \left( t \right) \rightarrow 0$. We showed that this is avoided if the ratio $\frac{1}{\psi} \, \frac{d U}{d \psi}$ remains finite as $\psi \rightarrow 0$. This is, for instance, the case if $U \approx \frac{1}{2} m_\psi^2 \, \psi^2$ at the origin. These considerations can be relevant for all the models in which symmetries are enhanced at some point during the cosmological evolution. For instance, we expect massive gauge modes to become massless when different branes move to the same bulk location as in the trapping mechanism of  \cite{Kofman:2004yc}. We also found that, for  $U \approx \frac{1}{2} m_\psi^2 \psi^2$, the longitudinal component is produced as much as each transverse component, and sources the same amount of gravity waves, for the most reasonable values of the mass $m_\psi$.
 
Finally, it is worth pointing out that our study applies to a single instance of particle production. Things may be different in cases of multiple bursts of particle production; for instance, if the potential $U$ 
does not flatten at large $\psi$, so that $\psi$ performs oscillations about its minimum, gauge quanta will be produced at each oscillation, in a regime of  parametric resonance \cite{Kofman:1997yn}. We have found that the gravity wave signal can reach an interesting level if the parametric resonance enhances the amount of produced quanta by a factor of $\gsim 200 \, \epsilon^{-1}$ with respect to the single episode of particle production ($\epsilon$ being the slow roll parameter). It would be interesting to study under which conditions this value can be reached in a concrete model.

\begin{enumerate}
\setcounter{enumi}{1}
\item {\bf Model II:} The amount of gravity waves produced in this model was already computed in 
\cite{Barnaby:2010vf,Sorbo:2011rz,Barnaby:2011vw}. The novel computation in this work is the amount of scalar perturbations produced in this model under the assumption that the inflaton is only gravitationally coupled to the gauge field, and the comparison of the two effects. We found that coupling the inflaton only gravitationally sufficiently suppresses the amount of scalar modes generated in this model, so that the limits from non-gaussianity are irrelevant when compared to those from gravity waves. Therefore,  particle production in this  model can lead to gravity waves observable at the CMB scales. Differently from Model I, the gravity waves in this model are produced at all scales, and not just with a localized bump. However, this signal may be distinguishable from the vacuum one since one gravity wave helicity is produced in a much stronger amount than the other one, and this can lead \cite{Sorbo:2011rz} to  observable nonvanishing TB and EB correlations in the CMB \cite{Saito:2007kt,Gluscevic:2010vv}. 
\end{enumerate}

Ref. \cite{Sorbo:2011rz} already studied whether the parity violation in the sourced gravity waves produced by this model can be observed. Also in that case, the problem was to suppress the non-gaussianity of the scalar perturbations produced by the gauge field  \cite{Barnaby:2010vf}. This was overcome in  \cite{Sorbo:2011rz} by assuming the presence of $\sim 1000$ sourcing gauge fields, or the curvaton mechanism for the generation of the scalar perturbations. It was shown in  \cite{Sorbo:2011rz}  that, under these assumptions, and for some values of parameters, the parity-violation signal can be above the $1 \sigma$ detection line for a  cosmic-variance limited experiment  \cite{Sorbo:2011rz}. We have seen that in our implementation of the mechanism (namely, by assuming that the gauge field is only gravitationally coupled to the inflaton) the parity violation can, for some choice of parameters, be observed already  by the ongoing / forthcoming Planck and SPIDER experiments.

We have seen that backreaction bounds from this mechanism are under control for an axion scale $f$ in the interval $ 10^{-4} M_p \lsim f \lsim  M_p $. Interestingly, the axion decay constant one typically finds in string theory is of the order of the GUT scale 
$f  \sim 10^{16}$ GeV (see, e.g., \cite{Banks:2003sx,Svrcek:2006yi})
which  fits comfortably within this window.
 Indeed, given the UV sensitivity of inflation, it is natural to ask whether one can
  realize our model in string theory.
The low energy spectrum of string theory contains generically many axion-like particles, which arise from the reduction of antisymmetric $p$-form fields on $p$-cycles of the internal space.\footnote{In addition to  these closed string axion-like particles, there are also open string axions but their presence is more model-dependent. For this discussion, we shall focus on closed string axions.
Moreover, we consider only those axions that are not projected out by discrete symmetries (e.g., orientifolding), and do not receive a Stuckelberg mass.}
These closed string axions have a pseudoscalar coupling $\psi F \tilde{F}$
to $U(1)$ gauge fields on the worldvolume of D-branes.\footnote{Such couplings arise from the reduction of Chern-Simons terms in the worldvolume action of D-branes, e.g., 
$\int_{D_{p+4}} C_p \wedge F \wedge \tilde{F}$.}
In such string theory setting, it is not difficult to find inflaton candidates with no direct coupling to the axion-gauge field sector.
For example, the inflaton can be another axion; the absence of direct couplings to the hidden sector follows from some topological and geometrical constraints. Consider a basis of $p$-cycles $\Sigma_i$, and their dual $p$-forms $\omega_j$ such that $\int_{\Sigma_i} \omega_j = \delta_i^j$. Two of such axions $\psi = \int_{\Sigma_i} C_p$, $\varphi = \int_{\Sigma_j} C_p$ do not have kinetic mixing if  $\int \omega_i \wedge \ast \omega_j=0$.
The absence of direct inflaton-gauge field $\varphi F \tilde{F}$  coupling is ensured by $\int \omega_i \wedge \tilde{\omega_j}=0$ where $\tilde{\omega_j}$ is the $6-p$ form dual to $\omega_j$, and $F$ is the gauge field to which the axion $\psi$ couples.
In fact, the shift symmetries enjoyed by the axions may provide a natural explanation for why they remain as the light dynamical fields during inflation.
The modest hierarchy of masses (see e.g.,  
eq.~(\ref{mod2-cond-mpsi})) can be reasonably accommodated without necessarily assuming an ``axiverse" \cite{Arvanitaki:2009fg}.\footnote{See \cite{Acharya:2010zx,Cicoli:2012sz} for some string theory ways of generating a logarithmic hierarchy of axion masses in an ``axiverse".}, though having a hierarchy of axion masses offer more flexibilities. 
Clearly, more model building possibilities (with the inflaton being an axion or not)
remain to be explored. 
We hope to return to such string theory realizations in the future.

\section*{Acknowledgments}

The work of N.B., R.N., and M.P. was supported in part by DOE grant DE-FG02-94ER-40823 at the University of Minnesota.  The work of JM, GS, and PZ was supported in part by DOE grant 
 DE-FG-02-95ER40896 at the University of Wisconsin.  GS would also like to thank the University of Amsterdam for hospitality during the final stage of this work, as he was visiting the Institute for Theoretical Physics as the Johannes Diderik van der Waals Chair.  We are grateful to 
 M.~Berg,
 P.~Camara,
M.~Haack,
 F.~Marchesano,
 L.~Sorbo, A.~Uranga, M.~Voloshin, and T.~Wiegand
  for helpful discussions.

\appendix

\section{Longitudinal vector mode}
\label{app-longA}

In the model (\ref{model1}) the vector has also a longitudinal mode, which was disregarded in the  computations performed in the main text. The longitudinal sector  is actually more subtle than the transverse one, as perturbation theory may break down  in the $m \rightarrow 0$ limit. This can be seen, for example, from the action of the longitudinal modes, as we now show. 

Let us start by disregarding the perturbations of $\Psi$, as we do in the main text. Also, for simplicity, we disregard the expansion of the universe. As we have discussed in the main text, this is a good approximation as long as $H \ll \sqrt{\dot{m}_*}$. Then the mass of the vector field is a classical function of time, and the longitudinal mode is encoded in
\begin{eqnarray}
&& A_\mu \left( x \right)  =  \int \frac{d^3 k}{\left( 2 \pi \right)^{3/2} } {\rm e}^{i {\bf x} \cdot {\bf k} } 
\left( {\tilde A}_0 ,\, i \, k_i \, {\tilde \chi} \right) \nonumber\\
&& \;\;\Rightarrow\;\; {\cal L} \supset - \frac{1}{4} F^2 - \frac{m^2}{2} \, A^2 = \frac{1}{2} \left[ 
k^2 \vert {\tilde \chi}' \vert^2 - k^2 {\tilde A}_0^\dagger {\tilde \chi}' - k^2 {\tilde \chi}^{'\dagger} {\tilde A}_0 + k^2 \vert {\tilde A}_0 \vert^2 - k^2 m^2 \vert {\tilde \chi} \vert^2 + m^2 \vert {\tilde A}_0 \vert^2 \right]
\label{lag-AL-1}
\end{eqnarray}
The component ${\tilde A}_0$ is non-dynamical (it enters in the action without time derivatives) and it can be integrated out. Namely, from (\ref{lag-AL-1}) we obtain the equation
\begin{equation}
{\tilde A}_0 = \frac{k^2}{k^2 + m^2} \, {\tilde \chi}'
\label{eqA0-AL}
\end{equation}
Plugging this solution back into  (\ref{lag-AL-1}) leads to the action of the longitudinal mode
\begin{equation}
S_{\rm long } = \frac{1}{2} \int d t  d^3 k   \left[ \vert {\tilde L}' \vert^2 - \left( k^2 + m^2 + \frac{ 3 k^2 m^{'2}}{\left( k^2 + m^2 \right)^2} - \frac{m''}{m} \, \frac{k^2}{k^2+m^2} \right) \vert {\tilde L} \vert^2 \right] \;\;\;,\;\;\; 
{\tilde L} \equiv \frac{k m}{\sqrt{k^2+m^2}} \, {\tilde \chi}
\label{act-AL}
\end{equation}
The field ${\tilde L}$ is the canonical field associated with the longitudinal mode. We see that its equation of motion has a term that formally diverges as $m \rightarrow 0$.~\footnote{Clearly, the same equation of motion can also be obtained by writing out the equations for the system in terms of the modes ${\tilde A}_0$ and ${\tilde \chi}$ and by eliminating ${\tilde A}_0$ from these equations.} The formal reason for the divergence is that the kinetic term for the original longitudinal mode vanishes in this limit, as it appears from the relation between ${\tilde L}$ and ${\tilde \chi}$  (\ref{act-AL}). This is not surprising, since a massless vector has only transverse modes.

When, as in the present case, the U(1) symmetry is broken spontaneously, there is actually not a decrease of the number of degrees of freedom when the classical background part  $\Psi^{(0)}$ vanishes. The physical mass term in the original action is obtained from ${\cal L} \supset e^2 \vert \Psi^{(0)} \left( t \right) + \delta \Psi \vert^2 \, A^2$, and the quantity that we have denoted by $m$ in the main text is only related to the classical part, $m^2 \equiv 2 e^2 \vert \Psi^{(0)} \left( t \right) \vert^2$. However, when  $ \Psi^{(0)} \left( t \right) = 0 $, the fluctuations of $\Psi$ cannot be disregarded, and one does not obtain a truly massless vector mode at this point (compare this with what would happen if $m$ was a hard and time dependent mass in the original theory; in this case there would be instead a discontinuity in the number of degrees of freedom at $m=0$).

In short, a full study of the longitudinal sector would require going beyond the linearized theory; this may also affect the transverse sector, since all the sectors are coupled to each other beyond the linearized level. Here we use a simpler approach, and discuss under which conditions the longitudinal mode does not blow up in the linearized theory; if this is the case, then one can expect that dealing with the full theory is unnecessary. We note that these considerations apply for a general class of model in which symmetries are enhanced at some point during the cosmological evolution. For instance we expect massive gauge modes to become massless when different branes move to the same bulk location as in the trapping mechanism of  \cite{Kofman:2004yc}.

Using the background equations of motion, the dangerous factor $m''/m$  in the linearized computation (\ref{act-AL}) rewrites
\begin{equation}
\frac{m''}{m} = \frac{\left( a \, \psi \right)''}{a \, \psi} = \frac{a''}{a} - \frac{a^2 \, U_{,\psi}}{\psi}
\simeq  - \frac{ U_{,\psi}}{\psi}
\end{equation}
where the first two equations are exact, and in the  final step  we instead disregard the expansion of the universe. 

Therefore, the equation of motion of ${\tilde L}$ remain finite provided that $U_{,\psi} / \psi$ does not diverge. This could be for instance the case for a quadratic potential $U \sim \frac{1}{2} m_\psi^2 \psi^2$ for $\psi \sim 0$. In our computations, we assumed a constant $\dot{\psi}^{(0)} \equiv \frac{\dot{m}_*}{e} \,$, corresponding to $\psi^{(0)} =  \frac{\dot{m}_*}{e} \left( t - t_* \right)$.  As we discussed in the main text, this implies that the gauge fields are produced during the time interval $\Delta t \sim \frac{1}{\sqrt{\dot{m}_*}}$ around $t = t_*$. Then, for a quadratic potential, imposing that the potential energy gained by $\psi^{(0)}$ during this interval is smaller than its kinetic energy at $t_*$ (so that  $\dot{\psi}^{(0)} $ can indeed be taken as constant)  amounts in requiring $m_\psi \ll \sqrt{\dot{m}_*}$. This also ensures that the period of oscillations of $\psi^{(0)}$ is much greater than the time in which particle production takes place, so that one can indeed treat $\alpha$ and $\beta$ as constant when computing the amount of perturbations sourced by the gauge modes.

Using (\ref{act-AL}), we computed the occupation number of longitudinal vector modes  with  $U \sim \frac{1}{2} m_\psi^2 \psi^2$, and  $m_\psi \ll \sqrt{\dot{m}_*}$. We found that in this regime the longitudinal mode is produced in essentially the same amount as each transverse vector mode (for instance, we found that for $m_\psi = 0.1 \,  \sqrt{\dot{m}_*}$ the total number densities of the longitudinal and of one transverse polarization differ from each other by less than $1 \%$). We stress however that this conclusion is model dependent, as the precise evolution of the effective frequency in  (\ref{act-AL}) depends on the details of $U \left( \psi \right)$. On the contrary, the amounts of the transverse modes produced is independent of $U$, provided that it remains sufficiently smaller than the kinetic energy during the time of production.

In the remainder of  this Appendix we study the spectrum of gravity waves produced in this model, once also the longitudinal modes are taken into account. When all modes are taken into account, we formally separate  the energy momentum tensor (\ref{Tmunu-mod1}) of the vector field into $T_{\mu \nu}^{TT} + T_{\mu \nu}^{LT} + T_{\mu \nu}^{LL}$, where the first term is quadratic in the transverse polarizations, and it is the only one used in the main text, while the third term is quadratic in the longitudinal polarization, and the second term is the ``mixed term''. Inserting the last two terms into (\ref{Jlambda-formal}) we obtain the two contributions
\begin{eqnarray}
J_{\lambda}^{LT} \left( \vec{k} \right) & = &  \frac{ \Pi_{ij,\lambda}^* }{a M_p } \left( {\hat k} \right) \int \frac{d^3 p}{\left( 2 \pi \right)^{3/2}} 2 i p_i M^2 \left[ - \frac{\partial_\tau^{(1)} \,  \partial_\tau^{(2)} }{p^2 + M^2} + 1 \right]
{\tilde \chi} \left( \vec{p} \right) {\tilde A}_j \left( \vec{k} - \vec{p} \right) \nonumber\\
J_{\lambda}^{LL} \left( \vec{k} \right) & = &  \frac{ \Pi_{ij,\lambda}^* }{a M_p } \left( {\hat k} \right) \int \frac{d^3 p}{\left( 2 \pi \right)^{3/2}} 
M^2 p_i    p_j \left[ - \frac{ M^2 \partial_\tau^{(1)} \,  \partial_\tau^{(2)}  }{ \left[ p^2 + M^2 \right] \left[ \left( k - p \right)^2 + M^2 \right] } + 1 \right] {\tilde \chi} \left( \vec{p} \right) {\tilde \chi} \left( \vec{k} - \vec{p} \right) \nonumber\\
\end{eqnarray}
to the gravity wave source in (\ref{hc-eq-formal}). These add up to the term $J_{\lambda}^{TT}$, which is the only one studied in Subsection \ref{subsec:GW1} of the main text (where it is denoted simply by $J_{\lambda}$). 

To obtain the total power spectrum we need to evaluate the correlator 
$\left\langle \left[  J_\lambda^{TT} +   J_\lambda^{LT} +   J_\lambda^{LL} \right]_{\tau_1 ,\, \vec{k}_1} 
 \left[  J_{\lambda'}^{TT} +   J_{\lambda'}^{LT} +   J_{\lambda'}^{LT} \right]_{\tau_2 ,\, \vec{k}_2} \right\rangle$ and insert it in (\ref{formal-QQ}). Each piece in $J_\lambda$ is correlated only with  the corresponding piece in   $J_{\lambda'}$.
 
Proceeding as in the main text, we obtain
\begin{eqnarray}
\left\langle J_{\lambda}^{LT} \left( \tau_1 ,\, \vec{k} \right) J_{\lambda}^{LT} \left( \tau_2 ,\, \vec{k}' \right) \right\rangle &\simeq&
 \frac{2}{5 \pi^2 M_p^2} \delta_{\lambda \lambda'} \delta^{(3)} \left( \vec{k} + \vec{k}' \right) \frac{M \left( \tau_1 \right)^2}{a \left( \tau_1 \right)}  \frac{M \left( \tau_2 \right)^2}{a \left( \tau_2 \right)} \int d p p^4 \nonumber\\
  && \!\!\!\!\!\!\!\! \!\!\!\!\!\!\!\! 
   \left[ - \frac{\delta_{a1}}{p^2 + M^2} + \delta_{a0} \right]_{\tau_1}  \left[ - \frac{\delta_{b1}}{p^2 + M^2} + \delta_{b0} \right]_{\tau_2} {\cal C}_{(a,b)} \left[ \tau_1 ,\, \tau_2 ;\, p \right]  {\cal D}_{(a,b)} \left[ \tau_1 ,\, \tau_2 ;\, p \right] 
\label{JLT-JLT}   
\end{eqnarray}
and
\begin{eqnarray}
\left\langle J_{\lambda}^{LL} \left( \tau_1 ,\, \vec{k} \right) J_{\lambda}^{LL} \left( \tau_2 ,\, \vec{k}' \right) \right\rangle &\simeq&
 \frac{2}{15 \pi^2 M_p^2} \delta_{\lambda \lambda'} \delta^{(3)} \left( \vec{k} + \vec{k}' \right) \frac{M \left( \tau_1 \right)^2}{a \left( \tau_1 \right)}  \frac{M \left( \tau_2 \right)^2}{a \left( \tau_2 \right)} \int d p p^6 \nonumber\\
  && \!\!\!\!\!\!\!\! \!\!\!\!\!\!\!\! 
   \left[ - \frac{M^2 \, \delta_{a1}}{ \left( p^2 + M^2 \right)^2 } + \delta_{a0} \right]_{\tau_1} 
   \left[ - \frac{M^2 \, \delta_{b1}}{ \left( p^2 + M^2 \right)^2 } + \delta_{b0} \right]_{\tau_2}  
    {\cal C}_{(a,b)}^2   \left[ \tau_1 ,\, \tau_2 ;\, p \right]  
\label{JLL-JLL}
\end{eqnarray}

In these expressions we have introduced the correlators 
\begin{equation}
\left\langle {\tilde \chi}  \left( \tau_1 ,\, \vec{p}_1 \right)  {\tilde \chi}  \left( \tau_2 ,\, \vec{p}_2 \right) \right\rangle  =  {\cal C}_{(0,0)}  \left[ \tau_1 ,\, \tau_2 ;\, p_1 \right] \, \delta^{(3)} \left( \vec{p}_1 + \vec{p}_2 \right) 
\end{equation} 
which evaluate to
\begin{equation}
{\cal C}_{(0,0)} \left[ \tau_1 ,\, \tau_2 ;\, p \right] =    \frac{\sqrt{p^2 + M^2 \left( \tau_1 \right)} \, \sqrt{p^2 + M^2 \left( \tau_2 \right)}}{p^2 M \left( \tau_1 \right) M \left( \tau_2 \right)}  \,  {\cal D}_{(0,0)}^L \left[ \tau_1 ,\, \tau_2 ;\, p \right]  
\end{equation}
where ${\cal D}_{(a,b)}^L$ are formally identical to the ${\cal D}_{(a,b)}^{(\sigma)}$ quantities given in eqs. (\ref{mod1-D00}) and (\ref{mod1-Dab}), with the only difference that in  ${\cal D}_{(a,b)}^L$ we use the Bogolyubov coefficients and the mode functions of the longitudinal mode:
\begin{equation}
{\tilde L} \left( \vec{k} \right) = \left[ \alpha_k^L \, g_k  +  \beta_k^L \, g_k^* \right] a_{\vec{k}}^L + {\rm h.c.}
\end{equation}
where $a_k^L$ is an annihilation operator, and
\begin{equation}
g_k \equiv \frac{{\rm e}^{- i \int^\tau d \tau' \, \omega_L \left( \tau' \right)} }{ \sqrt{2 \omega_L \left( \tau \right)} } \;\;\;,\;\;\;
\omega_L \equiv  \left( k^2 + M^2 + \frac{ 3 k^2 M^{'2}}{\left( k^2 + M^2 \right)^2} - \frac{M''}{M} \, \frac{k^2}{k^2+M^2} \right)^{1/2} 
\end{equation}

As in the computation of $\langle J_\lambda^{TT} J_{\lambda'}^{TT} \rangle$ presented in the main text, 
 the square of the correlators appearing in (\ref{JLT-JLT}) and (\ref{JLL-JLL}) contain terms proportional to fast oscillating phases, which give a negligible contribution to the final result. Disregarding these terms, we obtain
\begin{eqnarray}
\left\langle J_{\lambda}^{LT} \left( \tau_1 ,\, \vec{k} \right) 
J_{\lambda}^{LT} \left( \tau_2 ,\, \vec{k}' \right) \right\rangle \simeq
 \frac{- \delta_{\lambda \lambda'}}{20 \pi^2 M_p^2}  \delta^{(3)} \left( \vec{k} + \vec{k}' \right)  \;
  \int d p p^6   {\rm Re } 
  \left[ \alpha_p \alpha_p^{L*}   \beta_p^* \beta_p^L Q \left( \tau_1 \right)    Q \left( \tau_2 \right)   
  - \left\vert \beta_p \right\vert^2 \left\vert \beta_p \right\vert^2 Q \left( \tau_1 \right)  Q^* \left( \tau_1 \right) 
  \right] \nonumber\\
\label{JLTJLT-result}
\end{eqnarray} 
 where
\begin{equation}    
Q \left( \tau \right) \equiv \frac{1}{a M} \left[ \frac{2 M'}{M^2} - \frac{3 i M^{' 2}}{M^4} + \frac{i M''}{M^3} \right]
\end{equation} 
and
\begin{equation}
\left\langle J_{\lambda}^{LL} \left( \tau_1 ,\, \vec{k} \right) 
J_{\lambda}^{LL} \left( \tau_2 ,\, \vec{k}' \right) \right\rangle \simeq
 \frac{  \delta_{\lambda \lambda'}  }{15 \pi^2 M_p^2} \, \delta^{(3)} \left( \vec{k} + \vec{k}' \right) \frac{1}{a M \left( \tau_1 \right)}  \frac{1}{a M \left( \tau_2 \right)} \int d p p^6 
\;  \vert \beta_p^L \vert^2 \left( 
\vert \alpha_p^L \vert^2 +  \vert \beta_p^L \vert^2  \right)
\label{JLLJLL-result}
\end{equation}
In these expressions, we have retained only the dominant contributions to the $\omega_L$ in the adiabatic regime (namely, we have used $M' \ll M^2$ and $M'' \ll M^3$). As in the computation of the main text, the terms without oscillatory phases are of ${\rm O } \left( p^6 \right)$, as a consequence to a cancellation of the would be dominant terms in the transverse and traceless projection of the energy-momentum tensor. This cancellation is already visible in  (\ref{JLT-JLT}) and (\ref{JLL-JLL}), see the discussion after the analogous expression (\ref{mod1-ph-firstexp}) presented in the main text.

We see that the mixed term (\ref{JLTJLT-result})  is suppressed in the adiabatic regime, and can be disregarded. From the  (\ref{JLLJLL-result}) we instead obtain
\begin{equation}
P_{\lambda,{\rm s}} 
\big\vert_{{\rm from \; } A_L}
 \simeq  \frac{2 k^3}{15 \pi^4 a^2 M_p^4} 
\tilde{\cal T}_k^2  \int d p \,  p^6 \vert \beta_p^L \vert^2 \left( \vert \alpha_p^L \vert^2 + \vert \beta_p^L \vert^2 \right) 
\end{equation}
where $\tilde{\cal T}_k$ is defined in (\ref{def-TE-tilde}). This contribution adds up to the one of the transverse modes leading to the result (\ref{Plamda-TT-LL}) given in the main text.

\section{Source for $\zeta$ in Model I}
\label{app-mod1-sourcez}

In this appendix, we derive the approximated expression (\ref{mod1-source-z}) for the scalar source in Model I. The source we are interested in is the part of (\ref{mod1-sourceJJt}) that depends on the gauge field. From the three terms at the r.h.s. of  (\ref{mod1-eqphi2}) we obtain, respectively,
\begin{eqnarray}
&& J \left[ A_\mu^2 \right] =  J_1 \left( \tau , \vec{k} \right) +  J_2 \left( \tau , \vec{k} \right) +  J_3 \left( \tau , \vec{k} \right) \nonumber\\
&& J_1 \left( \tau , \vec{k} \right) = - \frac{\dot{\varphi}^{(0)}}{4 M_p^2 H a \left( \tau \right)} \int{\frac{d^3 p}{\left( 2 \pi \right)^{3/2}}} {\tilde A}_i' \left( \tau , \vec{p} \right) {\tilde A}_i' \left( \tau , \vec{k} - \vec{p} \right) \label{source_J1} \\
&& J_2 \left( \tau , \vec{k} \right) = \frac{\dot{\varphi}^{(0)}}{2 M_p^2 H a \left( \tau \right)} \frac{1}{k^2} \int{\frac{d^3 p}{\left( 2 \pi \right)^{3/2}}} \partial_\tau \left[ p_i \, k_i \, {\tilde A}_j \left( \tau , \vec{p} \right) {\tilde A}_j' \left( \tau , \vec{k} - \vec{p} \right) - k_i \, k_j \, {\tilde A}_i \left( \tau , \vec{p} \right) {\tilde A}_j' \left( \tau , \vec{k} - \vec{p} \right) \right] \label{source_J2} \\
&&  J_3 \left( \tau , \vec{k} \right) = - \frac{M^2 \dot{\varphi}^{(0)} \,}{4 M_p^2 H  a \left( \tau \right)} \int{\frac{d^3 p}{\left( 2 \pi \right)^{3/2}}} {\tilde A}_i \left( \tau , \vec{p} \right) {\tilde A}_i \left( \tau , \vec{k} - \vec{p} \right) . \label{source_J3}
\end{eqnarray}
In the part $J_1$ we have actually disregarded the ``magnetic'' contribution with respect to the ``electric'' one.\footnote{An analogous simplification cannot be done for the source of gravity waves, since in this case the dominant term cancels.} The relative contribution between the two terms in the integrand is $\frac{B^2}{E^2} \sim \frac{{\tilde A}^{' 2}}{{\tilde A}^2}  \sim \frac{k^2 ,\, k \, p ,\, p^2}{M^2}$. The quantity $k$ is the momentum of the cosmological perturbations that we are computing; we show in the main text that the signal from particle production is maximal at $k \sim H$ (we recall that the scale factor is normalized to one at the moment of particle production). We also show in the main text that the source integrand is peaked at $p \sim \sqrt{\dot{m}_*}$. We therefore have $k \ll p$ due to (\ref{mod1-bck-cond}). Finally $M = a m \geq m \gg  \sqrt{\dot{m}_*}$ after the particle production has taken place. As a consequence $p \ll M$ and the ``magnetic'' contribution can indeed be disregarded.

We perform the time derivative in $J_2$, and eliminate the second derivative through the equation of motion ${\tilde A}_i'' \simeq - M^2 \,  {\tilde A}_i$. We obtain
\begin{eqnarray}
 J_2 \simeq \frac{\dot{\phi}}{2 M_p^2 H a} \, \frac{1}{k^2} \int{\frac{d^3 p}{\left( 2 \pi \right)^{3/2}}} \left( \vec{k} \cdot \vec{p} \, \delta_{ij} - k_i \, k_j \right) 
\;  \left[ {\tilde A}_i' \left( \vec{p} \right) {\tilde A}_j' \left( \vec{k} - \vec{p} \right) -  M^2 \, {\tilde A}_i \left( \vec{p} \right) {\tilde A}_j \left( \vec{k} - \vec{p} \right) \right] .  \nonumber\\
\label{J2par}
\end{eqnarray}
We rewrite (\ref{J2par}) changing the integration variable $\vec{p} \rightarrow \vec{k} - \vec{p}$. We add the resulting expression to  (\ref{J2par}), and divide by two. We obtain
\begin{equation}
J_2 \simeq \frac{\dot{\phi}}{2 M_p^2 H a} \left( \frac{1}{2} \delta_{ij} - \hat{k}_i \hat{k}_j \right) \int{\frac{d^3 p}{\left( 2 \pi \right)^{3/2}}} \left[ {\tilde A}_i' \left( \vec{p} \right) {\tilde A}_j' \left( \vec{k} - \vec{p} \right) - M^2 {\tilde A}_i \left( \vec{p} \right) {\tilde A}_j \left( \vec{k} - \vec{p} \right) \right] \label{J2_app}
\end{equation}

It is worth noting that this expression explicitly shows that $J_2$ does not diverge in the $k \rightarrow 0$ limit. This is not immediately obvious from the original expression, since $J_2$ originates from the term with an inverse laplacian in   (\ref{mod1-eqphi2}). 

Adding this expression for $J_2$ to the expression for $J_1$ and $J_3$ given above, one readily obtains the result  (\ref{mod1-source-z}) given in the main text.

\section{Source for $\zeta$ in Model II}
\label{app:modelII}

In this appendix, we derive the  expression (\ref{mod2-explicitsource}) for the scalar source in Model II. In eq. (\ref{mod2-eqphi2}) we separated the source in the two parts  $J_1 + J_2$.

In Fourier space we have the relatively simple expression
\begin{equation}
 \label{mod2-J1}
  J_1\left(\tau,\vec{k}\right) = -\frac{\varphi'^{(0)} a^3}{4 M_p^2 \sH} \int\frac{d^3p}{(2\pi)^{3/2}}
  \left[\,\, \tilde{E}_i\left(\tau,\vec{p}\right) \tilde{E}_i\left(\tau,\vec{k}-\vec{p}\right) + \tilde{B}_i\left(\tau,\vec{p}\right)\tilde{B}_i\left(\tau,\vec{k}-\vec{p}\right) \,\,\right] \, ,
\end{equation}
for the first part. For the second part, we have instead
\begin{equation}
\label{mod2-J2-1}
J_2 \left( \tau ,\, \vec{k} \right) = \frac{\varphi^{'(0)}}{2 M_p^2 {\cal H} a} \frac{1}{k^2}
\int \frac{d^3 p}{\left( 2 \pi \right)^{3/2}} \partial_\tau \left[ p_i k_i {\tilde A}_j \left( \tau ,\, \vec{p} \right) {\tilde A}_j' \left( \tau ,\, \vec{k} - \vec{p} \right) - k_i k_j {\tilde A}_i \left( \tau ,\, \vec{p} \right) \, {\tilde A}_j' \left( \tau ,\, \vec{k} - \vec{p} \right) \right] \, .
\end{equation}
Some manipulations are instead necessary to put the non-local source term $J_2$ in a form that is amenable to computations and is manifestly infra-red finite.  We begin by noting that $\tilde{A}_i$ effectively only contains the ``$+$'' helicity state in this model.  This allows us to use the identity
\begin{equation}
  k_i \, k_j \, \epsilon_i^{(+)}\left(\vec{p}\right)\epsilon_j^{(+)}\left(\vec{k}-\vec{p}\right) = 
  \left[  p |\vec{k}-\vec{p}| + (\vec{k}-\vec{p})\cdot \vec{p}   \right] \, \epsilon_i^{(+)}\left(\vec{p}\right)\, \epsilon_i^{(+)}\left(\vec{k}-\vec{p}\right) \, ,
\end{equation}
to simplify the tensor structure in the second term of (\ref{mod2-J2-1}) and re-write $J_2$ as
\begin{equation}
  J_2\left(\tau,\vec{k}\right) = \frac{\varphi'^{(0)}}{2M_p^2 \sH a}\frac{1}{k^2} \int\frac{d^3p}{(2\pi)^{3/2}}\,\, p \left[ p - |\vec{k}-\vec{p}| \right] \partial_\tau
  \left[ \,\, \tilde{A}_i\left(\tau,\vec{p}\right) \tilde{A}_i'\left(\tau,\vec{k}-\vec{p}\right) \,\, \right]  \, .
\end{equation}
Next we perform the time derivative and use the equation of motion (\ref{pseudo_eom}) to eliminate $\tilde{A}''$ which gives the result
\begin{equation}
  J_2\left(\tau,\vec{k}\right) = \frac{\varphi'^{(0)} a^3}{2M_p^2 \sH}\frac{1}{k^2} \int\frac{d^3p}{(2\pi)^{3/2}} 
\left( |\vec{k}-\vec{p}|-p \right) 
\left[\,\, -p \tilde{E}_i\left(\tau,\vec{p}\right) \tilde{E}_i\left(\tau,\vec{k}-\vec{p}\right) 
         + \left(  |\vec{k}-\vec{p}|+\frac{2\xi}{\tau}  \right) \tilde{B}_i\left(\tau,\vec{p}\right)\tilde{B}_i\left(\tau,\vec{k}-\vec{p}\right)  \,\,\right] \, .
\label{J2:noflip}
\end{equation}
We rewrite (\ref{J2:noflip}) changing the integration variable $\vec{p} \rightarrow \vec{k} - \vec{p}$. We add the resulting expression to  (\ref{J2:noflip}), and divide by two. We obtain:
\begin{equation}
\label{mod2-J2-final_app}
 J_2\left(\tau,\vec{k}\right) = \frac{\varphi'^{(0)} a^3}{4M_p^2 \sH} \int\frac{d^3p}{(2\pi)^{3/2}}  \frac{(p-|\vec{k}-\vec{p}|)^2}{k^2}
 \left[ \,\, \tilde{E}_i\left(\tau,\vec{p}\right) \tilde{E}_i\left(\tau,\vec{k}-\vec{p}\right) + \tilde{B}_i\left(\tau,\vec{p}\right)\tilde{B}_i\left(\tau,\vec{k}-\vec{p}\right) \,\, \right] \, .
\end{equation}
We note that this expression is manifestly finite in the limit $k^2\rightarrow 0$, proving that the non-local source term in (\ref{mod2-eqphi2}) does not lead to any spurious effects in the infra-red.

Adding (\ref{mod2-J1}) and (\ref{mod2-J2-final_app}) leads to the expression (\ref{mod2-explicitsource}) reported in the main text.

\section{Fermionic production and gravity waves}
\label{app-fermions}

In this appendix, we outline the computation of gravity waves produced by a fermion with a mass varying as in the model I studied in the main text. If we denote by $X$ the fermion in the original action, and we rescale, $\chi = X \, a^{3/2}$, the action for the fermion field becomes identical  to the one of a massive fermion in Minkowski spacetime, whose mass is multiplied by the scale factor
\begin{equation}
S_f = \int d^4 x \, {\bar \chi} \left[ i \, \gamma^\mu \partial_\mu - M \left( t \right)\right] \chi
\;\;\;,\;\;\; M = g \, a \left( t \right) \, \left[ \psi^{(0)} \left( t \right) - \psi_* \right] 
\label{act-fermi}
\end{equation}
As in the main text, $\psi^{(0)}$ is  a homogeneous classical field that evaluates to $\psi_*$ at some given moment $t_*$ during inflation, while $g$ is the coupling of the Yukawa interaction in the original action. Without loss of generality we can choose $\psi_* = t_* = 0$. As in the main text, we consider a regime in which the expansion of the universe can be disregarded during the particle production, and we expand
\begin{equation}
g \, \psi^{(0)} \left( t \right) \equiv \dot{m}_* \, t \;\;\;,\;\;\; t \simeq 0
\label{fermion-mass}
\end{equation}
We instead include the expansion of the universe when we study the amount of gravity waves sourced by the fermionic quanta produced at $t \simeq 0$.

The fermionic production in this model was studied in \cite{preheating-fermions,Peloso:2000hy} for reheating after inflation and leptogenesis, and in \cite{Chung:1999ve} for  the imprint  on the scalar power spectrum. Here we follow the computations of \cite{Peloso:2000hy,Nilles:2001fg}, skipping some intermediate steps. We refer the reader to those works for details. One decomposes \footnote{We use 
\begin{equation}
\gamma^0 = \left( \begin{array}{cc} \identity & 0 \\ 0 & -\identity \end{array} \right) \;\;\;,\;\;\;
\gamma^i = \left( \begin{array}{cc} 0 & \sigma^i \\ -\sigma^i & 0 \end{array} \right)
\end{equation}
and, only in this Appendix, we switch to $+,-,-,-$ signature.}
\begin{eqnarray}
&& \chi  =  \int \frac{d^3 k}{\left( 2 \pi \right)^{3/2} } \, {\rm e}^{i \vec{k} \cdot \vec{x}} {\tilde \chi} \left( \tau ,\, \vec{k} \right) \;\;\;,\;\;\;  {\tilde \chi} \left( \tau ,\, \vec{k} \right) =  \sum_r
\left[ {\cal U}_r \left( \tau ,\, \vec{k} \right) \, a_r \left( \vec{k} \right) +  {\cal V}_r \left( \tau ,\, - \vec{k} \right) \, b_r^\dagger \left( - \vec{k} \right) \right] 
\end{eqnarray}
where, for $\vec{k}$ aligned along the $z-$axis,
\begin{eqnarray}
 {\cal U}_r  \left( \tau ,\, k_z \right) =  \frac{1}{\sqrt{2}} \left( \begin{array}{c} U_+ \left( \tau ,\, k \right) \, \psi_r \\     \;  U_- \left( \tau ,\, k \right) \; 
r \, \psi_r  \end{array} \right) \;\;\;,\;\;\;
 {\cal V}_r  \left( \tau ,\, - k_z \right) =  \frac{1}{\sqrt{2}} \left( \begin{array}{c} 
-  V_+ \left( \tau ,\,  k \right) \, \psi_{-r} \\     
-  V_- \left( \tau ,\,  k \right) \, r \, \psi_{-r}   \end{array} \right) 
\end{eqnarray}
where $\psi_+ =  \left( \begin{array}{c} 1 \\ 0  \end{array} \right) $ and  $\psi_- =  \left( \begin{array}{c} 0 \\ 1  \end{array} \right) $. The spinor ${\cal V}$ is related to ${\cal U}$ by charge conjugation, giving $V_\pm \left( \tau ,\, k \right) = U_\mp^* \left( \tau ,\, k \right)$.

It is convenient to decompose the spinors in terms of the Minkowski solutions (more precisely, the adiabatic solution, since in the current case $M$ is not constant)
\begin{equation}
U_\pm \left( \tau ,\, k \right) =  \alpha \left( t \right) f_\pm \left( \tau ,\, k \right)  \mp  \beta \left( t \right)  f_\mp^* \left( \tau ,\, k \right)    \;\;\;,\;\;\;
f_\pm \left( \tau ,\, k \right)  \equiv   \sqrt{ 1 \pm \frac{M}{E} }  {\rm e}^{ - i \int_{\tau_*}^\tau d \tau' \, E} \,
\label{fermi-deco}
\end{equation}
and $E \equiv \sqrt{k^2 +M^2}$. Starting from $\beta =0$ at asymptotically early times, and using (\ref{fermion-mass}) one finds, at late times (see, for instance,  \cite{Peloso:2000hy} for the computation)
\begin{equation}
\alpha \left( t \gg 0 \right)  \simeq  \frac{2 \sqrt{\pi}}{q} {\rm e}^{i q^2/2} {\rm e}^{i \pi/4} \left( \frac{q}{\sqrt{2}} \right)^{-i q^2} \, \frac{{\rm e}^{- \pi q^2/4}}{\Gamma \left( \frac{ - i q^2 }{ 2 } \right)} \;\;\;,\;\;\;
\beta \left( t \gg 0 \right) \simeq  - {\rm e}^{- \pi q^2/2} \;\;\;,\;\;\; q \equiv \frac{k}{\sqrt{\dot{m}_*}}
\label{late-ab-fermi}
\end{equation}
As in the bosonic case, $\vert \beta \vert^2$ is the occupation number, and Pauli blocking is ensured by the fact that, in the fermionic case, the  Bogolyubov coefficients  satisfy $\vert \alpha \vert^2 + \vert \beta \vert^2 = 1$.

To compute the gravity waves produced by these quanta, we proceed as in the main text, and obtain the formal solution (\ref{formal-sol2}) for the sourced part $Q_{\lambda,{\rm s}}$  of the canonical gravity wave modes  introduced in (\ref{formal-hc-def}). Using the fermionic  energy momentum tensor 
\begin{equation}
T_{\mu \nu} = \frac{i}{4 a^2} \left[ {\bar \chi} \gamma_\mu \partial_\nu \chi +  {\bar \chi} \gamma_\nu \partial_\mu \chi - \left( \partial_\mu {\bar \chi} \right) \gamma_\nu \chi -  \left( \partial_\nu {\bar \chi} \right) \gamma_\mu \chi  \right]
\end{equation}
we can cast the source appearing in  (\ref{formal-sol2}) in the form
\begin{equation}
J_\lambda \left( \tau,\, \vec{k} \right) = \frac{1}{2 a M_p} \,  \Pi_{ij,\lambda}^* \left( {\hat k} \right)  \int \frac{d^3 p}{\left( 2 \pi \right)^{3/2}} \bar{\tilde \chi} \left( \tau ,\, \vec{p} \right)  \left( \gamma_i \, p_j + \gamma_j \, p_i \right) \, {\tilde \chi } \left( \tau ,\, \vec{k} + \vec{p} \right) 
\end{equation}
After Wick contraction, the correlator of two sources acquires the form
\begin{eqnarray}
\left\langle J_\lambda \left( \tau ,\,  \vec{k} \right)  J_{\lambda'} \left( \tau' ,\,  \vec{k}' \right) \right\rangle
& = & \frac{1}{ M_p^2 a \left( \tau \right) a \left( \tau' \right) } \, \Pi_{ij,\lambda}^*
 \left( {\hat k} \right) 
\Pi_{lm,\lambda'}^* \left( {\hat k}' \right) \int \frac{d^3 p d^3 p'}{\left( 2 \pi \right)^3} \nonumber\\
&&  
{\rm tr } \left[  \gamma^i p^{j} 
 \left\langle   {\tilde \chi} \left( \tau ,\, \vec{k} + \vec{p} \right) \,  \bar{\tilde \chi} \left( \tau' ,\, \vec{p}' \right)  \right\rangle  \gamma^l p^{'m} 
 \left\langle  \bar{\tilde \chi} \left( \tau ,\, \vec{p} \right) \, {\tilde \chi} \left( \tau' ,\, \vec{k}' + \vec{p}'  \right) \right\rangle^T \right] 
\label{JJ-exp1-fermi}
\end{eqnarray}
where the transposition  acts on the spinor indices. The correlators appearing in this expression need to be regularized. We adopt the same prescription adopted in Section \ref{sec:model1} for the vector field correlators. Namely, we normal order the fields appearing in the correlators with respect to the time dependent operators
\begin{eqnarray}
{\hat a} \left( \vec{k} \right) & \equiv & \alpha \, a \left( \vec{k} \right) - \beta^* \, b^\dagger \left( - \vec{k} \right) \nonumber\\
{\hat b}^\dagger \left( - \vec{k} \right) & \equiv & \beta \, a \left( \vec{k} \right) + \alpha^* \, b^\dagger \left( - \vec{k} \right) 
\end{eqnarray}
that diagonalize the Hamiltonian at any given time. We then recall that the vacuum of the theory is annihilated by the original time-independent $a$ and $b$ operators. After some algebra, we obtain
\begin{eqnarray} 
 \left\langle  : {\tilde \chi} \left( \tau ,\, \vec{k} + \vec{p} \right) \,  \bar{\tilde \chi} \left( \tau' ,\, \vec{p}' \right)  : \right\rangle   & = & \delta^{(3)} \left( \vec{p}' - \vec{p} - \vec{k} \right) \;
  \left( \begin{array}{cc} 
 C_{11} \left[ \tau ,\, \tau' ;\, p' \right] \, \identity & C_{12}  \left[ \tau ,\, \tau' ;\, p' \right] \, \vec{\sigma} \cdot \hat{p}' \\
C_{21}  \left[ \tau ,\, \tau' ;\, p' \right] \, \vec{\sigma} \cdot \hat{p}' & C_{22} \ \left[ \tau ,\, \tau' ;\, p'\right] \, \identity 
\end{array} \right) \nonumber\\
&=& \delta^{(3)} \left( \vec{p}' - \vec{p} - \vec{k} \right) \;
 \left( \frac{  {\cal C}_{11} + {\cal C}_{22} }{2} \; \identity + 
 \frac{  {\cal C}_{11} - {\cal C}_{22} }{2}  \; \gamma^0 + 
 \frac{ {\cal C}_{12} + {\cal C}_{21} }{2} \; \gamma^0 \vec{\gamma} \cdot {\hat p}'
+ \frac{ {\cal C}_{12} - {\cal C}_{21} }{2}  \vec{\gamma} \cdot {\hat p}' \right) \nonumber\\
 \left\langle  \bar{\tilde \chi} \left( \tau ,\, \vec{p} \right) \, {\tilde \chi} \left( \tau' ,\, \vec{k}' + \vec{p}'  \right) \right\rangle^T  & = & \delta^{(3)} \left( \vec{p} - \vec{p}' - \vec{k}' \right) 
 \; \left( \begin{array}{cc}
- {\cal C}_{22} \left[ \tau ,\, \tau'  ;\, p \right] \; \identity &  {\cal C}_{21}^*  \left[ \tau ,\, \tau' ;\, p \right] \;
\vec{\sigma} \cdot {\hat p}  \\ 
{\cal C}_{12}^*  \left[ \tau ,\, \tau' ;\, p \right] \; \vec{\sigma} \cdot {\hat p} & - {\cal C}_{11}  \left[ \tau ,\, \tau' ;\, p  \right] \; \identity
\end{array} \right)  
\end{eqnarray}
where, when particle creation has completed, and (\ref{late-ab-fermi}) hold,
\begin{eqnarray}
{\cal C}_{11}  \left[ \tau ,\, \tau' ;\, p' \right] & =  &
\frac{1}{2}
\left\{ - \vert \beta \vert^2 \left[ f_+ \left( \tau , p' \right) f_+^* \left( \tau' ,\, p' \right)  - f_-^* \left( \tau , p' \right) f_- \left( \tau' ,\, p' \right)  \right] 
- \alpha \beta^* f_+ \left( \tau , p' \right) f_- \left( \tau' ,\, p' \right)  
- \alpha^* \beta f_-^* \left( \tau , p' \right) f_+^* \left( \tau' ,\, p' \right)  \right\} \nonumber\\
 {\cal C}_{12}  \left[ \tau ,\, \tau'  ;\, p' \right] & = &  \frac{1}{2}
\left\{  \vert \beta \vert^2 \left[ f_+ \left( \tau , p' \right) f_-^* \left( \tau' ,\, p' \right)  + f_-^* \left( \tau , p' \right) f_+ \left( \tau' ,\, p' \right)  \right] 
- \alpha \beta^* f_+ \left( \tau , p' \right) f_+ \left( \tau' ,\, p' \right)  
+ \alpha^* \beta f_-^* \left( \tau , p' \right) f_-^* \left( \tau' ,\, p' \right)  \right\}  \nonumber\\
{\cal C}_{21}  \left[ \tau ,\, \tau' ;\, p' \right] & = & - {\cal C}_{12}^*  \left[ \tau ,\, \tau' ;\, p' \right] 
  \nonumber\\
{\cal C}_{22}  \left[ \tau ,\, \tau' ;\, p' \right] & = & {\cal C}_{11}^*  \left[ \tau ,\, \tau' ;\, p' \right] 
\end{eqnarray}

We insert these expressions into (\ref{JJ-exp1-fermi}). We evaluate the trace, and (as in the analogous computation of the main text ) disregard $\vec{k}$ and $\vec{k}'$ in comparison to $\vec{p}$ and $\vec{p}'$. The  $d^3 p'$ integral can be performed using one $\delta-$function. We perform the angular part of the remaining  $d^3 p$ integral. Finally, we use the property of the polarization operators $\Pi_{ij}$.
We obtain
\begin{equation}
\left\langle J_\lambda \left( \tau ,\,  \vec{k} \right)  J_{\lambda'} \left( \tau' ,\,  \vec{k}' \right) \right\rangle
 =   \frac{2}{15 \, \pi^2} \frac{\delta^{(3)} \left( \vec{k} + \vec{k'} \right)
\delta_{\lambda \lambda'}
}{ M_p^2 a \left( \tau \right) a \left( \tau' \right) }  
 \int  d p \, p^4 \;   {\rm Re } \left[ 5 \, {\cal C}_{11}^2 + {\cal C}_{12}^2 \right]
\label{JJ-exp2-fermi}
\end{equation}
Squaring the correlators, we have
\begin{equation}
{\cal C}_{11}^2 =  \vert \beta \vert^2 \left( \vert \alpha \vert^2 - \vert \beta \vert^2 \right) \, \frac{p^2}{2 \, E \left( \tau \right) \, E \left( \tau' \right)} + {\rm oscillatory \; phases} \;\;\;,\;\;\;
{\cal C}_{12}^2 =  - {\cal C}_{11}^2  + {\rm oscillatory \; phases} 
\label{fermi-C11-C12-res}
\end{equation}
and we see that the non oscillatory part of the integrand in (\ref{JJ-exp2-fermi}) is of ${\cal O} \left( p^6 \right)$ as in the vector case studied in the main text. The oscillatory part gives a negligible contribution to the tensor power. Combining this result with (\ref{formal-QQ}) and (\ref{formal-Plambda-def}) we obtain
\begin{eqnarray}
P_{\lambda,{\rm s}} & \simeq & \frac{8 k^3}{15 \pi^4 a^2 M_p^4} 
\tilde{\cal T}_k^2  \int d p \,  p^6 \vert \beta \vert^2 \left( \vert \alpha \vert^2 - \vert \beta \vert^2 \right) 
\label{fermi-Plambda-beforpint}
 \end{eqnarray}
where $\tilde{\cal T}_k $ is defined in (\ref{def-TE-tilde}). 

Inserting the result (\ref{mod1-ttilde}) for this quantity, and performing the momentum integral, gives
\begin{equation}
P_{\lambda,{\rm s}}  \simeq  \frac{\left( 8 - \sqrt{2} \right) \, H^4 \, \dot{m}_*^{3/2}}{16 \, \pi^7 \, M_p^4 \, k^3} \,
\left[ \sin \left( \frac{k}{H} \right) - \frac{k}{H} \, \cos \left( \frac{k}{H} \right) \right]^2  \, {\rm ln }^2 \left( \frac{\sqrt{\dot{m}_*}}{H} \right) \;\;\;,\;\;\; k \ll \sqrt{\dot{m}_*}
\end{equation}
while the result is exponentially suppressed at higher momenta.

\end{document}